\documentclass{jnmp}

\usepackage{amsmath}
\usepackage{graphicx}

\JNMPnumberwithin{equation}{section}

\newtheorem{proposition}{Proposition}[section]

\theoremstyle{definition}
\newtheorem{remark}{Remark}[section]

\begin{document}

\renewcommand{\evenhead}{F~Calogero, J-P Fran\c{c}oise and M Sommacal}
\renewcommand{\oddhead}{Periodic Solutions of a Many-Rotator Problem in the Plane. II}

\thispagestyle{empty}

\FirstPageHead{10}{2}{2003}{\pageref{calogero-firstpage}--\pageref{calogero-lastpage}}{Article}

\copyrightnote{2003}{F~Calogero, J-P Fran\c{c}oise and M Sommacal}

\Name{Periodic Solutions of a Many-Rotator Problem\\
 in the Plane. II. Analysis of Various Motions}
\label{calogero-firstpage}

\Author{F~CALOGERO~$^{\dag\ddagger}$, J-P FRAN\c{C}OISE~$^{*}$
and M~SOMMACAL~$^{\#}$}

\Address
{$^{\dag}$~Dipartimento di Fisica, Universit\`a di Roma ``La Sapienza'', 00185
Roma, Italy\\
$\phantom{^{\dag^2}}$~E-mail: francesco.calogero@uniroma1.it\\[10pt]
$^{\ddagger}$~Istituto Nazionale di Fisica Nucleare, Sezione di Roma, Italy\\
$\phantom{^{\dag^1}}$~E-mail: francesco.calogero@roma1.infn.it\\[10pt]
$^{*}$~GSIB, 175 Rue du Chevaleret, Universit\'e Paris VI, 75013 Paris, France\\
$\phantom{^{\dag^3}}$~E-mail: jpf@ccr.jussieu.fr\\[10pt]
$^{\#}$~Dipartimento di Fisica, Universit\`a di Roma ``La Sapienza'', 00185
Roma, Italy\\
$\phantom{^{\dag^4}}$~E-mail: sommacal@sissa.it}

\Date{Received May 13, 2002;
Accepted August 2, 2002}

\begin{abstract}
\noindent
Various solutions are displayed and analyzed (both analytically
and numerically) of a~recently-introduced many-body problem in the
plane which includes both integrable and nonintegrable cases
(depending on the values of the coupling constants); in particular
the origin of certain periodic behaviors is explained. The light
thereby shone on the connection among \textit{integrability} and
\textit{analyticity} in (complex) time, as well as on the
emergence of a \textit{chaotic} behavior (in the guise of a
sensitive dependance on the initial data) not associated with any
local exponential divergence of trajectories in phase space, might
illuminate interesting phenomena of more general validity than for
the particular model considered herein.
\end{abstract}

\section{Introduction}

Recently it was pointed out [1, 2] that the $N$-body problem in the plane
characterized by the following Newtonian equations of motion possesses lots
of \textit{completely periodic} solutions:
\begin{subequations}
\begin{gather}
\ddot {\vec {r}}_{n} = \omega \hat {k} \wedge \dot {\vec {r}}_{n}
+ 2 \sum\limits_{m = 1;\,m \ne n}^{N} (r_{nm})^{-2}
( \alpha_{nm} + \alpha'_{nm} \hat{k} \wedge )\nonumber\\
\label{eq1.1}
\qquad{}\times \left[ \dot{\vec{r}}_{n} ( \dot{\vec{r}}_{m} \cdot \vec{r}_{nm} )
+ \dot{\vec{r}}_{m} ( \dot{\vec{r}}_{n} \cdot \vec{r}_{nm} )
- \vec{r}_{nm} ( \dot{\vec{r}}_{n} \cdot \dot{\vec{r}}_{m} ) \right].
\end{gather}
Here and below the subscripted indices run from $1$ to $N$ (unless otherwise
indicated), the number of moving particles $N$ is a positive integer, the
$N$ two-vectors $\vec {r}_{n} \equiv \vec {r}_{n} \left( {t} \right)$
identify the positions of the moving point-particles in a plane which for
notational convenience we imagine immersed in ordinary three-dimensional
space, so that $\vec {r}_{n} \equiv \left( {x_{n} ,y_{n} ,0} \right)$; $\hat
{k}$ is the unit three-vector orthogonal to that plane, $\hat {k} \equiv
\left( {0,0,1} \right)$, so that $\hat {k} \wedge \vec {r}_{n} \equiv \left(
{ - y_{n} ,x_{n} ,0} \right)$;
\begin{equation}
\label{eq1.2}
\vec {r}_{nm} \equiv \vec {r}_{n} - \vec {r}_{m} ,\qquad
r_{nm}^{2} = r_{n}^{2} + r_{m}^{2} - 2\vec {r}_{n} \cdot \vec {r}_{m} \mbox{ ;}
\end{equation}
\end{subequations}
superimposed dots denote of course time derivatives; $\omega $ is a
\textit{real} -- indeed, without loss of generality, \textit{positive} --
constant, $\omega > 0$, which sets the time scale, and to which we associate
the period
\begin{equation}
\label{eq1.3}
T = 2\pi /\omega
\end{equation}
(when discussing below specific examples we will conveniently set $\omega =
2\pi $, so that $T = 1$; but for the moment let us not do that); and
$\alpha _{nm} $ and ${\alpha} '_{nm} $ are \textit{real} ``coupling
constants''. Note that this $N$-body model in the plane is invariant under
both translations and rotations; as we shall see below, it is Hamiltonian;
it features one-body and two-body velocity-dependent forces; when the latter
are missing -- namely, when all the two-body coupling constants $\alpha
_{nm}$, ${\alpha} '_{nm} $ vanish,
\begin{subequations}
\begin{equation}
\label{eq1.4}
\alpha _{nm} = {\alpha} '_{nm} = 0,
\end{equation}
it represents the physical situation of $N$ electrically charged particles
moving in a plane under the influence of a constant magnetic field
orthogonal to that plane (``cyclotron''). In such a case of course every
particle moves uniformly, with the same period $T$, see (\ref{eq1.3}), on a circular
trajectory the center $\vec {c}_{n} $ and radius $\rho _{n} $ of which are
determined by the initial data (position and velocity) of each particle:
\begin{gather}
\label{eq1.5}
\vec {r}_{n} \left( {t} \right) = \vec {c}_{n} + \vec {\rho} _{n}
\sin\left( {\omega t} \right) - \hat {k} \wedge \vec {\rho} _{n}
\cos\left( {\omega t} \right),\\
\label{eq1.6}
\vec {c}_{n} = \vec {r}_{n} \left( {0} \right) + \hat {k} \wedge \vec {\rho}_{n},\qquad
\vec {\rho} _{n} = \dot {\vec {r}}_{n} \left( {0} \right)/\omega.
\end{gather}
\end{subequations}

Let us begin by tersely reviewing previous findings [1, 2]. First of all we
note that it is convenient to identify the \textit{physical} plane in which
the motion takes place with the \textit{complex} plane, via the relation
\begin{equation}
\label{eq1.7}
\vec {r}_{n} \equiv \left( {x_{n} ,y_{n} ,0} \right) \ \Leftrightarrow \ z_{n}
\equiv x_{n} + iy_{n},
\end{equation}
whereby the real Newtonian equations of motion in the plane (1.1) become the
following equations determining the motion of the $N$ points $z_{n} \equiv
z_{n} \left( {t} \right)$ in the complex $z$-plane:
\begin{subequations}
\begin{equation}
\label{eq1.8}
\ddot {z}_{n} = i\omega \dot {z}_{n} + 2\sum\limits_{m = 1;\,m \ne
n}^{N} {} a_{nm} \dot {z}_{n} \dot {z}_{m} /\left( {z_{n} - z_{m}}
\right),
\end{equation}
with
\begin{equation}
\label{eq1.9}
a_{nm} = \alpha _{nm} + i\alpha'_{nm} .
\end{equation}
\end{subequations}
Hereafter we always use this \textit{avatar}, (1.5), of the equations of
motion (1.1), and we moreover exploit the following key observation
[3, 4, 1, 2]: via the change of (independent) variable
\begin{equation}
\label{eq1.10}
\underline {z} \left( {t} \right) = \underline {\zeta}  \left( {\tau}  \right) ,\qquad
\tau = \left[ {\exp\left( {i\omega t} \right) - 1} \right]/\left(
{i\omega}  \right),
\end{equation}
the system (1.5) becomes
\begin{equation}
\label{eq1.11}
{\zeta} ''_{n} = 2\sum\limits_{m = 1;\,m \ne n}^{N} {} a_{nm} {\zeta}'_{n}
{\zeta} '_{m} /\left( {\zeta _{n} - \zeta _{m}}  \right).
\end{equation}
Here and below the underlined notation indicates an $N$-vector, say
$\underline{z} \equiv (z_{1},\ldots,z_{N})$, $\underline{\zeta} \equiv (\zeta_{1},
\ldots,\zeta_{N})$
and so on, and the primes denote of
course differentiations with respect to the independent variable $\tau $.
Note that the constant $\omega $ has completely disappeared from~(\ref{eq1.11}); nor
does it feature in the relations among the initial data for (1.5) and (\ref{eq1.11}),
which read simply
\begin{equation}
\label{eq1.12}
\underline {z} \left( {0} \right) = \underline {\zeta}  \left( {0} \right),\qquad
\underline {\dot {z}} \left( {0} \right) = \underline {{\zeta} '} \left( {0} \right).
\end{equation}

This of course entails that, to obtain the solution $\underline {z} \left(
{t} \right)$ of (1.5) corresponding to a given set of initial data
$\underline {z} \left( {0} \right)$, $\underline {\dot {z}} \left( {0}
\right)$, one can solve (\ref{eq1.11}) with the \textit{same} set of initial data,
see~(\ref{eq1.12}), thereby determine $\underline {\zeta}  \left( {\tau}  \right)$,
and then use (\ref{eq1.10}) to obtain $\underline {z} \left( {t} \right)$ (hence as
well, via (\ref{eq1.7}), the solution of the initial-value problem for (1.1)). This
possibility -- to infer the behavior of the evolution in the \textit{real
}time variable $t$ of the solutions $\underline {z} \left( {t} \right)$ of
(1.5) (namely as well of the solutions of the physical many-body problem in
the plane (1.1) ) from the properties of the solutions $\underline {\zeta}
\left( {\tau}  \right)$ of (\ref{eq1.11}) as functions of the \textit{complex}
variable $\tau $ -- is, as we will see, the main tool of our investigation.
Indeed, when the \textit{real} time variable $t$ evolves over one period $T$
-- say, from $t = 0$ to $t = T$ -- the \textit{complex} time-like variable
$\tau $ goes from $\tau = 0$ back to $\tau = 0$ by traveling
counter-clock-wise -- see (\ref{eq1.10}) -- on the circular contour~$\tilde {C}$
centered in the complex $\tau $-plane at $i/\omega $ and having radius
$1/\omega $. Hence whenever all the functions $\zeta _{n} \left( {\tau}
\right)$ -- obtained as solutions of (\ref{eq1.11}) -- are \textit{holomorphic}, as
functions of the complex variable $\tau $, inside that circular contour
$\tilde {C}$ -- or, more precisely, in the (closed) circular disk $C$
defined by that contour $\tilde {C}$ -- the corresponding functions $z_{n}
\left( {t} \right)$ -- namely the corresponding solutions of (1.5) -- are
\textit{nonsingular} and \textit{completely periodic} in the \textit{real}
time variable $t$, with period $T$, see (\ref{eq1.3}). (This also entails that, if
the only singularities of a solution $\underline {\zeta}  \left( {\tau}
\right)$ of (\ref{eq1.11}) inside the disk $C$ are a \textit{finite} number of
\textit{rational} branch points, then the corresponding solution $\underline
{z} \left( {t} \right)$ of (1.5) is again \textit{completely periodic} in
the \textit{real} time~$t$, albeit with a larger period which is then an
entire multiple of $T$ -- we shall discuss in detail this important point
below).

Now we note that \textit{all} solutions $\underline {\zeta}  \left( {\tau}
\right)$ of (\ref{eq1.11}), corresponding to \textit{arbitrary} initial data
$\underline {\zeta}  \left( {0} \right)$,
$\underline {{\zeta} '} \left( {0}
\right)$ (with the only restriction that these data be nonsingular, namely
$\left| {{\zeta} '_{n} \left( {0} \right)} \right| < \infty $ and $\zeta
_{n} \left( {0} \right) \ne \zeta _{m} \left( {0} \right)$, see the
right-hand side of (\ref{eq1.11})), yield solutions $\underline {\zeta}  \left( {\tau
} \right)$ which are \textit{holomorphic} in the neighborhood of $\tau = 0$,
as implied by the standard theorem which guarantees the existence,
uniqueness and analyticity of the solutions of analytic ODEs, in a
sufficiently small circular disk $D$ centered in the complex $\tau $-plane
at the origin, $\tau = 0$, where the initial data defining the solution are
given. The size of this disk $D$ is determined by the location of the
singularity of $\underline {\zeta}  \left( {\tau}  \right)$ closest to the
origin, and the structure of the right-hand side of (\ref{eq1.11}) clearly entails
that a \textit{lower} estimate of this distance -- namely of the radius
$\rho $ of $D$ -- reads as follows:
\begin{subequations}
\begin{equation}
\label{eq1.13}
\rho > R\tilde {\zeta} /\tilde {{\zeta} '}
\end{equation}
with $R$ a positive constant (dependent on the values of the coupling
constants $a_{nm} $ but not on the initial data) and $\tilde {\zeta} $
respectively $\tilde {{\zeta} '}$ providing \textit{lower} respectively
\textit{upper} estimates of the moduli of $\zeta _{n} \left( {0} \right) -
\zeta _{m} \left( {0} \right)$ respectively ${\zeta} '_{n} \left( {0}
\right)$, say
\begin{gather}
\label{eq1.14}
\tilde {\zeta}  = \min\limits_{n,m = 1,\ldots,N;\, n \ne m} \left| {\zeta
_{n} \left( {0} \right) - \zeta _{m} \left( {0} \right)} \right| ,\\
\label{eq1.15}
{\tilde {\zeta} }' = \max\limits_{n = 1,\ldots,N} \left| {{\zeta
}'_{n} \left( {0} \right)} \right|
\end{gather}
\end{subequations}
(for a derivation of this formula, including an explicit expression for $R$,
see Appendix~A). Let us now assume the initial data, see (\ref{eq1.12}), to entail
(via (1.9)) that $\rho > 2/\omega $. Then the disk $D$ includes the disk
$C$, and this entails that the solutions $\underline {\zeta}  \left( {\tau}
\right)$ of (\ref{eq1.11}) are \textit{holomorphic} functions of the \textit{complex}
variable $\tau $ in the (closed) disk $C$, hence (see (\ref{eq1.10})) the
corresponding solutions $\underline {z} \left( {t} \right)$ of (1.5) are
\textit{completely periodic} functions of the \textit{real} variable $t$,
with period $T$, see (\ref{eq1.3}).

This observation, together with the \textit{lower} estimate (1.9) of $\rho
$, imply the existence of a set, of \textit{nonvanishing} (in fact,
\textit{infinite}) measure in phase space, of initial data $\underline {z}
\left( {0} \right)$, $\underline {\dot {z}} \left( {0} \right)$ which yield
\textit{completely periodic} solutions of (1.5) (hence as well of (1.1)).
But before formulating this finding [1, 2] in the guise of the following
Proposition~\ref{prop-1.2}, let us interject the following obvious

\begin{remark} \label{remark-1.1} If $\underline {z} \left( {t} \right)$ is the solution
of (1.5) corresponding to initial data $\underline {z} \left( {0}
\right)$, $\underline {\dot {z}} \left( {0} \right)$, then $\underline
{\tilde {z}} \left( {t} \right) = c\underline {z} \left( {bt} \right)$
is the solution of the equations of motion that obtain from (1.5) by
replacing in it $\omega $ with $\tilde {\omega}  = b\omega $, and of
course the corresponding initial data read $\underline {\tilde {z}} \left(
{0} \right) = c\underline {z} \left( {0} \right)$, $\underline {\dot
{\tilde {z}}} \left( {0} \right) = bc\underline {\dot {z}} \left( {0}
\right)$. Here $b$, $c$ are of course arbitrary (nonvanishing!) rescaling
constants.
\end{remark}

Let us now formulate Proposition~\ref{prop-1.2} (correcting thereby the
partially incorrect formulation of this result given in Refs. [1, 2]).

\setcounter{proposition}{1}
\begin{proposition}\label{prop-1.2}
Let $\underline {z} \left( {t} \right)$ be the
solution of (1.5) with
\begin{subequations}
\begin{equation}
\label{eq1.16}
\omega = b\bar {\omega},
\end{equation}
corresponding to the assigned initial data
\begin{equation}
\underline{z} \left( {0} \right) = c\underline {u},
\qquad \underline {\dot {z}} \left( {0} \right) = \mu \underline {v} ,
\qquad \left[ \mbox{with} \ \ u_{n} \ne u_{m} \ \ \mbox{if } \ \ n \ne m \right] ,
\end{equation}
\end{subequations}
where the \textit{positive} numbers $b$, $c$, $\mu$ play the
role of scaling parameters (as we shall immediately see). Then the solution
$\underline {z} \left( {t} \right)$ is \textit{completely periodic} with
period $T$, see (\ref{eq1.3}),
\begin{equation}
\label{eq1.17}
\underline {z} \left( {t + T} \right) = \underline {z} \left( {t} \right),
\end{equation}
provided one of the following conditions hold:

\textit{(i)} for given $a_{nm}$, $\omega$, $\underline {z} \left( {0}
\right)$ and $\underline {v}$, the scaling number $\mu ,$ hence as well
the initial velocities $\dot {z}_{n} \left( {0} \right)$, are sufficiently
small: $0 \le \mu \le \mu _{c} $, where $\mu _{c} $ is a \textit{positive}
number, $\mu _{c} > 0$, whose value depends on the given quantities;

\textit{(ii)} for given $a_{nm}$, $\omega$, $\underline {\dot {z}}
\left( {0} \right)$ and $\underline {u}$, the scaling number $c$
is sufficiently large, $c > c_{c}$ (hence the initial positions of
the $N$ particles in the plane are sufficiently well separated),
with $c_{c}$ a \textit{positive} number, $c_{c} > 0$, whose value
depends on the given quantities;

\textit{(iii)} for given $a_{nm}$, $\bar{\omega} $, $\underline
{z} \left( {0} \right)$ and $\underline {\dot {z}} \left( {0}
\right)$, the scaling number $b$, hence as well the circular
frequency $\omega = b\bar {\omega }$, is sufficiently large, $b
> b_{c} $, where $b_{c} $ is a \textit{positive} number, $b_{c} >
0$, whose value depends on the given quantities.
\end{proposition}

\setcounter{remark}{2}
\begin{remark} \label{remark-1.3} The first two formulations (items \textit{(i)} and
\textit{(ii)}) of Proposition~\ref{prop-1.2} refer to the \textit{same}
equations of motion, with modified (rescaled) initial conditions; the third
formulation (item \textit{(iii)}) refers to \textit{different} equations of
motion (due to the change via rescaling of the constant $\omega $) with the
\textit{same} initial conditions. But in fact these 3 formulations are
\textit{completely equivalent}, see~Remark~\ref{remark-1.1}.
\end{remark}

We have thereby reviewed (and rectified!) the results of References~[1,~2],
to an extent sufficient to make this paper self-contained. But before
focusing on the specific, novel, contributions of this paper three
additional facts are now recalled. The first two are relevant to the case --
to which attention will be hereafter restricted -- in which the (possibly
complex) coupling constants $a_{nm} $ depend \textit{symmetrically} on their
two indices:
\begin{equation}
\label{eq1.18}
a_{nm} = a_{mn}.
\end{equation}
Then the equations of motion (1.5) are Hamiltonian, being derivable in the
standard manner from the Hamiltonian
\begin{equation}
\label{eq1.19}
H\left( {\underline {z} ,\underline {p}}  \right) = \sum\limits_{n = 1}^{N}
 \left\{ i\omega z_{n} /c + \exp\left( {cp_{n}}
\right)\prod\limits_{m = 1;\,m \ne n}^{N} {} \left[ {z_{n} - z_{m}}
\right]^{ - a_{nm}}  \right\}
\end{equation}
(and let us recall that this entails that the ``physical'' equations of motion
(1.1) are as well Hamiltonian [2]). Note the presence, in this expression of
the Hamiltonian function $H( \underline{z}, \underline{p} )$,
of the arbitrary (nonvanishing) constant $c$, which does not
feature in the equations of motion (1.5). Also note that $H( \underline{z}, \underline{p} )$,
in contrast to the equations of
motion (1.5), is \textit{not} quite invariant under translations ($z_{n} \to
\tilde {z}_{n} = z_{n} + z_{0} $), although the only effect of such a
translation on $H( \underline{z}, \underline{p} )$ is
addition of a constant.

Moreover, as clearly implied by the equations of motion (1.5) with (\ref{eq1.18}),
the center of mass,
\begin{equation}
\label{eq1.20}
Z\left( {t} \right) = N^{ - 1}\sum\limits_{n = 1}^{N} {} z_{n} \left( {t} \right) ,
\end{equation}
moves \textit{periodically} (with period $T$, see (\ref{eq1.3})) on a circular
trajectory (in the complex $z$-plane):
\begin{equation}
\label{eq1.21}
Z\left( {t} \right) = Z\left( {0} \right) + \dot {Z}\left( {0}
\right)\left[ {\exp\left( {i\omega t} \right) - 1} \right]/\left(
{i\omega}  \right).
\end{equation}

The third and last fact we like to recall is that, if all the coupling
constants in (1.5) are unity,
\begin{equation}
\label{eq1.22}
a_{nm} = 1,
\end{equation}
then the equations of motion (1.5) are \textit{integrable} indeed
\textit{solvable} [5,~2], the solution of the corresponding initial-value
problem being given by the following neat prescription: \textit{the}
$N$ \textit{coordinates} $z_{n} \left( {t} \right)$ \textit{which
constitute the solution} $\underline {z} \left( {t} \right)$
\textit{of the equations of motion (1.5) corresponding to the initial data} $\underline {z}
\left( {0} \right)$, $\underline {\dot {z}} \left( {0} \right)$
\textit{are the} $N$ \textit{roots of the following algebraic equation in the variable}
$z$\textit{:}
\begin{equation}
\label{eq1.23}
\sum\limits_{m = 1}^{N} \dot {z}_{m} \left( {0} \right)/\left[ {z -
z_{m} \left( {0} \right)} \right] = i\omega /\left[ {\exp\left(
{i\omega t} \right) - 1} \right] .
\end{equation}
Note that, after elimination of all denominators, this is a polynomial
equation of degree~$N$ for the variable $z$, with all coefficients of the
polynomial \textit{periodic} in $t$ with period~$T$, see~(\ref{eq1.3}). Hence the
set $\underline {z} \left( {t} \right)$ of its $N$ zeros is as well
\textit{periodic} with period $T$. It is therefore clear that, in this
special \textit{solvable} case, see~(\ref{eq1.22}), \textit{all} nonsingular
solutions of~(1.5) are \textit{completely periodic}, with a period that is
either $T$ or an integer multiple of $T$ (the latter possibility arises
because the zeros of the polynomial -- which is itself periodic with
period~$T$~-- may get reshuffled through the motion -- but of course that integer
multiple cannot exceed $N!$ -- indeed it cannot exceed the combinatorial
factor $p\left( {N} \right)$ defined as the smallest integer $p\left( {N}
\right)$ such that the iteration $p\left( {N} \right)$ times of any
permutation of $N$ objects yield unity, and of course $p\left( {N} \right) \ll N!$ for large $N$).

As we just saw, Proposition \ref{prop-1.2} entails that the equations of
motion (1.5) (as well~as the equivalent ``physical'' equations of motion (1.1))
possess a lot of \textit{nonsingular} and \textit{completely periodic}
solutions. This finding~[1,~2] was (of course!) validated by simulations
performed via a computer program created by one of us (MS) to solve
numerically the equations of motion (1.5). The new findings discussed in
this paper emerged from the (rather successful!) effort to understand
certain remarkable features of these numerical simulations -- in particular
the existence, in the cases characterized by \textit{real} and
\textit{rational} values of the coupling constants $a_{nm} $ in (1.5), of
\textit{additional} sets of initial data, also of \textit{nonvanishing}
measure in phase space, yielding \textit{nonsingular} and
\textit{completely periodic} solutions with periods which are
\textit{integer multiples} of $T$, see (\ref{eq1.3}) -- via a mechanism, at which we
already hinted above, associated with \textit{rational} branch points of the
solutions $\underline {\zeta}  \left( {\tau}  \right)$, which is indeed also
responsible in the \textit{solvable} case (with $a_{nm} = 1$) to yield
\textit{completely periodic} solutions with such larger periods (integer
multiples of~$T$). These new findings also suggest the existence of other
\textit{integrable} -- but presumably not \textit{solvable} -- cases of the
many-body problem (1.1), as well as of \textit{nonintegrable} cases for
which however \textit{integrable behaviors}, such as those associated with
\textit{completely periodic} motions, emerge from sectors of initial data
having nonvanishing measure in phase space; and they display as well a
mechanism for the emergence of a \textit{chaotic behavior} (for other
regions of phase space, of course only in the \textit{nonintegrable} cases)
characterized by a sensitive dependence on the initial data (the signature
of chaos!) which is \textit{not} associated with any local exponential
divergence of trajectories over time -- in analogy to the type of
\textit{chaotic behavior} that ensues when a particle moves \textit{freely}
(except for the reflections on the borders) inside, say, a triangular plane
billiard whose angles are \textit{irrational} fractions of~$\pi$.

These findings shall be reviewed at the end of this paper, after they have
been discussed in the body of it. In particular, in the following Section~2
we discuss the \textit{conservation laws} associated with the equations of
motion~(\ref{eq1.11}); in Section~3 we exhibit certain \textit{similarity solutions}
of these equations of motion; in Section~4 we discuss the solution of the
\textit{two-body problem}, in somewhat more detail than it was done in~[1];
in Section~5 we investigate the \textit{analytic structure} of the solutions
of (\ref{eq1.11}), in particular the nature of the branch points, in the complex
$\tau $-plane, featured by the solutions of these equations of motion, and
the structure of the associated Riemann surfaces, which is crucial to
determine the behavior~-- be it \textit{completely periodic} or
\textit{chaotic} -- of the solutions of the ``physical'' equations of motion
(1.5) namely (1.1) as the \textit{real} time $t$ unfolds. The relevance of
these findings for the motions of the ``physical'' problem (1.5) or (1.1) are
then discussed, and also illustrated via the display of numerical
simulations, in Section~6. But -- to avoid this paper becoming excessively
long -- the treatments of Sections~5 and~6 are restricted to the case in
which the real parts of the coupling constants $a_{nm} $ are
\textit{nonnegative},
\begin{equation}
\label{eq1.24}
\mbox{Re}\left( {a_{nm}}  \right) \ge 0,
\end{equation}
a condition which is \textit{sufficient} (albeit not
\textit{necessary}; see below, in particular (5.8)) to exclude
the phenomenon of ``escape to infinity''; the general case shall be
treated in a~subsequent paper. A final Section~7 outlines how the
study of this model can and should be pursued. To avoid
interrupting the flow of the discourse certain technical
developments are relegated to four Appendices.

\section{Conserved quantities}

The equations of motion (1.7) are of course (see (1.13)) implied by the
Hamiltonian
\begin{equation}
\label{eq2.1}
H( \underline{\zeta}, \underline{\pi} ) = \sum\limits_{n =
1}^{N} {} \left\{ \exp\left( {c\pi _{n}}
\right)\prod\limits_{m = 1;\, m \ne n}^{N} {} \left[ {\zeta _{n} - \zeta _{m}
} \right]^{ - a_{nm}} \right\} .
\end{equation}
Indeed this yields the Hamiltonian equations of motion
\begin{subequations}
\begin{gather}
\label{eq2.2}
{\zeta} '_{n} = \partial H/\partial \pi _{n} = c\exp\left( {c\pi
_{n}}  \right)\prod\limits_{m = 1;\,m \ne n}^{N}  \left[ {\zeta _{n} -
\zeta _{m}}  \right]^{ - a_{nm}},\\
\label{eq2.3}
{\pi} '_{n} = - \partial H/\partial\zeta _{n} = c^{-
1}\sum\limits_{m = 1;\,m \ne n}^{N} { \left( {\zeta}'_{n} + {\zeta}'_{m} \right)
\left( \zeta_{n} - \zeta_{m} \right) },
\end{gather}
\end{subequations}
where to write in more convenient form the second set of equations, (\ref{eq2.3}),
we used the first set, (\ref{eq2.2}) (and we of course assumed validity of the
symmetry property (1.12)). Clearly logarithmic differentiation of (\ref{eq2.2})
yields
\begin{equation}
\label{eq2.4}
{\zeta} ''_{n} /{\zeta} '_{n} = c{\pi} '_{n} - \sum\limits_{m = 1;\,m
\ne n}^{N} a_{nm} \left( {{\zeta} '_{n} - {\zeta} '_{m}}
\right)/\left( {\zeta _{n} - \zeta _{m}}  \right) ,
\end{equation}
and these equations, via (\ref{eq2.3}), clearly yields the equations of motion (1.7).

It is thus seen, from (\ref{eq2.1}) and (\ref{eq2.2}), that
\begin{equation}
\label{eq2.5}
H = c^{- 1}\sum\limits_{n = 1}^{N} {\zeta} '_{n},
\end{equation}
hence in this case the fact that the Hamiltonian is a constant of the motion
coincides with the time-independence of the ``center-of-mass'' velocity,
\begin{equation}
\label{eq2.6}
\left( {d/d\tau}  \right)\sum\limits_{n = 1}^{N} {\zeta} '_{n} = 0,
\end{equation}
which is itself an immediate consequence of the equations of motion (1.7)
(indeed, their sum over the index $n$ from $1$ to $N$ implies the vanishing
of the right-hand side due to the antisymmetry under exchange of the two
dummy indices $n$, $m$ of the summand in the double sum -- since we always
assume (1.12) to hold).

A second conserved quantity, associated with the translation-invariant
character of the Hamiltonian (\ref{eq2.1}), is the total momentum $\Pi $,
\begin{equation}
\label{eq2.7}
\Pi = \sum\limits_{n = 1}^{N} \pi _{n},
\end{equation}
as it is clear by summing over the index $n$ from $1$ to $N$ the equations
of motion (\ref{eq2.3}).

By exponentiation and via (\ref{eq2.2}) this entails the following version of the
second conserved quantity:
\begin{equation}
\label{eq2.8}
K = \prod\limits_{n = 1}^{N} \left[ {\zeta} '_{n}
\prod\limits_{m = 1;\,m \ne n}^{N} \left( {\zeta _{n} - \zeta _{m}}
\right)^{a_{nm}} \right].
\end{equation}

The fact that this quantity, $K$, is a constant of motion can also be
verified as follows. Logarithmic differentiation of (\ref{eq2.8}) yields
\begin{subequations}
\begin{gather}
\label{eq2.9}
{K}'/K = \sum\limits_{n = 1}^{N} \left[{\zeta} ''_{n}
/{\zeta} '_{n} + \sum\limits_{m = 1;\,m \ne n}^{N} a_{nm} \left(
{{\zeta} '_{n} - {\zeta} '_{m}}  \right)/\left( {\zeta _{n} - \zeta _{m} }
\right) \right],\\
\label{eq2.10}
{K}'/K = \sum\limits_{n = 1}^{N} \left[{\zeta} ''_{n}
/{\zeta} '_{n} - 2\sum\limits_{m = 1;\,m \ne n}^{N} a_{nm} {\zeta
}'_{m} /\left( {\zeta _{n} - \zeta _{m}}  \right) \right] .
\end{gather}
\end{subequations}
The second of these equations, (\ref{eq2.10}), follows from the first, (\ref{eq2.9}), by
exchanging the role of the two dummy indices $n$ and $m$ in the first term
of the double sum (and of course by taking advantage of (1.12)). It is then
obvious, see (1.7), that ${K}'$ vanishes, namely that~$K$ is a constant of
motion.

\section{Similarity solutions}

In this section we report certain special solutions of the evolution
equations (1.7), hence as well, via (1.6), of the Newtonian equations of
motion (1.5) and (1.1). They read (obviously, up to arbitrary permutations)
\begin{subequations}
\begin{gather}
\label{eq3.1}
\zeta _{n} \left( {\tau}  \right) = c_{0} + c_{n} \left( {\tau - \tau _{b}}
\right)^{\Gamma} ,\qquad n = 1,2,\ldots,M ,\\
\label{eq3.2}
\zeta _{n} \left( {\tau}  \right) = C_{n} ,\qquad n = M + 1,M + 2,\ldots,N,
\end{gather}
\end{subequations}
with $M$ a positive integer, $2 \le M \le N$, the complex constant $\tau
_{b} $ arbitrary (it identifies the value of $\tau $ at which this solution
has, generally, a branch point; see (\ref{eq3.1}) and below), the constants $C_{n}
$ arbitrary (and largely irrelevant: particles that do not move, see (\ref{eq3.2}),
can be altogether ignored, since the structure of the right-hand side of
(1.7) entails that they neither feel nor cause any force), the constant
$c_{0} $ also arbitrary (reflecting the translation-invariant character of
the model), while the constants $\Gamma $ and $c_{n} $ are determined (the
latter up to an arbitrary common multiplicative factor) by the following
equations, which clearly obtain by inserting (3.1) in (1.7):
\begin{subequations}
\begin{gather}
\label{eq3.3}
\Gamma \left( {\Gamma - 1} \right) = - \Gamma ^{2}A,
\end{gather}
\end{subequations}
\begin{subequations}
\begin{gather}
\label{eq3.4}
A = - 2\sum\limits_{m = 1;\,m \ne n}^{M} {} a_{nm} c_{m} /(c_{n} - c_{m}),\qquad n = 1,2,\ldots,M.
\end{gather}
\end{subequations}

The first of these equations, (\ref{eq3.3}), entails
\setcounter{equation}{1}
\begin{subequations}
\begin{gather}\setcounter{equation}{1}
\label{eq3.5}
\Gamma = \left( {1 + A} \right)^{ - 1}
\end{gather}
\end{subequations}
(provided $A \ne - 1$; if instead $A = - 1$ the function $\left( {\tau -
\tau _{b}}  \right)^{\Gamma} $ in the right-hand side of (\ref{eq3.1}) must be
replaced by $\exp\left( \beta \tau  \right)$, with $\beta $ an arbitrary
nonvanishing constant; see below).

The $M$ equations (\ref{eq3.4}) determine $A$ (note the assumed independence of
this quantity from the index $n$) and, up to a common rescaling factor, the
$M$ constants $c_{m} $ with $m = 1,2,\ldots,M$. The expressions of these $M$
constants cannot be generally obtained (for arbitrary $a_{nm} $) in explicit
form from (\ref{eq3.4}), but this is instead possible for $A$, hence, via~(\ref{eq3.5}),
for $\Gamma $ (provided, as we always assume, the symmetry property (1.12)
of the coupling constants $a_{nm} $ hold). Indeed by replacing $c_{m} $ in
the numerator in the right-hand side of (\ref{eq3.4}) with $\left( {c_{m} - c_{n}}
\right) + c_{n} $ we get
\setcounter{equation}{2}
\begin{subequations}
\begin{gather}\setcounter{equation}{1}
\label{eq3.6}
A = 2\sum\limits_{m = 1;\,m \ne n}^{M} {} a_{nm} - 2\sum\limits_{m = 1;\,m
\ne n}^{M} {} a_{nm}c_{n}/(c_{n} - c_{m}),
\end{gather}
hence, by summing over $n$ from $1$ to $M$,
\begin{equation}
\label{eq3.7}
MA = 2\sum\limits_{m,n = 1;\, m \ne n}^{M} a_{nm} - MA.
\end{equation}

To obtain this equation, (\ref{eq3.7}), we exchanged the order of the second double
sum over~$n$ and $m$ in the right-hand side, we took advantage of the
symmetry property (1.12), and we used again (\ref{eq3.4}). From (\ref{eq3.7}) (and taking
again advantage of the symmetry property (1.12) so as to avoid any double
counting in the sum) we finally get
\begin{equation}
\label{eq3.8}
A = \left( {2/M} \right)\sum\limits_{n,m = 1;\, n > m}^{M}a_{nm}.
\end{equation}
\end{subequations}

In conclusion we see that the system (1.7), for any arbitrary choice of the
coupling constants $a_{nm} $, possesses the (exact) solution (3.1), with an
arbitrary choice of the positive integer $M$ in the range $2 \le M \le N$
and of the $N - M$ (complex) constants $C_{m} $ (for~$m$ in the range $M + 1
\le m \le N$), and with the $M$ constants $c_{n} $ (for $n$ in the range $1
\le n \le M$) determined, up to a common rescaling factor, by the system of
$M$ algebraic equations~(\ref{eq3.4}), where the constant $A$ is given (provided,
as we always assume, there holds the symmetry property (1.12)) by the simple
formula (\ref{eq3.8}). The (generally complex) constant~$\tau _{b} $ in (\ref{eq3.1}) is
also an arbitrary constant, and its significance is that, at $\tau = \tau
_{b} $, the $M$ particles labeled from $1$ to $M$ all ``collide'' if $\mbox{Re}\left(
{\Gamma}  \right) > 0$ or ``escape to infinity'' if $\mbox{Re}\left( {\Gamma}
\right) < 0$, yielding of course a singularity of the equations of motion,
as evidenced by the right-hand side of (1.7). Note that the solutions $\zeta
_{n} \left( {\tau}  \right)$ possess, at this point $\tau = \tau _{b} $,
a~branch point singularity, whose nature is characterized by the exponent
$\Gamma $, see (\ref{eq3.1}) and (\ref{eq3.5}) with (\ref{eq3.8}). Of course analogous solutions
exist for any arbitrary permutation of the $N$ indices $n$. (Note that above
we put inverted commas around the word ``collide'' or ``escape to infinity'' --
to underline that this is not, generally, a true ``physical'' phenomenon,
since it occurs at some \textit{complex} value of the~$\tau $ variable; see
below).

The corresponding solutions $\underline {z} \left( {t} \right)$ of the
equations of motion (1.5), hence as well (via (1.4)) of the Newtonian
equations of motion (1.1), obtain from these solutions $\underline {\zeta}
\left( {\tau}  \right)$ via~(1.6). Therefore when the complex constant $\tau
_{b} $ falls \textit{outside} the disk $C$ (centered, in the complex $\tau
$-plane, at $\tau = i/\omega $, and with radius $1/\omega $), these
solutions $\underline {z} \left( {t} \right)$ are \textit{nonsingular} and
\textit{completely periodic} with period $T$, see (1.2), while if the
complex point $\tau _{b} $ falls \textit{inside} the disk $C$, the solutions
$\underline {z} \left( {t} \right)$ are again \textit{nonsingular} but
generally \textit{not periodic}, unless the exponent $\Gamma $, see (\ref{eq3.1})
and (\ref{eq3.5}) with (\ref{eq3.8}) is \textit{rational}, $\Gamma = p/q$ (with~$p$ and
$q$ two coprime \textit{integers}, and, say, $q$ positive, $q \ge 1$), in
which case the solutions $\underline {z} \left( {t} \right)$ are again
\textit{completely periodic}, but with period $\tilde {T} = qT$ (as clearly
entailed by (\ref{eq3.1}) with (1.6)). In the special case in which the point $\tau
_{b} $ falls exactly on the circle $\tilde {C}$ that constitutes the boundary of
the disk $C$, so that the formula $\tau _{b} = \left[ {\exp\left( {i\omega
t_{c}}  \right) - 1} \right]/\left( {i\omega}  \right)$ defines
${\rm mod}\left( {T} \right)$ a~\textit{real} time $t_{c} $, then the solution
$\underline {z} \left( {t} \right)$ corresponds to a special Newtonian motion
in which, at the ``collision time''~$t_{c} $, $M$ particles simultaneously
collide (if $\mbox{Re}\left( {\Gamma}  \right) > 0$) or escape to infinity (if
$\mbox{Re}\left( {\Gamma}  \right) < 0$).

This analysis applies provided $A \ne - 1$ (see (\ref{eq3.8})), so that $\Gamma $
is well defined (see (\ref{eq3.5})). If instead $A = - 1$, the similarity solutions
$\underline {\zeta}  \left( {\tau}  \right)$ are entire functions of $\tau
$, since the expression (\ref{eq3.1}) is then replaced by the formula
\begin{equation}
\label{eq3.9}
\zeta _{n} \left( {\tau}  \right) = c_{n} \exp\left( {\beta \tau} \right),
\end{equation}
where $\beta $ is an arbitrary constant (just as $\tau _{b} $ is an
arbitrary constant in (\ref{eq3.1})). Hence in this case the corresponding
similarity solutions $\underline {z} \left( {t} \right)$ are always
\textit{nonsingular} and \textit{completely periodic} with period $T$, see
(1.2), as indeed clearly entailed by (\ref{eq3.9}) with (1.5).

The actual behavior of these similarity solutions $\underline {z} \left( {t}
\right)$ need not be discussed any further, nor explicitly displayed, since
the interested reader will have no difficulty to figure it out from (\ref{eq3.1})
(or (\ref{eq3.9})) and (1.6). Let us only note that, in the ``equal particle'' case in
which all the coupling constants $a_{nm} $ coincide,
\begin{subequations}
\begin{equation}
\label{eq3.10}
a_{nm} = a ,
\end{equation}
the expression (\ref{eq3.8}) yields
\begin{equation}
\label{eq3.11}
A = a\left( {M - 1} \right),
\end{equation}
hence (\ref{eq3.5}) yields (provided $A \ne - 1$; otherwise, see the previous
paragraph)
\begin{equation}
\label{eq3.12}
\Gamma = \left[ {1 + a\left( {M - 1} \right)} \right]^{- 1},
\end{equation}
and in this special case the constants $c_{n} $ can be explicitly exhibited,
\begin{equation}
\label{eq3.13}
c_{n} = c\exp\left( {2i\pi n/M} \right),
\end{equation}
\end{subequations}
consistently, see (\ref{eq3.4}) and (\ref{eq3.11}), with the \textit{identities}
\begin{gather}
\sum\limits_{m = 1;\,m \ne n}^{N} \exp ( 2i\pi m/M ) /
\left[ \exp ( 2i\pi n/M ) - \exp ( 2i\pi m/M ) \right] = - (M - 1)/2,\nonumber\\
\qquad \qquad \qquad
n = 1,2,\ldots,M.\label{eq3.14}
\end{gather}
The treatment of this special ``equal particle'' case is not new [4, 2].

Let us finally note that, even though in general the constants $c_{m} $
cannot be determined in explicit closed form in the general case with
arbitrary $N$ and different coupling constants~$a_{nm} $, whenever these
depend symmetrically from their two indices, see (1.12), there holds the sum
rule
\begin{equation}
\label{eq3.15}
\sum\limits_{n = 1}^{N} c_{n} = 0,
\end{equation}
which clearly obtains by multiplying (\ref{eq3.4}) by $c_{n} $ and summing over the
index $n$ from $1$ to $N$ (since the double sum in the right-hand side then
vanishes due to the antisymmetry of the summand). For $N = 2$ this sum rule
is sufficient to determine the constants $c_{n} $ (up to a common rescaling
constants), since it clearly entails $c_{1} = c$,
$c_{2} = - c$.

\section{The two-body problem}

In this section we provide a somewhat more complete treatment of the
two-body problem than given in [1]. This is of interest in itself, but even
more so for the insight it provides, not only for the two-body case but as
well for the $N$-body case (as discussed in the following Section~5), on the
nature of the singularities of the solutions $\underline {\zeta}  \left( {t}
\right)$ of (1.7) as functions of the complex variable $\tau $, hence on the
periodicity of the corresponding solutions $\underline {z} \left( {t}
\right)$ of the ``physical'' equations of motion (1.5).

For $N = 2$ the equations of motion (1.7) are consistent with the
assignment (corresponding to the standard separation of the center of mass
and relative motions)
\begin{subequations}
\begin{gather}
\label{eq4.1}
\zeta _{1} \left( {\tau}  \right) + \zeta _{2} \left( {\tau}  \right) =
\zeta _{1} \left( {0} \right) + \zeta _{2} \left( {0} \right) + V\tau ,\\
\label{eq4.2}
\zeta _{1} \left( {\tau}  \right) - \zeta _{2} \left( {\tau}  \right) =
\zeta \left( {\tau}  \right),
\end{gather}
namely
\begin{gather}
\label{eq4.3}
\zeta _{1} \left( {\tau}  \right) = \left[ {\zeta _{1} \left( {0} \right) +
\zeta _{2} \left( {0} \right) + V\tau + \zeta \left( {\tau}  \right)}
\right]/2 ,\\
\label{eq4.4}
\zeta _{2} \left( {\tau}  \right) = \left[ {\zeta _{1} \left( {0} \right) +
\zeta _{2} \left( {0} \right) + V\tau - \zeta \left( {\tau}  \right)}
\right]/2,
\end{gather}
\end{subequations}
where (see (1.14), (1.8) and (\ref{eq4.1}))
\begin{equation}
\label{eq4.5}
V = {\zeta} '_{1} \left( {0} \right) + {\zeta} '_{2} \left( {0} \right) =
{\zeta} '_{1} \left( {\tau}  \right) + {\zeta} '_{2} \left( {\tau}  \right)
= 2\dot {Z}\left( {0} \right)
\end{equation}
is a (generally complex) constant and the difference $\zeta \left( {\tau}
\right)$ satisfies the second-order ODE
\begin{equation}
\label{eq4.6}
{\zeta} '' = a\left[V^{2} - \left( {{\zeta} '}
\right)^{2} \right]/\zeta.
\end{equation}
Here and throughout this section $a = a_{12} = a_{21} $ is the relevant
``coupling constant'', and primes denote of course differentiations with
respect to $\tau $.

This ODE is easily integrated once (after multiplying it by the factor
$2{\zeta}'/[V^{2}-({\zeta}')^{2}]$), and one gets thereby
\begin{subequations}
\begin{gather}
\label{eq4.7}
\left( {{\zeta} '} \right)^{2} = V^{2} + B\zeta ^{ - 2a},\\
\label{eq4.8}
\left( {{\zeta} '} \right)^{2} = V^{2}\left[1 + \left( {\zeta
/L} \right)^{ - 2a} \right],
\end{gather}
\end{subequations}
with $B = - 4{\zeta} '_{1} \left( {0} \right){\zeta} '_{2} \left( {0}
\right)\left[ {\zeta \left( {0} \right)} \right]^{2a}$ -- or, when $V
\ne 0$, equivalently but notationally more conveniently, $L = \left(
{B/V^{2}} \right)^{1/\left( {2a} \right)} = \zeta \left( {0}
\right)\left\{ \left[ {{\zeta} '\left( {0} \right)/V}
\right]^{2} - 1\right\}^{1/\left( {2a} \right)}$ -- a~(generally complex) integration constant.

In the special case $V = 0$ the first-order ODE (\ref{eq4.7}) can be easily
integrated once more, and one obtains thereby [1,~2] the solution of (\ref{eq4.6}) in
closed form: for $a \ne - 1$,
\begin{subequations}
\begin{equation}
\label{eq4.9}
\zeta \left( {\tau}  \right) = c\left( {\tau - \tau _{b}}  \right)^{\gamma}
\end{equation}
\end{subequations}
with
\begin{equation}
\label{eq4.10}
\gamma = 1/\left( {1 + a} \right);
\end{equation}
for $a = - 1$,
\begin{subequations}
\begin{equation}
\label{eq4.11}
\zeta \left( {\tau}  \right) = c\exp\left( {\beta \tau}  \right).
\end{equation}
\end{subequations}
Here $c$, $\tau _{b} $ and $\beta $ are arbitrary (complex) constants, and it
is easily seen that $\tau _{b} $ respectively $\beta $ are related to the
initial data by the relations
\setcounter{equation}{4}
\begin{subequations}
\setcounter{equation}{1}
\begin{equation}
\label{eq4.12}
\tau _{b} = - \left( {1 + a} \right)^{ - a}\zeta \left( {0}
\right)/{\zeta} '\left( {0} \right)
\end{equation}
\end{subequations}
respectively
\setcounter{equation}{6}
\begin{subequations}
\setcounter{equation}{1}
\begin{equation}
\label{eq4.13}
\beta = {\zeta} '\left( {0} \right)/\zeta \left( {0} \right) .
\end{equation}
\end{subequations}
Note that this solution coincides with the similarity solution (3.1) (with
$N = M = 2$, $\gamma = \Gamma$).

For $V \ne 0$ the ODE (4.4) can of course be generally integrated by a
further quadrature, but the corresponding formula in terms of the
hypergeometric function $F\left( {A,B;C;Z} \right)$ [6],
\begin{equation}
\label{eq4.14}
\zeta F\left( {1/2, - 1/\left( {2a} \right);1 - 1/\left( {2a}
\right);\left[ {\zeta /L} \right]^{- 2a}} \right) = V\left( {\tau -
\tau _{b}}  \right) ,
\end{equation}
is not particularly enlightening, except in the special cases listed below
in which the hypergeometric function reduces to elementary functions
(actually in these cases direct integration of (4.4) is the neatest way to
get the solution, without going through the hypergeometric function).

For $a = - 1$, one easily finds
\begin{equation}
\label{eq4.15}
\zeta \left( {\tau}  \right) = L\sinh\left[ {\left( {V/L}
\right)\left( {\tau - \tau _{0}}  \right)} \right],
\end{equation}
with $\tau_{0} $ a (complex) constant (related of course
to the initial data: $\zeta(0)\!= \!-L\sinh[(V/L)\tau_{0}]).\!$ Hence in this case $\zeta \left(
{\tau} \right)$ is an entire function of $\tau $, and via (1.6)
this entails that \textit{all} solutions of the Newtonian
equations of motion (1.5) (with $N = 2$) are in this case
\textit{nonsingular} and \textit{completely periodic} with
period~$T$, see (1.2) (as entailed by (4.1) with (4.7) or
(\ref{eq4.15})).

For $a = 1$ (namely, for the special value of the coupling constant that
corresponds, in the equal-particle $N$-body problem, to the \textit{solvable}
case, see (1.17)), it is as well easy to get
\begin{subequations}
\begin{equation}
\label{eq4.16}
\zeta \left( {\tau}  \right) = V\left[ {\left( {\tau - \tau _{ +} }
\right)\left( {\tau - \tau _{ -} }  \right)} \right]^{1/2}
\end{equation}
with the two (complex) constants $\tau_{\pm}$ related to the initial
values by the formula
\begin{equation}
\label{eq4.17}
\tau _{ \pm}  = \left[ {\zeta _{2} \left( {0} \right) - \zeta _{1} \left(
{0} \right)} \right]\left\{\left[ {{\zeta} '_{1} \left( {0}
\right)} \right]^{1/2} \pm i\left[ {{\zeta} '_{2} \left( {0} \right)}
\right]^{1/2}\right\}^{2}/V^{2}.
\end{equation}
\end{subequations}
Note the square-root branch point in the right-hand side of (\ref{eq4.16}), and its
consistency with (4.5) and (\ref{eq4.10}) (of course with $a = 1$). It is easily
seen, via (1.6), that these findings entail the following results for the
solutions $z_{n} \left( {t} \right)$ of the Newtonian equations of motion~(1.5) (with $N = 2$ and $a = 1$). If $V = 0$ (namely, if the center of mass
does not move -- initially, hence as well throughout the motion, see
(1.15)), the solutions are \textit{nonsingular} and \textit{completely
periodic} with period $T$, see (1.2), if (the initial data entail, see
(\ref{eq4.12}) and (1.8), that) $\tau _{b} $ falls, in the complex $\tau $-plane,
\textit{outside} the circular disk $C$ (centered at $\tau = i/\omega $ and
of radius $1/\omega $); they are \textit{nonsingular} and \textit{completely
periodic} with period $\tilde {T} = 2T$ if $\tau _{b} $ instead falls
\textit{inside} the circular disk $C$; while they are \textit{singular} (due
to the occurrence of a two-body collision) at the time $t_{c} $ defined
$mod\left( {T} \right)$ by the formula $\tau _{b} = \left[ {\exp\left(
{i\omega t_{c}}  \right) - 1} \right]/\left( {i\omega}  \right)$, if
$\tau _{b} $ falls just on the boundary of $C$, namely on the circular
contour $\tilde {C}$ (which is just the condition necessary and sufficient
to entail that $t_{c} $, as defined above, be \textit{real}). Likewise, if
instead $V \ne 0$ (in which case the center of mass moves on a circular
orbit, see (1.15)), the solutions are \textit{nonsingular} and
\textit{completely periodic} with period $T$, see (1.2), if (the initial
data entail, see (\ref{eq4.17}) and (1.8), that) the two complex constants~$\tau _{
\pm}  $ fall either both \textit{inside} or both \textit{outside} the
circular disk $C$; they are \textit{nonsingular} and \textit{completely
periodic} with period $\tilde {T} = 2T$ if one of the two constants~$\tau
_{ \pm}  $ falls \textit{inside} the circular disk $C$ and the other
\textit{outside} it; while they are \textit{singular} (due to the occurrence
of a two-body collision) if $\tau _{ +}  $ or $\tau _{ -}  $ falls just on
the boundary of $C$, namely on the circular contour $\tilde {C}$. These
results of course confirm those entailed by the resolvent formula (1.17), as
described after that formula, see above. (If one considers initial data of
type (1.10b) with fixed $z_{n} \left( {0} \right) = \zeta _{n} \left( {0}
\right)$ and $\dot {z}_{n} \left( {0} \right) = {\zeta} '_{n} \left( {0}
\right) = \mu v_{n} $ entailing $V = \mu \left( {v_{1} + v_{2}}
\right)$ with fixed $v_{n} $, then (\ref{eq4.17}) yields $\tau _{ \pm}  \equiv \tau
_{ \pm}  \left( {\mu}  \right) = \mu ^{ - 1}\tau _{ \pm}  \left( {1}
\right)$, hence the motion is \textit{completely periodic} with period~$T$,
see (1.2), both for sufficiently small and for sufficiently large values of
the positive scaling parameter $\mu $; it is instead \textit{completely
periodic} with period $\tilde {T} = 2T$ for a finite intermediate interval
of values of $\mu $).

A third neatly solvable case obtains if $a = - 1/2$. Then
\begin{equation}
\label{eq4.18}
\zeta \left( {\tau}  \right) = L\left\{ \left[ {\left( {V/L}
\right)\left( {\tau - \tau _{0}}  \right)/2} \right]^{2} - 1 \right\},
\end{equation}
which clearly entails that $\zeta \left( {\tau}  \right)$ is \textit{entire}
(indeed, just a second-degree polynomial) in $\tau $. Note that the same
conclusion is implied in this case by (\ref{eq4.9}) with (\ref{eq4.10}). Hence we conclude
that in this case \textit{all} solutions of the Newtonian equations of
motion (1.5) (can be exhibited in explicit closed form and) are
\textit{nonsingular} and \textit{completely periodic} with period~$T$, see~(1.2).

The next two cases the solutions of which we report explicitly are
characterized by $a = - 2$ and $a = - 3$, respectively (the alert reader can
easily check these results, or derive them via the approach of Appendix~B).

For $a = - 2$ the solution reads
\begin{subequations}
\begin{equation}
\label{eq4.19}
\zeta \left( {\tau}  \right) = \lambda \zeta \left( {0} \right)\left\{
\mbox{cn}\left( {\left\{ {V\left( {\tau - \tau _{0}}
\right)/\left[ {\lambda \zeta \left( {0} \right)} \right]}
\right\},2^{- 1/2}} \right) \right\}^{- 1}.
\end{equation}
Here and below $\mbox{cn}\left( {u,k} \right)$ is the standard Jacobian elliptic
function (see for instance~[6]), and the two constants $\lambda$ and $\tau_{0}$
are related to the initial data as follows:
\begin{gather}
\label{eq4.20}
\lambda = 2^{- 1/4}\left[ {{\zeta} '\left( {0} \right)/V}
\right]^{ - 1/2},\\
\label{eq4.21}
\mbox{cn}\left( {\left\{ {V\tau _{0} /\left[ {\lambda \zeta \left( {0}
\right)} \right]} \right\},2^{- 1/2}} \right) = \lambda.
\end{gather}
\end{subequations}
These formulas entail that, in this case with $N = 2$ and $a =
a_{12} = a_{21} = - 2$, \textit{all} solutions $\zeta _{n} \left(
{\tau}  \right)$ of the equations of motion (1.7) are
\textit{meromorphic} functions of $\tau$ (note that this is also
the case if $V = 0$, because $a = - 2$ entails $\gamma = - 1$; see
(4.5) and (\ref{eq4.10})). Hence, in this case with $N = 2$ and $a
= a_{12} = a_{21} = - 2$, \textit{all nonsingular} solutions
$z_{n} \left( {t} \right)$ of the ``physical'' equations of motion
(1.5) or (1.1) are \textit{completely periodic} functions of the
time~$t$, with period $T$ , see (1.2).

For $a = - 3$ the solution reads
\begin{subequations}
\begin{gather}
\label{eq4.22}
\zeta \left( {\tau}  \right) = \lambda \zeta \left( {0} \right)\left\{
\wp \left( {\left\{ {V\left( {\tau - \tau _{0}}
\right)/\left[ {\lambda \zeta \left( {0} \right)} \right]}
\right\};0, - 4} \right) \right\}^{ - 1/2},\\
\label{eq4.23}
\zeta \left( {\tau}  \right) = V\left[\wp \left( {\left( {\tau
- \tau _{0}}  \right);0, - 4\left[ {V/\zeta \left( {0} \right)}
\right]^{6}\left\{ - 1 + \left[ {{\zeta} '\left( {0}
\right)/V} \right]^{2} \right\}} \right) \right]^{ -1/2}.
\end{gather}
Here and below $\wp \left( {u;g_{2} ,g_{3}}  \right)$ is the standard
Weierstrass elliptic function (see for instan\-ce~[6]), and the two constants
$\lambda $ and $\tau _{0} $ are related to the initial data as follows:
\begin{gather}
\label{eq4.24}
\lambda = \left\{ - 1 + \left[ {{\zeta} '\left( {0}
\right)/V} \right]^{2} \right\}^{ - 1/6},\\
\label{eq4.25}
\lambda ^{2}\wp \left( {\left\{ {V\tau _{0} /\left[ {\lambda
\zeta \left( {0} \right)} \right]} \right\};0, - 4} \right) = 1
\end{gather}
\end{subequations}
(the equivalence of (\ref{eq4.22}) and (\ref{eq4.23}) is entailed by the rescaling
properties of the Weierstrass function, see for instance eq.~14.13(2)
of~[6]). These formulas entail that, in this case with $N = 2$ and $a = a_{12}
= a_{21} = - 3$, the only singularities featured by the solutions
$\underline {\zeta}  \left( {\tau}  \right)$ of the equations of motion
(1.7) as functions of $\tau $ are branch points with exponent~$ - 1/2$ (note
that this is also the case if $V = 0$, because $a = - 3$ entails $\gamma = -
1/2$; see (4.5) and~(\ref{eq4.10})). Hence, in this case with $N = 2$ and $a = a_{12}
= a_{21} = - 3$, \textit{all} nonsingular solutions $\underline {z} \left(
{t} \right)$ of the ``physical'' equations of motion (1.5) or (1.1) are
\textit{completely periodic} functions of the time\textit{} $t$, with period
$T$, see (1.2), or $\tilde {T} = 2T$.

The last case we treat in detail is characterized by $a = - 1/4$. It can
then be shown (again, via simple calculations analogous to those carried out
in Appendix B) that in this case the general solution of (\ref{eq4.8}) (with $V \ne
0$) is given by the (implicit) formula
\begin{subequations}
\begin{equation}
\label{eq4.26}
\left[ {\zeta \left( {\tau}  \right)/L} \right]^{3/2} - 3\left[
{\zeta \left( {\tau}  \right)/L} \right] = \left( {9/16} \right)\left[
{\left( {V/L} \right)\left( {\tau - \tau _{b}}  \right)} \right]^{2}
- 3\left( {V/L} \right)\left( {\tau - \tau _{b}}  \right),
\end{equation}
entailing
\begin{equation}
\label{eq4.27}
\zeta \left( {\tau}  \right) = V\left( {\tau - \tau _{b}}
\right)\sum\limits_{l = 0}^{\infty}  \alpha _{l} \left[ {\left(
{V/L} \right)\left( {\tau - \tau _{b}}  \right)} \right]^{l/2}
\end{equation}
with
\begin{equation}
\label{eq4.28}
\alpha _{0} = \alpha _{1} = 1,\qquad \alpha _{2} = - 3/16, \quad \ldots.
\end{equation}
\end{subequations}
It is thus seen that, in this case with $a = - 1/4$, $\zeta \left( {\tau}
\right)$ has a \textit{square-root} branch point at $\tau = \tau _{b} $ if
the center of mass moves ($V \ne 0$), while it has a \textit{cubic-root}
branch point if the center of mass does not move ($V = 0$; indeed $a = -
1/4$ entails $\gamma = 4/3$, see (\ref{eq4.9}) and~(\ref{eq4.10})). Hence, in this case with
$a = - 1/4$, the solutions $z_{n} \left( {t} \right)$ of the physical
two-body problem (1.5) are \textit{nonsingular} and \textit{completely
periodic} with period $T$, see (1.2), if the branch point $\tau _{b} $
(whose location depends of course on the initial conditions, see (\ref{eq4.12}) or
(\ref{eq4.26})) fall \textit{outside} the circular disk $C$ (centered in the
complex $\tau $-plane at $\tau = i/\omega $ and of radius $1/\omega $); they
are of course \textit{singular} at a finite real time $t = t_{c} $
(when the two particles collide) if $\tau _{b} $ falls just on the boundary
of $C$; while if the branch point $\tau _{b} $ falls \textit{inside} the
circular disk $C$ they are \textit{nonsingular} and \textit{completely
periodic} with period $\tilde {T} = 2T$ or $\tilde {T} = 3T$ depending
whether the center of mass does ($V \ne 0$) or does not ($V = 0$) move.

As implied by our general treatment and exemplified by these examples, an
analogous analysis of the emergence of periodic motions in the two-body
problem (1.5) can also be made for an \textit{arbitrary} value of the
coupling constant~$a$. The conclusions of such an analysis depend
essentially on the nature, and location, of the branch points featured by
the solutions of the first-order ODE (4.4) (with $V \ne 0$; the $V = 0$ case
is completely illuminated by the explicit solution (4.5) with (\ref{eq4.10}), hence
our treatment below refers exclusively to the $V \ne 0$ case). A detailed
treatment of this question is given in Appendix~B, and it leads to the
following results.

Three cases must be distinguished, depending on the value of the real part
of the coupling constant $a$.

Case \textit{(i)}:
\begin{subequations}
\begin{equation}
\label{eq4.29}
\mbox{Re}\left( {a} \right) > 0.
\end{equation}
In this case the branch point at, say, $\tau = \tau _{b}
$, is characterized by the exponents $\gamma $ and $1 - \gamma $,
see (\ref{eq4.10}), with the behavior of $\zeta \left( {\tau}
\right)$ for $\tau \approx \tau _{b} $ given by the formula
\begin{gather}
\zeta(\tau) = L\gamma ^{-\gamma}
[(V/L)(\tau-\tau_{b})]^{\gamma}\nonumber\\
\label{eq4.30}\phantom{\zeta(\tau) =} {}\times
\left\{ 1 + \sum\limits_{l = 1}^{\infty} \sum\limits_{k =0}^{l} g_{kl}
[(V/L)(\tau-\tau_{b})]^{k\gamma}
[(V/L)(\tau-\tau_{b})]^{2l(1-\gamma)}
\right\}
\end{gather}
\end{subequations}
(for a justification of this formula, including the significance of the
coefficients $g_{kl} $, see (B.20) with (B.2)). Note that this formula
entails $\zeta \left( {\tau _{b}}  \right) = 0$ (as well as $\left| {{\zeta
}'\left( {\tau _{b}}  \right)} \right| = \infty $, since (\ref{eq4.10}) and (\ref{eq4.29})
entail $0 < \mbox{Re}\left( {\gamma}  \right) < 1$), namely this singularity is
associated with a ``collision'' of the two particles (see (\ref{eq4.2})), which both
move with infinite speed when they collide. This singularity is analogous to
that which obtains in the $V = 0$ case, see~(\ref{eq4.9}).

Case \textit{(ii)}:
\begin{subequations}
\begin{equation}
\label{eq4.31}
- 1 < \mbox{Re}\left( {a} \right) < 0.
\end{equation}
In this case the branch point at, say, $\tau = \tau _{b} $, is characterized
by the exponent
\begin{equation}
\label{eq4.32}
\beta = - 2a,
\end{equation}
with the behavior of $\zeta \left( {\tau}  \right)$ for $\tau \approx \tau_{b}$
given by the formula
\begin{equation}
\label{eq4.33}
\zeta \left( {\tau}  \right) = \pm V\left( {\tau - \tau _{b}}
\right)\left\{1 + \sum\limits_{l = 1}^{\infty}  {} g_{l}
\left[ { \pm \left( {V/L} \right)\left( {\tau - \tau _{b}}  \right)}
\right]^{l\beta} \right\}.
\end{equation}
This singularity, however, is generally \textit{not} of the same type as
that which obtains in the $V = 0$ case, see (\ref{eq4.9}) (except for those values
of $a$ such that $\gamma = \left( {1 + a} \right)^{ - 1}$ and $\beta = -
2a$ differ by an integer). In both cases ($V = 0$, see (\ref{eq4.9}) with (\ref{eq4.10}) and
(\ref{eq4.31}); $V \ne 0$, see (4.16)) $\zeta \left( {\tau _{b}}  \right) = 0$,
namely this singularity is again associated with a ``collision'' of the two
particles (see (\ref{eq4.2})); however, in the $V = 0$ case $\left| {{\zeta
}'\left( {\tau _{b}}  \right)} \right| = \infty $, while in the $V \ne 0$
case ${\zeta} '\left( {\tau _{b}}  \right) = \pm V$ and this entails (see
(4.1)) that either ${\zeta} '_{1} \left( {\tau}  \right)$ or ${\zeta} '_{2}
\left( {\tau}  \right)$ vanishes at $\tau = \tau _{b} $ so that
\begin{equation}
\label{eq4.34}
{\zeta} '_{1} \left( {\tau _{b}}  \right){\zeta} '_{2} \left( {\tau _{b}} \right) = 0.
\end{equation}
\end{subequations}

Case \textit{(iii)}:
\begin{equation}
\label{eq4.35}
\mbox{Re}\left( {a} \right) < - 1.
\end{equation}
In this case both behaviors, (\ref{eq4.30}) respectively (\ref{eq4.33}), are possible, so
both type of branch points, characterized by the exponents $\gamma = \left(
{1 + a} \right)^{ - 1}$ respectively $\beta = - 2a$, may be present: but
the first type of branch point, characterized by the exponent $\gamma $
(which obtains now for all values of $V$) corresponds now to the phenomenon
of ``escape to infinity'' ($\left| {\zeta \left( {\tau _{b}}  \right)} \right|
= \left| {{\zeta} '\left( {\tau _{b}}  \right)} \right| = \infty $), while
the second type of branch point, characterized by the exponent $\beta $
(which obtains only if $V$ does \textit{not} vanish, $V \ne 0$) corresponds
to the phenomenon of ``two-body collision'' ($\zeta \left( {\tau _{b}}
\right) = 0$, with ${\zeta} '\left( {\tau _{b}}  \right) = \pm V$ entailing
(\ref{eq4.34})).

The behaviors of the corresponding solutions, $\underline {z} \left( {t}
\right)$, of the ``physical'' equations of motion (1.5) depends on the
locations of the branch points $\tau _{b} $ in the complex $\tau $-plane,
which of course depend themselves on the initial data (and on the coupling
constant $a$). (Note that the above analysis is local, namely it applies in
the neighborhood of each branch point; nothing excludes that there be
several, or possibly an infinity, of them). If none of these branch points
is located inside the disk $C$ (centered in the complex $\tau $-plane at
$\tau = i/\omega $ and of radius $1/\omega $) nor on its boundary $\tilde
{C}$, then the solutions $z_{n} \left( {t} \right)$ are \textit{nonsingular}
and \textit{completely periodic} with period $T$, see (1.2) (irrespective of
the value of the coupling constant $a$). If instead one branch point, say at
$\tau = \tau _{b} $, falls inside the disk~$C$, then the motion is again
\textit{nonsingular} but generally \textit{not} periodic, unless the branch
point exponent is (\textit{real} and) \textit{rational}, in which case the
motion \textit{may} again be \textit{nonsingular} and \textit{completely
periodic} (see below for a justification of the conditional), but with a
larger period $\tilde {T} = qT$, where $q$ is the (positive) integer in
the denominator of the rational exponent characterizing the branch point,
which depends of course on the two-body ``coupling constant'', as discussed
above (indeed a necessary and sufficient condition for the branch point
exponent to be \textit{rational} is that the coupling constant $a$ be itself
\textit{rational} -- except for the special case $a = - 1$, when the motion,
see (4.7) and (\ref{eq4.15}), is \textit{always} nonsingular and completely periodic
with period $T$, see (1.2)). And if $\tau _{b} $ falls just on the boundary
of~$C$, namely on the circular contour $\tilde {C}$, then at a finite
\textit{real} time $t_{c} $ defined $mod\left( {T} \right)$ by the formula
$\tau _{b} = \left[ {\exp\left( {i\omega t_{c}}  \right) - 1}
\right]/\left( {i\omega}  \right)$ the ``physical'' equations of motion
(1.5) become generally \textit{singular}, either because the two particles
collide, or because they escape to infinity (as the case may be, see above).
Note however that, if $a$ is a \textit{negative integer} or a
\textit{negative half-integer}, in the case in which the two particles
collide there is no singularity; this is entailed by (\ref{eq4.33}), it is
consistent with the occurrence of a collision thanks to (\ref{eq4.34}), and it is
indeed verified in the special cases $a = - 2$,
$a = - 3$, $a = - 1/2$, see
(4.12), (4.13), (4.14). If more than one branch point occurs inside the disk
$C$ the analysis must be adjusted accordingly; of course the outcome is
critically affected not only by the presence of such branch points, but as
well by which sheets they are located on, namely it depends on the overall
structure of the Riemann surface associated with $\underline {\zeta}  \left(
{\tau}  \right)$, the key element being always the path traveled on that
surface by the complex point $\tau = \left[ {\exp\left( {i\omega t}
\right) - 1} \right]/\left( {i\omega}  \right)$ as the \textit{real}
variable $t$ (``time'') evolves onward from the initial moment $t = 0$.

In conclusion let us emphasize that our findings identify the following
reasoning as instrumental to understand the motion of the particles in the
plane entailed, for given initial data, by the Newtonian equations of motion
(1.1), or equivalently by the complex equations of motion (1.5). The idea is
to fix attention on the solution of (1.7) (rather than~(1.5)) corresponding
to the same initial data (see (1.8)). This defines the solution $\zeta _{n}
\left( {\tau}  \right)$ (with $n = 1,2$, since we are now restricting
attention to the two-body case; \textit{but we will see in the next section
that essentially the same reasoning applies in the} $N$\textit{-body case}),
to which is generally associated a multi-sheeted Riemann surface in the
complex $\tau $-plane. The behavior of the solution of the ``physical''
equations of motion (1.5) as a function of the, of course \textit{real},
time $t$ is then obtained by traveling on that Riemann surface following the
circular contour $\tilde {C}$ defined by (1.6). Depending on the structure
of the Riemann surface, this may entail a motion that is \textit{nonsingular
and completely periodic} (with period~$T$, see~(1.2), or with a period which
is an \textit{integer multiple} of $T$), that is \textit{singular} (if a
branch point happens to sit just on the contour $\tilde {C}$), or that is
\textit{nonsingular but not periodic}. Two mechanisms may give rise to the
latter outcome (no periodicity): \textit{(i)} the nature of the branch
points, if they are characterized by an exponent that is not a \textit{real
rational number} (whether this is going or not to happen is immediately
predictable, since it depends on whether the coupling constant $a$ is or is
not itself a \textit{real rational number}, see (\ref{eq4.10}) and (\ref{eq4.32}));
\textit{(ii)} even if the coupling constant is a \textit{real rational
number}, so that each branch point yields only a \textit{finite} number of
sheets, there still may be an \textit{infinite} number of sheets due to an
\textit{infinity} of branch points (of course not all of them occurring
necessarily on the same sheet, but possibly in a nested fashion), and it may
then happen that by traveling along the contour $\tilde {C}$ an
\textit{endless} sequence of \textit{new} sheets is accessed (a necessary
condition for this to happen is that the Riemann surface feature an
\textit{infinity} of branch points \textit{inside} the contour $\tilde {C}$
-- of course, on different sheets). Of course both mechanisms could be at
work simultaneously.

Clearly the \textit{completely periodic} motions correspond to an
\textit{integrable} \textit{behavior} of the system; and, as shown above,
\textit{whatever} the value of the coupling constant $a$, there
\textit{always} exist at least a set (having \textit{nonvanishing},
indeed \textit{infinite}, measure in phase space) of initial data which
yield such a behavior (\textit{nonsingular} and \textit{completely periodic}
motions with period~$T$, see (1.2)). As for the motions which are instead
\textit{not periodic}, one may well ask whether they should be categorized
as \textit{chaotic} or as \textit{not chaotic}. It seems natural to suggest
that, if the \textit{chaotic} behavior is characterized by the fact that two
trajectories, however close the initial data are that define them, give
eventually rise to completely different motions (``sensitive dependence on
initial conditions''), then clearly the \textit{nonperiodic} motions will be
\textit{chaotic} respectively \textit{not chaotic} according to whether it
is \textit{infinite} respectively \textit{ finite} the number of branch
points which, by falling just inside or just outside the circular
contour~$\tilde {C}$, determine the (\textit{infinite}) number of sheets that are
accessed by a path following that contour~$\tilde {C}$ on the Riemann
surface associated with the solution $\underline {\zeta}  \left( {\tau}
\right)$ of (1.7). Indeed, if that number of branch point is
\textit{finite}, two sets of initial data that are sufficiently close to
each other (in phase space) yield two Riemann surfaces which are
sufficiently similar to each other so that, throughout the two motions
corresponding to the two sets of initial data, the \textit{same sequence} of
sheets is traveled. But if that number of branch points is
\textit{infinite}, then even two solutions $\underline {\zeta}  \left( {\tau
} \right)$ of (1.7) that are initially very close will inevitably be
associated with two Riemann surfaces in which one relevant branch points
falls in one case \textit{inside}, and in the other \textit{outside}, the
circular contour~$\tilde {C}$, hence, by traveling on $\tilde {C}$, after
that point has been passed the corresponding trajectories of the physical
problem (1.5) (or, equivalently, (1.1)) become different, because from that
moment a \textit{different sequence} of sheets is accessed of the two
Riemann surfaces associated with the corresponding solutions~$\underline
{\zeta}  \left( {\tau}  \right)$ of the evolution equations (1.7). This is
then to be interpreted as the cause of \textit{chaos}. Note however that,
even in the \textit{chaotic} case, there is no local exponential divergence
of trajectories; the mechanism that causes the onset of chaos in this case
is rather analogous to that which characterizes the \textit{nonperiodic}
free motion of a point in, say, a triangular plane billiard with angles
which are \textit{irrational} fractions of~$\pi $ (then any two
trajectories, however close they initially are -- and for however long they
remain close -- eventually become topologically different because one of the
two misses a reflection that the other one takes, and from that moment
onwards their evolutions become quite different).

\section{Branch points of the solutions of (1.7) in the $\boldsymbol{N}$-body case}

In this Section we investigate the branch point structure in the complex
$\tau$-plane of the solutions $\underline {\zeta}  \left( {\tau}  \right)$
of the equations of motion (1.7), since -- as discussed above -- the nature
and location of these branch points determine the behavior of the solutions
$\underline {z} \left( {t} \right)$ of the equations of motion (1.5), namely
of the Newtonian equations of motion (1.1), as functions of the \textit{real
}time variable $t$. Let us re-emphasize that it is indeed the structure of
the Riemann surface associated with the solution $\underline {\zeta}  \left(
{\tau}  \right)$ of the equations of motion (1.7) that determines whether
the corresponding solution $\underline {z} \left( {t} \right)$ of the
``physical'' equations of motion~(1.5) namely (1.1) does or does not become
\textit{singular} as function of the \textit{real} time variable $t$, and if
it is not singular throughout time whether or not it is \textit{completely
periodic}, and if it is periodic then with what period (whether $T$, see
(1.2), or an integer multiple of $T$). The rule to evince these conclusions
is quite simple, see (1.6): to obtain $\underline {z} \left( {t} \right)$ as
function of the \textit{real} variable $t$ one must follow the corresponding
solution $\underline {\zeta}  \left( {\tau}  \right)$ (namely, that
characterized by the \textit{same} initial data, see (1.8)) as the
\textit{complex} ``time-like'' variable $\tau $ travels, \textit{on the
Riemann surface associated with that solution} $\underline {\zeta}  \left(
{\tau}  \right)$, around and around counterclockwise along the circular
contour $\tilde {C}$ centered at $i/\omega $ and of radius $1/\omega $.

The implications of this analysis have already been discussed in the
preceding Section~4 in the context of the two-body problem. The situation in
the $N$-body case is somewhat analogous. Indeed the structure of the
evolution equations (1.7) entails that the same two mechanisms discussed in
the two-body case are \textit{generically} responsible for the emergence of
singularities, at some complex values $\tau _{b}$, of the solutions of
these equations of motion, (1.7), for an arbitrary number $N$ of particles:
namely, singularities arise either from the ``collision'' of two particles,
characterized by the relation $\zeta _{1} \left( {\tau _{b}}  \right) =
\zeta _{2} \left( {\tau _{b}}  \right)$ (with $\left| {\zeta _{1} \left(
{\tau _{b}}  \right)} \right| = \left| {\zeta _{2} \left( {\tau _{b}}
\right)} \right| < \infty $, where we assign, without loss of generality,
the labels $1$ and $2$ to the two ``colliding'' particles), or from the
simultaneous ``escape to infinity'' of two or more (say, $M$) particles,
characterized by the relation $\left| {\zeta _{1} \left( {\tau _{b}}
\right)} \right| = \left| {\zeta _{2} \left( {\tau _{b}}  \right)} \right| =
\cdots = \left| {\zeta _{M} \left( {\tau _{b}}  \right)} \right| = \infty $
(where, without loss of generality, we assumed the $M$ ``particles'' $\zeta
_{n} \left( {\tau}  \right)$ that escape to infinity as $\tau \to \tau _{b}
$ to be labeled by the first $M$ indices, $n = 1,\ldots,M$). Note that here we
use again inverted commas around the word ``collision'' and ``escape to
infinity'' to underline that only in the special cases in which, via the
transformation (1.6), to the value~$\tau _{b} $ there corresponds a
\textit{real} value~$t_{c} $ (namely, $\tau _{b} = \left[ {\exp\left(
{i\omega t_{c}}  \right) - 1} \right]/\left( {i\omega}  \right)$
with $t_{c} $ \textit{real}; see (1.6)), the ``collision'', or the ``escape to
infinity'', corresponds indeed to a real event for the physical problem (1.5)
namely~(1.1).

The statement we just made is not meant to exclude the possibility that
``collisions'' involving simultaneously more than two particles occur and
cause correspondingly a singularity: indeed the exact similarity solutions
(with $M > 2$) presented in Section~3 provide examples of solutions
characterized by such phenomena. But it stands to reason -- and it is
confirmed by our analysis, see below -- that $M$-particle ``collisions'' with
$M > 2$ are \textit{not generic}, namely they are \textit{not} associated to
solutions emerging from the assignment of \textit{generic} initial data:
indeed for a \textit{generic} solution of the equations of motion (1.7) the
\textit{complex} equation $\zeta _{1} \left( {\tau _{b}}  \right) = \zeta
_{2} \left( {\tau _{b}}  \right)$ generally has at least one \textit{complex}
solution $\tau _{b} $ (and more likely many such solutions, indeed quite
possibly an infinity of them), while one should not expect the $M - 1$
complex equations $\zeta _{1} \left( {\tau _{b}}  \right) = \zeta _{2}
\left( {\tau _{b}}  \right) = \cdots = \zeta _{M} \left( {\tau _{b}}  \right)$,
with $M > 2$, to possess any solution at all (although there are of course
\textit{special} solutions $\underline {\zeta}  \left( {\tau}  \right)$ of
(1.7) for which such multiple equations do possess solutions, see for
instance Section~3). Let us then understand the type of singularity
associated to these two types of ``events''. To this end the discussion of the
preceding Section~4 is helpful (especially to guess at the nature of the
singularity associated with such ``events''), but a new treatment in the
$N$-body context is nevertheless necessary.

We analyze firstly the singularities associated with ``two-body collisions''.

For notational convenience let us assume, without loss of generality, that
the two particles involved in the two-body event -- which happens at $\tau =
\tau _{b} $ -- carry the labels~$1$ and~$2$, and let us call~$a$ the coupling
constant associated with this particle pair,
\begin{equation}
\label{eq5.1}
a = a_{12} = a_{21}.
\end{equation}

Let us firstly assume that
\begin{subequations}
\begin{equation}
\label{eq5.2}
\mbox{Re}\left( {a} \right) > 0,
\end{equation}
\end{subequations}
so that the real part of the branch-point exponent $\gamma $ (see (4.6), and
below),
\begin{equation}
\label{eq5.3}
\gamma = 1/\left( {1 + a} \right),
\end{equation}
satisfies the restriction
\setcounter{equation}{1}
\begin{subequations}
\setcounter{equation}{1}
\begin{equation}
\label{eq5.4}
0 < \mbox{Re}\left( {\gamma}  \right) < 1 .
\end{equation}
\end{subequations}
It can then be shown (see Appendix C) that in the neighborhood of a
``two-body collision'' occurring at $\tau = \tau _{b} $ the solution
$\underline {\zeta}  \left( {\tau}  \right)$ of (1.7) features the following
behavior:
\setcounter{equation}{3}
\begin{subequations}
\begin{gather}
\zeta _{s} \left( {\tau}  \right) = b + \left( { - 1} \right)^{s -
1} c\left( {\tau - \tau _{b}}  \right)^{\gamma}  + v\left( {\tau -
\tau _{b}}  \right)\nonumber\\
\label{eq5.5}
\phantom{\zeta _{s} \left( {\tau}  \right) =}{} + \sum\limits_{k = 1}^{\infty}
\sum\limits_{l,m = 0; \,l + m \ge
1}^{\infty} g_{klm}^{\left( {s} \right)} \left( {\tau - \tau _{b}}
\right)^{k + l\gamma + m\left( {1 - \gamma}  \right)},\qquad
s = 1,2,\\
\zeta _{n} \left( {\tau}  \right) = b_{n} + v_{n} \left( {\tau - \tau _{b}
} \right)
\nonumber\\
\label{eq5.6}
\phantom{\zeta _{n} \left( {\tau}  \right) =}{}
+ \sum\limits_{k = 1}^{\infty}  {} \sum\limits_{l = \delta _{k1}} ^{\infty
} {} \sum\limits_{m = 0}^{\infty}  {} g_{klm}^{\left( {n} \right)} \left(
{\tau - \tau _{b}}  \right)^{k + l\gamma + m\left( {1 - \gamma}
\right)},\qquad n = 3,\ldots,N.
\end{gather}
In these formulas the $2N$ (complex) constants $b$,
$c$, $v$, $\tau _{b} $
and $b_{n}$, $v_{n} $ (with $n = 3,\ldots,N$) are arbitrary, except for the
inequalities
\begin{equation}
\label{eq5.7}
b \ne b_{n} ,\qquad b_{n} \ne b_{m} ,\qquad n,m = 3,\ldots,N,
\end{equation}
\end{subequations}
while the coefficients of the sums are determined in terms of these
constants (see Appendix~C). Clearly these formulas, (5.4), entail $\zeta
_{1} \left( {\tau _{b}}  \right) = \zeta _{2} \left( {\tau _{b}}  \right) =
b \ne \zeta _{n} \left( {\tau _{b}}  \right) = b_{n} \ne \zeta _{m} \left(
{\tau _{b}}  \right) = b_{m} $ for $n \ne m$, $3 \le n,m \le N$ with
$\left| {\zeta _{1} \left( {\tau _{b}}  \right)} \right| = \left| {\zeta
_{2} \left( {\tau _{b}}  \right)} \right| = \left| {b} \right| < \infty $
but (see (\ref{eq5.5}) and (\ref{eq5.4})) $\left| {{\zeta} '_{1} \left( {\tau _{b}}
\right)} \right| = \left| {{\zeta} '_{2} \left( {\tau _{b}}  \right)}
\right| = \infty $, while $\left| {\zeta _{n} \left( {\tau _{b}}  \right)}
\right| = \left| {b_{n}}  \right| < \infty $ and (see (\ref{eq5.6}) and (\ref{eq5.4}))
$\left| {{\zeta} '_{n} \left( {\tau _{b}}  \right)} \right| = \left| {v_{n}
} \right| < \infty$, $n = 3,\ldots,N$. This confirms that the corresponding
event is to be interpreted as a ``collision'' of the two particles $1$ and
$2$, with both colliding particles moving infinitely fast at the collision
time $\tau = \tau _{b} $. What interests us most is the nature of the
corresponding singularity: a \textit{branch point} characterized by the
exponents $\gamma $ and $1 - \gamma $, see (\ref{eq5.3}). And the fact that such a
singularity is associated with an expression of the solution $\underline
{\zeta}  \left( {\tau}  \right)$ of (1.7) that features, see (5.4), the
\textit{maximal} number, $2N$, of arbitrary constants, demonstrates the
\textit{generic} character of such singularities, which are therefore likely
to be featured by the solutions $\underline {\zeta}  \left( {\tau}  \right)$
corresponding to a \textit{generic} set of initial data.

Of course a more complete treatment shall have to face the nontrivial task
to prove that the iteration used in Appendix~C to justify (5.4) does
converge and thereby yields a solution which has indeed the analytic
structure suggested by the formulas written above. In this paper we limit
our consideration, both here and below, to this incomplete treatment, which
is however sufficient to guess the character of the branch points.

Likewise, if the inequality (\ref{eq5.2}) is reversed,
\begin{subequations}
\begin{equation}
\label{eq5.8}
\mbox{Re}\left( {a} \right) < 0,
\end{equation}
\end{subequations}
it can be shown (see Appendix~C) that in the neighborhood of a ``two-body
collision'' the behavior of the solutions $\underline {\zeta}  \left( {\tau}
\right)$ is characterized, rather than by (5.4), by the following
expressions:
\begin{subequations}
\begin{gather}
\label{eq5.9}
\zeta _{1} \left( {\tau}  \right) = b + c\left( {\tau - \tau _{b}}
\right)^{1 + \beta}  + \sum\limits_{k,l = 1;\, k + l \ge 3}^{\infty}
g_{kl}^{\left( {1} \right)} \left( {\tau - \tau _{b}}  \right)^{k + l\beta},\\
\label{eq5.10}
\zeta _{2} \left( {\tau}  \right) = b + v_{2} \left( {\tau - \tau _{b}}
\right) - c\left( {\tau - \tau _{b}}  \right)^{1 + \beta}
 + \sum\limits_{k = 1}^{\infty}  {} \sum\limits_{l = 2\delta _{k1}} ^{\infty
} {} g_{kl}^{\left( {2} \right)} \left( {\tau - \tau _{b}}  \right)^{k +
l\beta},\\
\label{eq5.11}
\zeta _{n} \left( {\tau}  \right) = b_{n} + v_{n} \left( {\tau - \tau _{b}
} \right)
 + \sum\limits_{k = 2}^{\infty}  {} \sum\limits_{l = 0}^{\infty}  {}
g_{kl}^{\left( {n} \right)} \left( {\tau - \tau _{b}}  \right)^{k +
l\beta} ,\qquad n = 3,\ldots,N,
\end{gather}
\end{subequations}
with
\begin{equation}
\label{eq5.12}
\beta = - 2a
\end{equation}
so that (see (\ref{eq5.8}))
\setcounter{equation}{4}
\begin{subequations}
\setcounter{equation}{1}
\begin{equation}
\label{eq5.13}
\mbox{Re}\left( {\beta}  \right) > 0.
\end{equation}
\end{subequations}
In these formulas the $2N$ (complex) constants $b$,
$c$, $v_{2}$, $\tau _{b}
$ and $b_{n}$, $v_{n} $ (with $n = 3,\ldots,N$) are \textit{arbitrary} (except,
again, for the inequalities (\ref{eq5.7})), while the coefficients of the sums are
determined in terms of these constants (see Appendix~C). Clearly these
formulas (see~(5.6) and (\ref{eq5.7})) entail again $\zeta _{1} \left( {\tau _{b}}
\right) = \zeta _{2} \left( {\tau _{b}}  \right) = b \ne \zeta _{n} \left(
{\tau _{b}}  \right) = b_{n} \ne \zeta _{m} \left( {\tau _{b}}  \right) =
b_{m} $ for $n \ne m$,
$3 \le n,m \le N$ with $\left| {\zeta _{1} \left(
{\tau _{b}}  \right)} \right| = \left| {\zeta _{2} \left( {\tau _{b}}
\right)} \right| < \infty $, but now (see (\ref{eq5.9}) with (\ref{eq5.13}) and (\ref{eq5.10}))
${\zeta} '_{1} \left( {\tau _{b}}  \right) = 0$, ${\zeta} '_{2} \left(
{\tau _{b}}  \right) = v_{2} $ (so that ${\zeta} '_{1} \left( {\tau _{b}}
\right){\zeta} '_{1} \left( {\tau _{b}}  \right) = 0$; see the right-hand
side of~(1.7)), while of course again $\left| {\zeta _{n} \left( {\tau _{b}
} \right)} \right| = \left| {b_{n}}  \right| < \infty $ and (see (\ref{eq5.4}))
$\left| {{\zeta} '_{n} \left( {\tau _{b}}  \right)} \right| = \left| {v_{n}
} \right| < \infty $ for $n = 3,\ldots,N$. Hence the corresponding event is
again to be interpreted as a ``collision'' of the two particles $1$ and $2$,
but now with particle~1 having \textit{zero} velocity at the time of the
collision, in contrast to particle~2 which moves with velocity $v_{2} $
(note the notational distinction thereby introduced among the two colliding
particles); while for our purposes the interpretation of these formulas,
(5.6a,b,c), is that the nature of the corresponding singularity is a
\textit{branch point} characterized by the exponent $\beta $, see (\ref{eq5.12}). And
again the fact that such a singularity is associated with an expression of
the solution $\zeta _{n} \left( {\tau}  \right)$ of (1.7) that features, see
(5.6a,b,c), the \textit{maximal} number, $2N$, of \textit{arbitrary}
constants, demonstrates the \textit{generic} character of such
singularities, which are therefore likely to be featured by the solutions
$\underline {\zeta}  \left( {\tau}  \right)$ corresponding to a
\textit{generic} set of initial data.

This concludes our discussion of the singularities associated with ``two-body
collisions''. Those associated with ``collisions'' involving more than two
particles can be discussed in an analogous manner (also taking advantage of
the results of Section~3), but in view of their lack of genericity we
forsake their treatment here, and we rather proceed to discuss the
singularities associated with ``escapes to infinity''.

This phenomenon can only occur if, for some group of the interacting
particles, which without loss of generality is hereafter assumed to be
formed by the $M$ particles with labels from $1$ to $M$ (where $2 \le M \le
N$), the quantity $A$,
\setcounter{equation}{7}
\begin{subequations}
\begin{equation}
\label{eq5.14}
A = \left( {2/M} \right)\sum\limits_{n,m = 1;\, n > m}^{M} {} a_{nm},
\end{equation}
has real part \textit{less than negative unity},
\begin{equation}
\label{eq5.15}
\mbox{Re}\left( {A} \right) < - 1,
\end{equation}
\end{subequations}
so that the corresponding quantity $\Gamma $ (see (3.2b)),
\begin{subequations}
\begin{equation}
\label{eq5.16}
\Gamma = \left( {1 + A} \right)^{ - 1},
\end{equation}
has \textit{negative} real part,
\begin{equation}
\label{eq5.17}
\mbox{Re}\left( {\Gamma}  \right) < 0.
\end{equation}
\end{subequations}
The quantities $a_{nm} $ appearing in the right-hand side of (\ref{eq5.14}) are of
course the coupling constants that characterize the two-body interactions
acting among the particles belonging to this group of $M$ particles, see
(1.7) or (1.5); and there may of course be, for a given $N$-body problem,
several subgroups of particles such that the corresponding quantity $\Gamma
$, defined according to the above prescription, has \textit{negative} real
part, see (\ref{eq5.17}). Let us assume for simplicity that there is just one such
group. It is then easily seen that the dominant term at $\tau \approx \tau
_{b} $ of the solution $\underline {\zeta}  \left( {\tau}  \right)$
representing the ``escape to infinity'' of the~$M$ coordinates $\zeta _{n}
\left( {\tau}  \right)$, $n = 1,\ldots,M$, reads
\begin{subequations}
\begin{gather}
\label{eq5.18}
\zeta _{n} \left( {\tau}  \right) \approx c_{n} \left( {\tau - \tau _{b}}
\right)^{\Gamma} ,\qquad n = 1,\ldots,M,\\
\label{eq5.19}
\zeta _{n} \left( {\tau}  \right) = b_{n},\qquad n = M + 1,\ldots,N.
\end{gather}
\end{subequations}
In these formulas the $N$ (complex) constants $b_{n} $ are
\textit{arbitrary,} while the coefficients $c_{n}$,
$n = 1,\ldots,M$ are
determined (up to a common rescaling factor) in terms of the coupling
constants $a_{nm} $ with $n,m = 1,\ldots,M$ by (3.3a) with (\ref{eq5.14}). Clearly
these formulas (see (5.10) and (\ref{eq5.15})) entail $\left| {\zeta _{n} \left(
{\tau _{b}}  \right)} \right| = \left| {{\zeta} '_{n} \left( {\tau _{b}}
\right)} \right| = \infty $ for $n = 1,\ldots,M$ and $\zeta _{n} \left( {\tau
_{b}}  \right) = b_{n} $ for $n = M + 1,\ldots,N$. Hence the corresponding
``event'' at $\tau = \tau _{b} $ is indeed to be interpreted as the ``escape to
infinity'' of the $M$ particles with (conveniently chosen) labels from $1$
to~$M$. But, as already indicated in the introductory Section~1, a discussion
of the analytic structure of this solution $\underline {\zeta}  \left( {\tau
} \right)$ at $\tau \approx \tau _{b} $ is postponed to a separate paper.

\section{Analysis of various motions}

In this section we survey the implications of the findings discussed above
(especially in Section~5) for the motions of the physical models (1.1)
characterized by various choices of the number $N$ of particles and of the
coupling constants $a_{nm} $. We always refer, for notational convenience,
to the complex \textit{avatar} (1.5) of these equations of motion, and we
display the trajectories of solutions produced via a computer code
manufactured by one of us (MS) for the numerical integration of these
equations of motion, (1.5). Additional numerical examples, as well as the
discussion and display of the numerical code, will be reported elsewhere~[8],
enabling thereby any interested party to enjoy an ampler personal
exploration of analogous numerical results (easily generated on a personal
computer).

The basic mode of presentation of our results begins by choosing a specific
model, characterized by the number $N$ of particles (for simplicity we
restrict consideration below to three- and four-body cases, $N = 3$ and $N =
4$, although the simulation program can actually handle an arbitrary number
$N$ of moving points) and by a set of coupling constants~$a_{nm} $, see
(1.5), which we always assume to satisfy the symmetry property (1.12). We
moreover always set, for notational convenience (and of course without loss
of generality, see the Remark~\ref{remark-1.1})
\begin{subequations}
\begin{equation}
\label{eq6.1}
\omega = 2\pi ,
\end{equation}
entailing an assignment of the time scale such that the basic period be
unity,
\begin{equation}
\label{eq6.2}
T = 1,
\end{equation}
\end{subequations}
see (1.2). We then generally consider, for each model, a sequence of
solutions of the equations of motions (1.5) corresponding to a sequence of
initial data characterized by the \textit{same} initial configuration of the
particle positions in the plane and by sets of initial velocities modified,
from one solution to the next one considered, via multiplication by an
increasing sequence of \textit{positive} factors $\mu $ (by analogy with the
notation in (1.10b)):
\begin{equation}
\label{eq6.3}
z_{n} \left( {0} \right) = z_{n}^{\left( {0} \right)},
\qquad \dot{z}_{n} \left( {0} \right) = \mu \dot {z}_{n}^{\left( {0} \right)}.
\end{equation}
As implied by the treatment of the preceding sections (see in particular
Section~1 and the Remarks~\ref{remark-1.1} and~\ref{remark-1.3}) this entails that
all the sequentially analyzed solutions $\underline {z} \left( {t} \right)$
of the ``physical'' equations of motion (in their \textit{avatar} (1.5), and
characterized by a specific assignment of $N$ and $a_{nm} $) correspond, via
the fundamental relation (1.6) (with an appropriately chosen value of the
constant $\omega $ in this formula (1.6), namely $\omega = 2\pi /\mu $),
to the \textit{same} solution $\underline {\zeta}  \left( {\tau}
\right)$ of (1.7), characterized by the \textit{same} values of $N$ and
$a_{nm} $, and by the \textit{same} initial data,
\begin{equation}
\label{eq6.4}
\zeta _{n} \left( {0} \right) = z_{n}^{\left( {0} \right)},\qquad
{\zeta} '_{n} \left( {0} \right) = \dot {z}_{n}^{\left( {0} \right)} ;
\end{equation}
hence the solutions $\underline {z} \left( {t} \right)$ of the equations of
motion (1.5) discussed below are obtained via~(1.6) from the \textit{same}
solution $\underline {\zeta}  \left( {\tau}  \right)$ of (1.7) by traveling
on the Riemann surface associated with this solution of (1.7) on circular
contours $\tilde {C} = \tilde {C}\left( {\mu}  \right)$, in the complex
$\tau $-plane, the diameters of which always lie on the upper imaginary axis
of the complex $\tau $-plane, with their lower ends fixed at the origin,
$\tau = 0$, while the size of these diameters is $\mu /\pi $, namely it
gets increased via multiplication by the same rescaling factor, $\mu $, that
yielded by multiplication the set of initial velocities. The sequence of
solutions we consider starts generally with a \textit{completely periodic}
one having the basic period $T = 1$; note that, as implied by
Proposition~\ref{prop-1.2}, and as of course confirmed by the numerical
simulations, such solutions $\underline {z} \left( {t} \right)$
\textit{always} exist for sufficiently small initial velocities, namely
sufficiently small values of $\mu $ (they correspond to solutions
$\underline {\zeta}  \left( {\tau}  \right)$ of (1.7) which are singularity
free, on the main sheet of their Riemann surface, inside the circle $\tilde
{C}\left( {\mu}  \right)$ -- which, for very small~$\mu $, becomes very
small, and gets very close to the origin, $\tau = 0$, where the solution
$\underline {\zeta}  \left( {\tau}  \right)$ is of course
\textit{holomorphic}). The sequence of solutions $\underline {z} \left( {t}
\right)$ identified according to the procedure indicated above corresponds
therefore generally to the sequential coming into play of different
\textit{branch points} (see Section~5) of the \textit{same} solution
$\underline {\zeta}  \left( {\tau}  \right)$ of (1.7), as they get enclosed
inside the circular contours $\tilde {C} = \tilde {C}\left( {\mu}  \right)$
when the size of these contours gets enlarged proportionally to the
parameter $\mu $ (which is sequentially increased from one simulation to the
next one). The inclusion of such \textit{branch points} may entail the
transition -- for models with \textit{real rational} coupling constants, to
which our consideration is mainly restricted below, as they display the most
interesting phenomenology (aside from the fact that, in numerical
computations, \textit{all} numbers are \textit{rational}!) -- to solutions
which are again \textit{completely periodic} but with \textit{larger}
periods, or which possibly are no more periodic (although of course it is
hardly possible to distinguish numerically a nonperiodic solution from one
which is periodic but with a very large period, see below), according to a
phenomenology which can be understood on the basis of the treatment of the
preceding Sections~4 and~5 and the discussion at the end of Section~1 -- an
analysis which turns out to be indeed validated by the numerical results
reported below (and see also the analogous discussion in Ref.~[9]). We also
present below some results with \textit{complex} coupling constants, and
tersely discuss the associated phenomenology. On the other hand, as already
mentioned above, our consideration is restricted to coupling constants the
real part of which is \textit{positive}, $\mbox{Re}\left( {a_{nm}}  \right) > 0$.

Enough of preliminary remarks: let us proceed to display and comment a few
representative examples. In the graphs displayed below, particle~1 will be
shown in \textit{Red}, particle~2 in \textit{Green}, particle~3 in
\textit{Blue} and particle~4 (if present) in \textit{Yellow}. Whenever we
felt such an additional indication might be usefully displayed we indicated
with a black diamond the \textit{initial} position of each particle (at $t =
0$), with a black dot the position at a subsequent time $t = t_{1} $, and
with smaller black dots the position at every subsequent integer multiple of
$t_{1} $ (namely at $t = t_{k} = kt_{1}$, $k = 2,3,\ldots$); in this manner
the direction of the motion along the trajectories can be inferred (from the
relative positions along the trajectories of the diamond and the larger
dot), as well as some indication of the positions of the particles over
time, as they move (by counting the dots along the trajectory). Of course a
much more satisfactory visualization of the behavior of the many-body system
is provided by simulations in which the particle motions are displayed as
they unfold over time (as in a movie); it is planned to make available soon,
via the web, the numerical code suitable to perform such simulations on
personal computers~[8]. Let us emphasize that such simulations are
particularly stunning to watch in the case of high-period trajectories,
which are very complicated (see below), so that the fact that the particles
return eventually \textit{exactly} on their tracks appears quite miraculous
and is indeed a remarkable proof of the reliability of the numerics.

The first example we report is characterized by the following parameters:
\begin{subequations}
\begin{gather}
N=3; \qquad a_{12} = a_{21} = a_{RG} = 1, \qquad
a_{13} = a_{31} = a_{RB} = 1/2,\nonumber\\
a_{23} = a_{32} = a_{GB} = 3/2 ,\label{eq6.5}
\end{gather}
and by the following values of the parameters
$x_{n}^{\left( {0} \right)}$, $y_{n}^{\left( {0} \right)}$,
$\dot {x}_{n}^{\left( {0} \right)}$, $\dot {y}_{n}^{\left( {0} \right)}$
characterizing the initial data via (\ref{eq6.3}):
\begin{gather}
 x_{1}^{\left( {0} \right)} = 0,\qquad y_{1}^{\left( {0} \right)} =
0,\qquad \dot {x}_{1}^{\left( {0} \right)} = - 1,\qquad \dot
{y}_{1}^{\left( {0} \right)} = - 0.5; \nonumber\\
 x_{2}^{\left( {0} \right)} = 1,\qquad y_{2}^{\left( {0} \right)} =
0,\qquad \dot {x}_{2}^{\left( {0} \right)} = 0.5,\qquad \dot
{y}_{2}^{\left( {0} \right)} = 1; \nonumber\\
 x_{3}^{\left( {0} \right)} = 2,\qquad y_{3}^{\left( {0} \right)} =
0,\qquad \dot {x}_{3}^{\left( {0} \right)} = 0.5,\qquad \dot
{y}_{3}^{\left( {0} \right)} = - 1 .\label{eq6.6}
\end{gather}
\end{subequations}

Our numerical findings are synthesized in the Tables 6.1a,b, and in the
sequence of Figures~6.1 (the contents of which are identified in
Table~6.1a). Here and below we employ the acronym HSL (\textit{Hic Sunt Leones})
to denote a (presumably) \textit{chaotic} behavior. We trust the
significance of this material -- as presented here and in the rest of this
section -- to be self-evident, but we nevertheless add a few comments.

\begin{center}
Table 6.1a
\medskip

\small{
\begin{tabular}{|c|c|c|c|c|c|c|c|c|c|c|c|}
\hline
$\mu$ & 0.8 & 1 & 1.301 & 1.302 & 5.28 & 5.29 & 5.93 & 5.94 & 6.82 & 6.83 & 50 \\
\hline
Period & 1 & 1 & 1 & 2 & 2 & 6 & 6 & 7 & 7 & HSL & HSL \\
\hline
Fig. 6.1 & a & -- & b & c & d & e & f & g & h & i & -- \\
\hline
\end{tabular}
}

\bigskip

Table 6.1b
\medskip

\small{
\begin{tabular}{|c|c|c|c|c|}
\hline
Period Change & 1 $\to$ 2 & 2 $\to$ 6 & 6 $\to$ 7 & 7 $\to$ \textup{HSL} \\
\hline
Colliding Particles & R--G & G--B & R--G & R--B \\
\hline
Collision Time & 0.80 & 0.77 & 2.87 & 3.72 \\
\hline
\end{tabular}
}
\end{center}

Clearly for $\mu \le 1.301$ the solution $\underline {\zeta}  \left( {\tau}
\right)$ of (1.7) (with (\ref{eq6.4}) and (\ref{eq6.6})) is \textit{holomorphic} inside the
circle $\tilde {C}\left( {\mu}  \right)$, hence the (corresponding) solution
$\underline {z} \left( {t} \right)$ of (1.5) (with (6.1), (\ref{eq6.3}) and (6.3b))
is \textit{completely periodic} with period $T = 1$ (see the first 3 columns
of Table 6.1a, and Figs. 6.1a,b). For $1.302 \le \mu \le 5.28$ this (same)
solution $\underline {\zeta}  \left( {\tau}  \right)$ of (1.7) (with (\ref{eq6.4})
and (\ref{eq6.6})) has a single singularity inside the circle $\tilde {C}\left(
{\mu}  \right)$, a \textit{branch point} with exponent
\begin{subequations}
\begin{equation}
\label{eq6.7}
\gamma _{RG} = 1/\left( {a_{RG} + 1} \right) = 1/2;
\end{equation}
hence by traveling on the Riemann surface associated with $\underline {\zeta}
\left( {\tau}  \right)$ on the circle $\tilde {C}\left( {\mu}  \right)$
\textit{two} sheets are now visited before returning to the point of
departure, and therefore the (corresponding) solution $\underline {z} \left(
{t} \right)$ of (1.5) (with (6.1), (\ref{eq6.3}) and (6.3b)) is \textit{completely
periodic} with period $T = 2$ (see the fourth and fifth columns of Table
6.1a, and Figs. 6.1c,d). The transition from the regime with period $T = 1$
to that with period $T = 2$ occurs due to the entrance of the relevant
\textit{branch point} from outside to inside the circle $\tilde {C}\left(
{\mu}  \right)$; this corresponds to the occurrence of a collision among the
\textit{Red} and \textit{Green} particles (see Figs. 6.1b and 6.1c, and the
first column of Table 6.1b), consistently -- according to the analysis of
Section 5 -- with the presence of the coupling constant $a_{RG} = 1$ in
(\ref{eq6.7}).

For $5.29 \le \mu \le 5.93$ this (same) solution $\underline {\zeta}  \left(
{\tau}  \right)$ of (1.7) (with (\ref{eq6.4}) and (\ref{eq6.6})) has \textit{two}
singularities inside the circle $\tilde {C}\left( {\mu}  \right)$, namely
the \textit{branch point} with the exponent (\ref{eq6.7}) and a second one (located
either on the main sheet, or possibly on the second sheet accessed via the
cut associated with the first branch point, with exponent (\ref{eq6.7})), with
exponent
\begin{equation}
\label{eq6.8}
\gamma _{GB} = 1/\left( {a_{GB} + 1} \right) = 2/5 ,
\end{equation}
which opens the way to \textit{four} additional sheets; hence by traveling
on the Riemann surface associated with $\underline {\zeta}  \left( {\tau}
\right)$ on the circle $\tilde {C}\left( {\mu}  \right)$ altogether
\textit{six} sheets are now visited before returning to the point of
departure, and therefore the (corresponding) solution $\underline {z} \left(
{t} \right)$ of (1.5) (with (6.1), (\ref{eq6.3}) and (6.3b)) is \textit{completely
periodic} with period $T = 6$ (see the sixth and seventh columns of Table
6.1a, and Figs. 6.1e,f). The transition from the regime with period $T = 2$
to that with period $T = 6$ occurs due to the entrance (in one of the two
sheets of the Riemann surface) of the (second) \textit{branch point},
characterized by the exponent (\ref{eq6.8}), from outside to inside the circle
$\tilde {C}\left( {\mu}  \right)$; this corresponds to the occurrence of a
collision among the \textit{Green} and \textit{Blue} particles (see Figs.
6.1d and 6.1e, and the second column of Table 6.1b), consistently -- again
according to the analysis of Section 5 -- with the presence of the coupling
constant $a_{GB} = 3/2$ in (\ref{eq6.8}).

Another transition occurs for a value of $\mu $ between $5.93$ and $5.94$
(see the seventh and eighth columns of Table 6.1, and Figs. 6.1f,g),
associated to another collision among the \textit{Red} and \textit{Green}
particles (see the third column of Table 6.2) and bringing inside the circle
$\tilde {C}\left( {\mu}  \right)$ another branch point of square-root type
(see (\ref{eq6.7})), that opens the way to \textit{one} additional sheet of the
Riemann surface, and thereby causes the solutions $\underline {z} \left( {t}
\right)$ of (1.5) (with (6.1), (\ref{eq6.3}) and (6.3b)) to be \textit{completely
periodic} with period $T = 7$ (see the eighth and ninth columns of Table
6.1a, and Figs. 6.1g,h) for $5.94 \le \mu \le 6.82$.

Finally another collision occurs, for $\mu $ between $6.82$ and $6.83$ (see
the ninth and tenth column of Table 6.1a, and Figs. 6.1h,i), among the
\textit{Blue} and \textit{Red} particles (see the last column of Table
6.1b), which brings (on one of the \textit{seven} relevant sheets of the
Riemann surface associated with the solution $\underline {\zeta}  \left(
{\tau}  \right)$ of (1.7) with (\ref{eq6.4}) and (\ref{eq6.6})) a \textit{branch point}
with exponent
\begin{equation}
\label{eq6.9}
\gamma _{RB} = 1/\left( {1 + a_{RB}}  \right) = 2/3
\end{equation}
\end{subequations}
inside the circle $\tilde {C}\left( {\mu}  \right)$. This opens the way to
\textit{two} additional sheets of the Riemann surface (associated with the
solution $\underline {\zeta}  \left( {\tau}  \right)$ of (1.7) with (\ref{eq6.4})
and (\ref{eq6.6})); but clearly on (at least) one of these two sheets there is (at
least) one additional \textit{branch point} (characterized by one of the
three exponents (6.5a,b,c)) which is already located \textit{inside} the
circle $\tilde {C}\left( {\mu}  \right)$, and this opens the way to
additional sheets on (some of) which there also are additional branch points
\textit{inside} the circle $\tilde {C}\left( {\mu}  \right)$, and so on and
on, causing presumably a transition to \textit{chaos} (see the last two
columns of Table 6.1a, and Fig. 6.1i) according to the mechanism outlined at
the end of the introductory Section 1 and discussed in Section 4 (and also
discussed, in an analogous context, in Ref.~[9]).

The second example we report is much richer. It is characterized by the
following parameters:
\begin{subequations}
\begin{gather}
N = 3; \qquad  a_{12} = a_{21} = a_{RG} = 1,\qquad
a_{13} = a_{31} = a_{RB} = 2,\nonumber\\
 a_{23} = a_{32} = a_{GB} = 3,\label{eq6.10}
\end{gather}
and by the following values of the parameters
$x_{n}^{\left( {0} \right)}$, $y_{n}^{\left( {0} \right)}$, $\dot
{x}_{n}^{\left( {0} \right)}$, $\dot {y}_{n}^{\left( {0} \right)}
$ characterizing the initial data via (\ref{eq6.3}):
\begin{gather}
 x_{1}^{\left( {0} \right)} = 0,\qquad y_{1}^{\left( {0} \right)} =
0,\qquad \dot {x}_{1}^{\left( {0} \right)} = - 1,\qquad \dot
{y}_{1}^{\left( {0} \right)} = 1; \nonumber\\
 x_{2}^{\left( {0} \right)} = 0,\qquad y_{2}^{\left( {0} \right)} =
1,\qquad \dot {x}_{2}^{\left( {0} \right)} = 1,\qquad \dot
{y}_{2}^{\left( {0} \right)} = 0; \nonumber\\
 x_{3}^{\left( {0} \right)} = - 1,\qquad y_{3}^{\left( {0} \right)} =
0,\qquad \dot {x}_{3}^{\left( {0} \right)} = - 0.5,\qquad \dot
{y}_{3}^{\left( {0} \right)} = - 0.5 .\label{eq6.11}
\end{gather}
\end{subequations}

Our numerical findings are now synthesized in the Tables 6.2a,b, and in the
sequence of Figs.~6.2 (the contents of which are identified in Table
6.2a). We urge the alert reader to repeat the analysis we detailed in the
preceding example: the fact that the branch point exponents take in this
case the values $\gamma _{RG} = 1/2$, $\gamma _{BR} = 1/3$ and $\gamma _{GB}
= 1/4$ is of course relevant, and clearly some transitions from a periodic
regime to another periodic regime with higher period are the effect of a
single identifiable branch point, while others are due to the synergistic
effect of several branch points coming into play simultaneously. The
presence of trajectories with stunningly large periods is of course
noteworthy.

\begin{center}
Table 6.2a
\medskip

\small{
\begin{tabular}{|c|c|c|c|c|c|c|c|c|c|}
\hline
$\mu$ & 0.5 & 0.780 & 0.781 & 1 & 1.213 & 1.214 & 1.219 & 1.220 & 1.241 \\
\hline
Period & 1 & 1 & 6 & 6 & 6 & 49 & 49 & 50 & 50 \\
\hline
Fig. 6.2 & a & b & c & -- & d & e & -- & -- & -- \\
\hline
\end{tabular}

\medskip

\begin{tabular}{|c|c|c|c|c|c|c|c|c|c|}
\hline
$\mu$ & 1.242 & 1.25 & 1.293 & 1.294 & 1.295 & 1.400 & 1.401 & 1.41 & 1.442 \\
\hline
Period & 51 & 51 & 51 & 57 & 58 & 58 & 59 & 59 & 59 \\
\hline
Fig. 6.2 & -- & f & -- & g & -- & -- & -- & h & -- \\
\hline
\end{tabular}

\medskip

\begin{tabular}{|c|c|c|c|c|c|c|c|c|c|}
\hline
$\mu$ & 1.443 & 1.721 & 1.722 & 1.8 & 1.944 & 1.945 & 2.053 & 2.054 & 2.1 \\
\hline
Period & 65 & 65 & 66 & 66 & 66 & 68 & 68 & 71 & 71 \\
\hline
Fig. 6.2 & -- & -- & -- & i & -- & -- & -- & -- & j \\
\hline
\end{tabular}

\medskip

\begin{tabular}{|c|c|c|c|c|c|c|c|c|c|}
\hline
$\mu$ & 2.108 & 2.109 & 2.164 & 2.165 & 2.17 & 2.171 & 2.172 & 2.2 & 2.5 \\
\hline
Period & 71 & 72 & 72 & 74 & 74 & 74 & HSL & HSL & HSL \\
\hline
Fig. 6.2 & -- & -- & -- & -- & k & -- & -- & m & -- \\
\hline
\end{tabular}
}

\bigskip

Table 6.2b
\medskip

\small{
\begin{tabular}{|c|c|c|c|c|c|}
\hline
Period Change & 1 $\to$ 6 & 6 $\to$ 49 & 49 $\to$ 50 & 50 $\to$ 51 & 51 $\to$ 57 \\
\hline
Colliding Particles & R--B & R--G & R--G & R--G & R--G \\
\hline
Collision Time & 0.48 & 3.54 & 24.72 & 20.59 & 43.75 \\
\hline
\end{tabular}

\medskip

\begin{tabular}{|c|c|c|c|c|c|}
\hline
Period Change & 57 $\to$ 58 & 58 $\to$ 59 & 59 $\to$ 65 & 65 $\to$ 66 & 66 $\to$ 68 \\
\hline
Colliding Particles & R--G & R--G & R--G & R--G & R--B \\
\hline
Collision Time & 31.49 & 27.37 & 8.35 & 44.25 & 54.41 \\
\hline
\end{tabular}

\medskip

\begin{tabular}{|c|c|c|c|c|}
\hline
Period Change & 68 $\to$ 71 & 71 $\to$ 72 & 72 $\to$ 74 & 74 $\to$ \textup{HSL} \\
\hline
Colliding Particles & G--B & R--G & R--B & R--G \\
\hline
Collision Time & 56.26 & 5.63 & 13.75 & 24.905 \\
\hline
\end{tabular}
}
\end{center}

The third example we report is the last one with coupling constants which
are both \textit{real} and \textit{rational} -- namely the last one belonging
to the class that can give rise to \textit{completely periodic} motions with
periods \textit{larger than unity}; and it features \textit{four} particles.
It is characterized by the following parameters:
\begin{subequations}
\begin{gather}
N = 4 ;\qquad  a_{12} = a_{21} = a_{RG} = 2,\qquad
a_{13} = a_{31} = a_{RB} = 5/2,\nonumber\\  a_{14} = a_{41} = a_{RY} = 1,\qquad
a_{23} = a_{32} = a_{GB} = 1/2,\qquad  a_{24} =
a_{42} = a_{GY} = 3,\nonumber\\
 a_{34} = a_{43} = a_{BY} = 3/2 ,\label{eq6.12}
\end{gather}
and by the following values of the parameters
$x_{n}^{\left( {0} \right)}$, $y_{n}^{\left( {0} \right)}$,
$\dot {x}_{n}^{\left( {0} \right)}$, $\dot {y}_{n}^{\left( {0} \right)}$
characterizing the initial data via (\ref{eq6.3}):
\begin{gather}
 x_{1}^{\left( {0} \right)} = 0,\qquad y_{1}^{\left( {0} \right)} =
1,\qquad \dot {x}_{1}^{\left( {0} \right)} = 1,\qquad \dot
{y}_{1}^{\left( {0} \right)} = - 1; \nonumber\\
 x_{2}^{\left( {0} \right)} = 0,\qquad y_{2}^{\left( {0} \right)} =
0,\qquad \dot {x}_{2}^{\left( {0} \right)} = - 0.5,\qquad \dot
{y}_{2}^{\left( {0} \right)} = 0;\nonumber \\
 x_{3}^{\left( {0} \right)} = - 1.5,\qquad y_{3}^{\left( {0} \right)} =
0,\qquad \dot {x}_{3}^{\left( {0} \right)} = 0,\qquad \dot
{y}_{3}^{\left( {0} \right)} = - 0.5;\nonumber\\
 x_{4}^{\left( {0} \right)} = 1,\qquad y_{4}^{\left( {0} \right)} =
0,\qquad \dot {x}_{4}^{\left( {0} \right)} = - 1,\qquad \dot
{y}_{4}^{\left( {0} \right)} = 1 .\label{eq13}
\end{gather}
\end{subequations}

The corresponding numerical results are synthesized in the Tables 6.3a,b,
and in the sequence of Figs.~6.3 (the contents of which are identified in
Table 6.3a). We again urge the alert reader to repeat the analysis we
detailed in the first example, taking of course into account that,
corresponding to the \textit{six} different coupling constants (see (\ref{eq6.12})),
there are in this case \textit{six} different branch point exponents:
$\gamma _{RG} = 1/3$, $\gamma _{BR} = 2/7$, $\gamma _{RY} = 1/2$, $\gamma_{GB} = 2/3$,
$\gamma _{GY} = 1/4$ and $\gamma _{BY} = 2/5$. And the
presence of trajectories with remarkably large periods should again be
noted.

\begin{center}
Table 6.3a
\medskip

\small{
\begin{tabular}{|c|c|c|c|c|c|c|c|c|}
\hline
$\mu$ & 0.4 & 0.626 & 0.627 & 0.692 & 0.693 & 1 & 1.037 & 1.038 \\
\hline
Period & 1 & 1 & 4 & 4 & 6 & 6 & 6 & 22 \\
\hline
Fig. 6.3 & a & b & c & d & e & -- & f & g \\
\hline
\end{tabular}

\medskip

\begin{tabular}{|c|c|c|c|c|c|c|c|c|c|}
\hline
$\mu$ & 1.156 & 1.157 & 1.207 & 1.208 & 1.3 & 1.318 & 1.319 & 1.32 & 1.4 \\
\hline
Period & 22 & 23 & 23 & 26 & 26 & 26 & HSL & HSL & HSL \\
\hline
Fig. 6.3 & -- & -- & -- & -- & h & -- & -- & i & -- \\
\hline
\end{tabular}
}

\bigskip

Table 6.3b
\medskip

\small{
\begin{tabular}{|c|c|c|c|c|c|c|}
\hline
Period Change & 1 $\to$ 4 & 4 $\to$ 6 & 6 $\to$ 22 & 22 $\to$ 23 & 23 $\to$ 26 & 26 $\to$ \textup{HSL} \\
\hline
Colliding Particles & G--Y & R--G & R--B & R--Y & G--Y & B--Y \\
\hline
Collision Time & 0.44 & 3.33 & 3.33 & 9.26 & 12.36 & 14.53 \\
\hline
\end{tabular}
}
\end{center}

Finally let us display two examples with \textit{complex} coupling
constants, which can both be considered deformations of the first example
considered above, obtained by adding to the three coupling constants
characterizing that model, see (\ref{eq6.5}), (not too large) imaginary parts,
either all of them \textit{negative} (first example below), or all of them
\textit{positive} (second example below). We also take, in both cases, the
same assignment (\ref{eq6.6}) of the parameters that determine the initial data via
(\ref{eq6.3}).

The first of these last two examples is characterized by the parameters
\begin{gather}
N = 3 ;\qquad  a_{12} = a_{21} = a_{RG} = 1 - 0.3i ,
\nonumber\\
\label{eq14}
a_{13} = a_{31} = a_{RB} = 0.5 - 0.4i, \qquad a_{23} = a_{32} = a_{GB} = 1.5 - 0.2i
\end{gather}
and the corresponding trajectories are displayed in the sequence of Figs.
6.4a-f; in each figure the value is now indicated of the scaling factor $\mu$
that determines via (\ref{eq6.3}) with (\ref{eq6.6}) the initial data for the displayed
motions, as well as the period -- which is of course always $T = 1$ -- in
the case of \textit{periodic} motions (see Figs. 6.4a,b) or instead the
final time~$t$ of the displayed trajectories in the case of
\textit{nonperiodic} motions (see Figs. 6.4c-f).

The qualitative understanding of these results is easy, and it displays an
interesting phenomenology -- \textit{limit cycles}, with (asymptotic) period
$T = 1$ -- that is presumably characteristic of all models of type (1.5)
with complex coupling constants the imaginary parts of which are
\textit{negative}, $\mbox{Im}\left( {a_{nm}}  \right) < 0$. Indeed the (first) two
Figs. 6.4a,b display the \textit{completely periodic} motions, with period
$T = 1$, which as we know are always featured by all models of type (1.5),
as guaranteed by Proposition~\ref{prop-1.2}; and let us recall that our
understanding of the existence of these \textit{completely periodic
}solutions $\underline {z} \left( {t} \right)$ of (1.5) with period $T = 1$
is in terms of the corresponding solution $\underline {\zeta}  \left( {\tau
} \right)$ of (1.7) being singularity free inside the corresponding circles
$\tilde {C}\left( {\mu}  \right)$ in the complex $\tau $-plane. We also know
that, by increasing $\mu $ -- hence, by increasing the size of the circle
$\tilde {C}\left( {\mu}  \right)$, of diameter $\mu /\pi $ --
\textit{branch points} (of the corresponding solution $\underline {\zeta}
\left( {\tau}  \right)$ of (1.7)) are eventually enclosed inside it. These
\textit{branch points} are characterized by exponents $\gamma _{nm} =
1/\left( {1 + a_{nm}}  \right)$, hence by exponents that, for this example,
have a \textit{positive} imaginary part, $\mbox{Im}\left( {\gamma _{nm}}  \right) >
0$, since the imaginary part of the coupling constant $a_{nm} $ is
\textit{negative}, $\mbox{Im}\left( {a_{nm}}  \right) < 0$. In this case the main
effect of such a branch point, when one travels over the circle $\tilde
{C}\left( {\mu}  \right)$ namely follows the evolution of $\underline {\zeta
} \left( {\tau}  \right)$ with $\tau $ given by the formula $\tau = \mu
\left[ {\exp\left( {2\pi it} \right) - 1} \right]/\left( {2\pi
i} \right)$ (see (1.6) and (6.1)), is to yield the \textit{shrinking}
factor $\exp\left[ { - 2\pi \,\mbox{Im}\left( {\gamma _{nm}}  \right)t}
\right]$, as entailed by the formula
\begin{gather}
\left( {\tau - \tau _{b}}  \right)^{\gamma _{nm}}  = \left\{ \mu
\left[ {\exp\left( {2\pi it} \right) - 1} \right]/\left( {2\pi
i} \right) - \tau _{b} \right\}^{\gamma _{nm}}
 = \exp\left[ { - 2\pi \,\mbox{Im}\left( {\gamma _{nm}}  \right)t} \right]
\nonumber\\
\qquad {}\times \exp\left[ {2\pi i\,\mbox{Re}\left( {\gamma _{nm}}  \right)t}
\right]\left\{ \mu \left[ {1 - \exp\left( {-2\pi it} \right)} \right]/
(2\pi i)-\tau_{b}\exp\left( {-2\pi it} \right)\right\}.\label{eq6.15}
\end{gather}
The effect of such shrinking, as long as it is relevant, is to reduce --
qualitatively -- by the same amount the magnitude of the (nonlinear)
right-hand side of the equations of motion (1.5) or (1.7); but we know (see
Section~1) that when this happens eventually one reaches the regime
characterized by the \textit{completely periodic} solutions of period $T =
1$ of~(1.5), the existence of which is guaranteed by Proposition~\ref{prop-1.2}.
This is what one sees in the trajectories displayed in the Figs.
(6.4c-f): in fact, given the relative scale of the motions of the various
particles, what one apparently sees is that one of the three particles
eventually stops (asymptotically; or it might actually tend to a periodic
trajectory of very small size, moving along it very slowly since the period
of that asymptotic motion is always $T = 1$; in either case, because of its
slow motion, it essentially decouples hence moves on a circular orbit,
neither feeling the presence of the other particles nor affecting their
motion, due to the velocity-dependence of the two-body forces, see the
right-hand side of (1.5)), while the other two particles approach
asymptotically (in fact, given the graphical limitations, seem to eventually
follow) periodic trajectories, always of course with period $T = 1$.
Specifically, the first transition from the \textit{completely periodic}
regime to that affected by the presence of a \textit{branch point} occurs
at a value of $\mu $ between $1.682$ and $1.683$, due to a collision among
the \textit{Red} and \textit{Green} particles occurring approximately at $t
= 0.85$ (see Figs. 6.4b,c; note that the latter of these two figures
displays a motion in which the \textit{Green} and \textit{Blue} particles
follow asymptotically periodic trajectories, while the \textit{Red} particle
seems to stop asymptotically). Several other transitions occur for larger
values of $\mu $. A representative instance, due to a collision among the
\textit{Green} and \textit{Blue} particles occurring approximately at $t =
0.85$, is displayed by Figs. 6.4d,e: note that Fig. 6.4d displays a motion
in which asymptotically all three particles move on periodic trajectories,
but the \textit{Red} one moves on a circle of very small size; while Fig.
6.4e displays as well a motion with all three particles asymptotically on
periodic trajectories, but now with the \textit{Blue} one approaching
asymptotically a circular trajectory of very small size, in fact so small
that it appears merely as a point in the figure. Finally Fig. 6.4f displays
a motion in which the \textit{Blue} particle seems to stop asymptotically,
while the other two approach periodic trajectories.

The interpretation of this phenomenology as outlined above is confirmed by
the numerical results of the last example we report, in which the imaginary
parts of the coupling constants $a_{nm} $ are instead \textit{positive},
$\mbox{Im}\left( {a_{nm}}  \right) > 0$. This example is characterized by the
parameters
\begin{gather}
N = 3 ;\qquad a_{12} = a_{21} = a_{RG} = 1 + 0.4i ,\nonumber\\
\label{eq6.16}
a_{13} = a_{31} = a_{RB} = 0.5 + 0.3i ,
\qquad a_{23} = a_{32} = a_{GB} = 1.5 + 0.2i,
\end{gather}
and the corresponding trajectories are displayed in the sequence of Figs.
6.5a-f; as above, in each figure the corresponding value is indicated of the
scaling factor $\mu $ that determines the initial data via (\ref{eq6.3}) with
(\ref{eq6.6}), as well as the period -- which is of course always $T = 1$ -- in the
case of \textit{periodic} motions (see Figs. 6.5a,b) or instead the final
time $t$ of the displayed trajectories in the case of \textit{nonperiodic}
motions (see Figs. 6.5c-f). As in the preceding example, the
\textit{completely periodic} motions of period $T = 1$ displayed in the
first two figures correspond to the solutions predicted by
Proposition~\ref{prop-1.2}, namely they correspond to the solution $\underline
{\zeta}  \left( {\tau}  \right)$ of (1.7) being singularity-free inside
$\tilde {C}\left( {\mu}  \right)$; while the behavior (\textit{spiraling
outwardly}) displayed by the other four figures is consistent with the
coming into play of an \textit{expanding} factor, the origin of which can be
explained exactly as above (see (\ref{eq6.15})), the difference among
\textit{expanding} and \textit{shrinking} behavior being of course
consistent with the change of sign of the imaginary parts of the coupling
constants $a_{nm} $. Specifically one sees from Figs. 6.5b,c that the first
transition from the \textit{completely periodic} regime is due to a
collision among the \textit{Red} and \textit{Green} particles (occurring
approximately at $t = 0.74$), and another transition is apparent from Figs.
6.5d,e, due a collision among the \textit{Green} and \textit{Blue} particles
(occurring approximately at $t = 0,69$). Several other transitions occur
between these, as well as for values of $\mu $ between $1.010$ (see Fig.
6.5c) and $3.474$ (see Fig. 6.5d), between $3.475$ (see Fig. 6.5e) and $20$
(see Fig. 6.5f), and beyond $20$. Note that in all these simulations -- with
$\mu \ge 1.010$, namely when the motions are no more periodic -- all three
particles eventually spiral out. This, however, does not seem to represent a
universal behavior; other (three-body) cases have been observed, with all
coupling constants having \textit{positive} imaginary parts, in which only
two particles spiral out asymptotically, while the third one seems to stop.
Clearly a more complete analysis of this phenomenology will have to be made;
indeed while the discussion of the last two examples seems to provide a
satisfactory qualitative understanding of the numerical results reported, it
also confirms the richness of behaviors displayed by the many-body problem
in the plane (1.1), or equivalently (1.5), leaving an ample scope for
additional study (for instance, of cases with coupling constant some of
which have \textit{positive}, and others \textit{negative}, imaginary parts)
-- and of course there is moreover the case with coupling constants having a
negative \textit{real} part, the investigation of which shall be reported
separately.

Let us finally emphasize that the findings reported above are merely a tiny
subset -- selected on the basis of their representative character -- of the
data we collected in our numerical explorations. Indeed we expect that every
interested reader will eventually like to engage in personal
experimentations, as soon as the computer program~[8] which yielded these
results shall become available (it is planned to put it on the web soon). In
particular we recommend once again as a thrilling experience the observation
as they evolve over time of periodic motions with \textit{large} periods,
such as some of those which have been displayed above (but only in the guise
of trajectories: see for instance Figs. 6.1e-h and especially Figs. 6.2c-k,
6.3c-h).

\newpage

\begin {center}
\includegraphics[width=15cm] {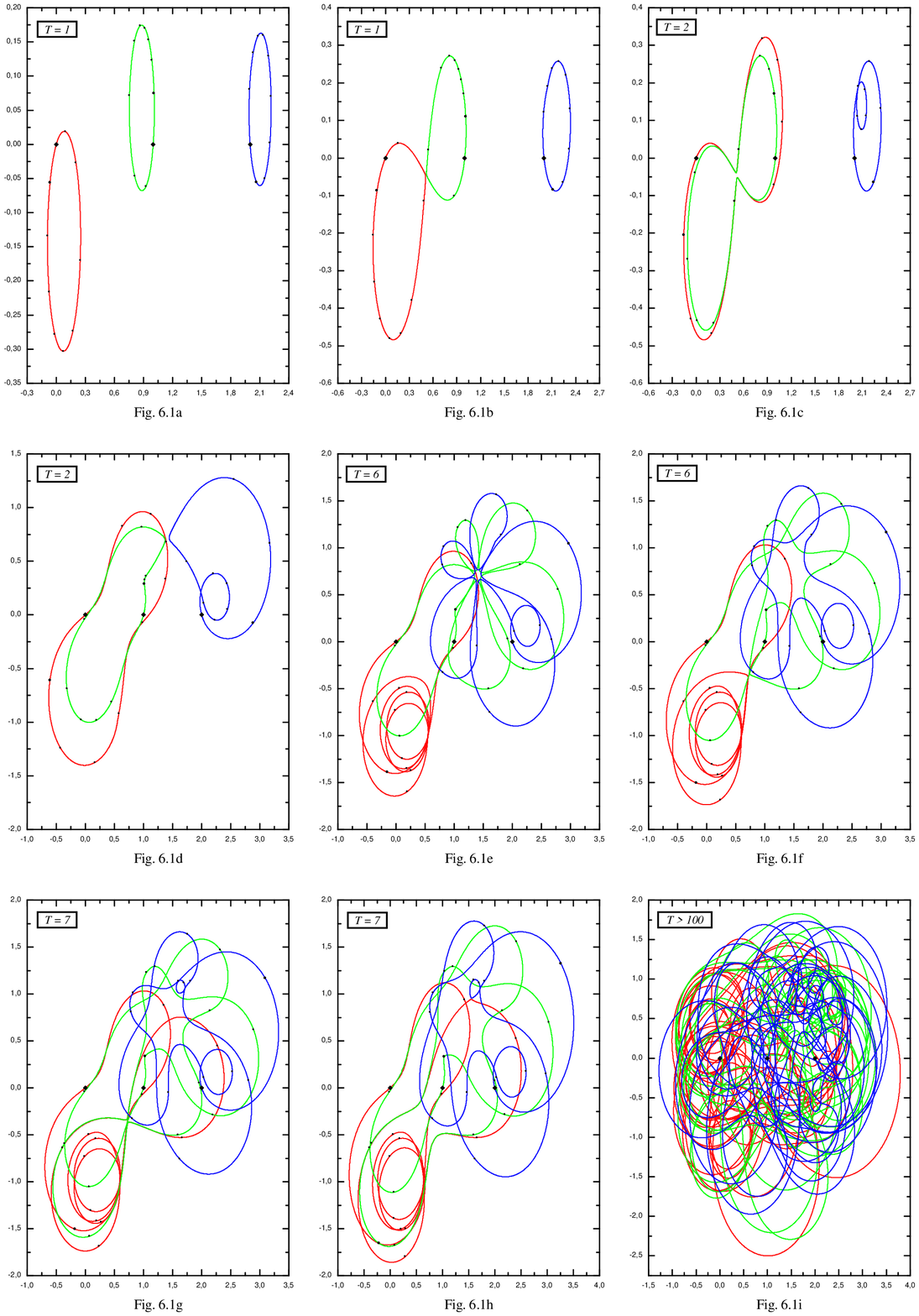}
\end {center}

\newpage

\begin {center}
\includegraphics[width=15cm] {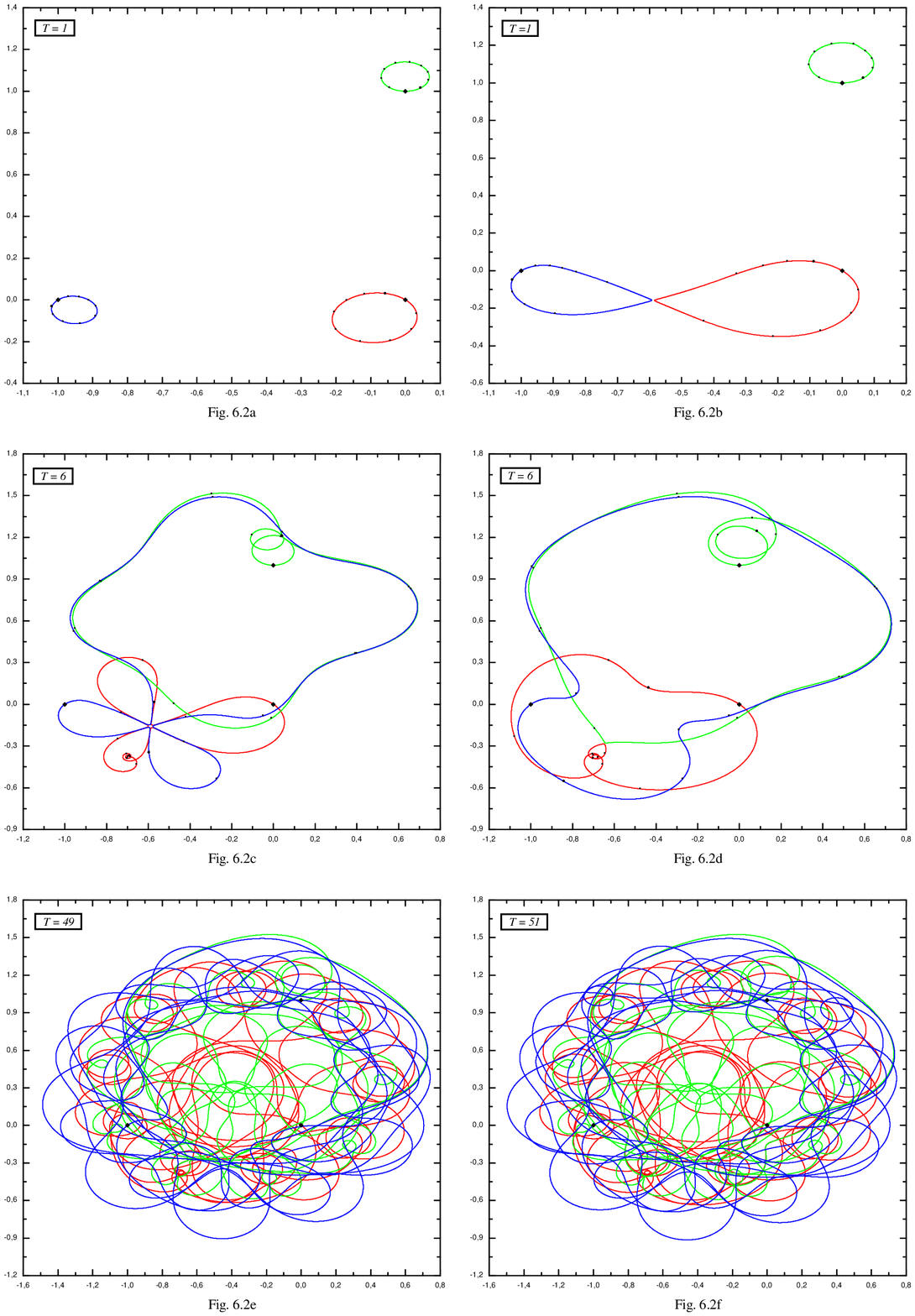}
\end {center}

\newpage

\begin {center}
\includegraphics[width=15cm] {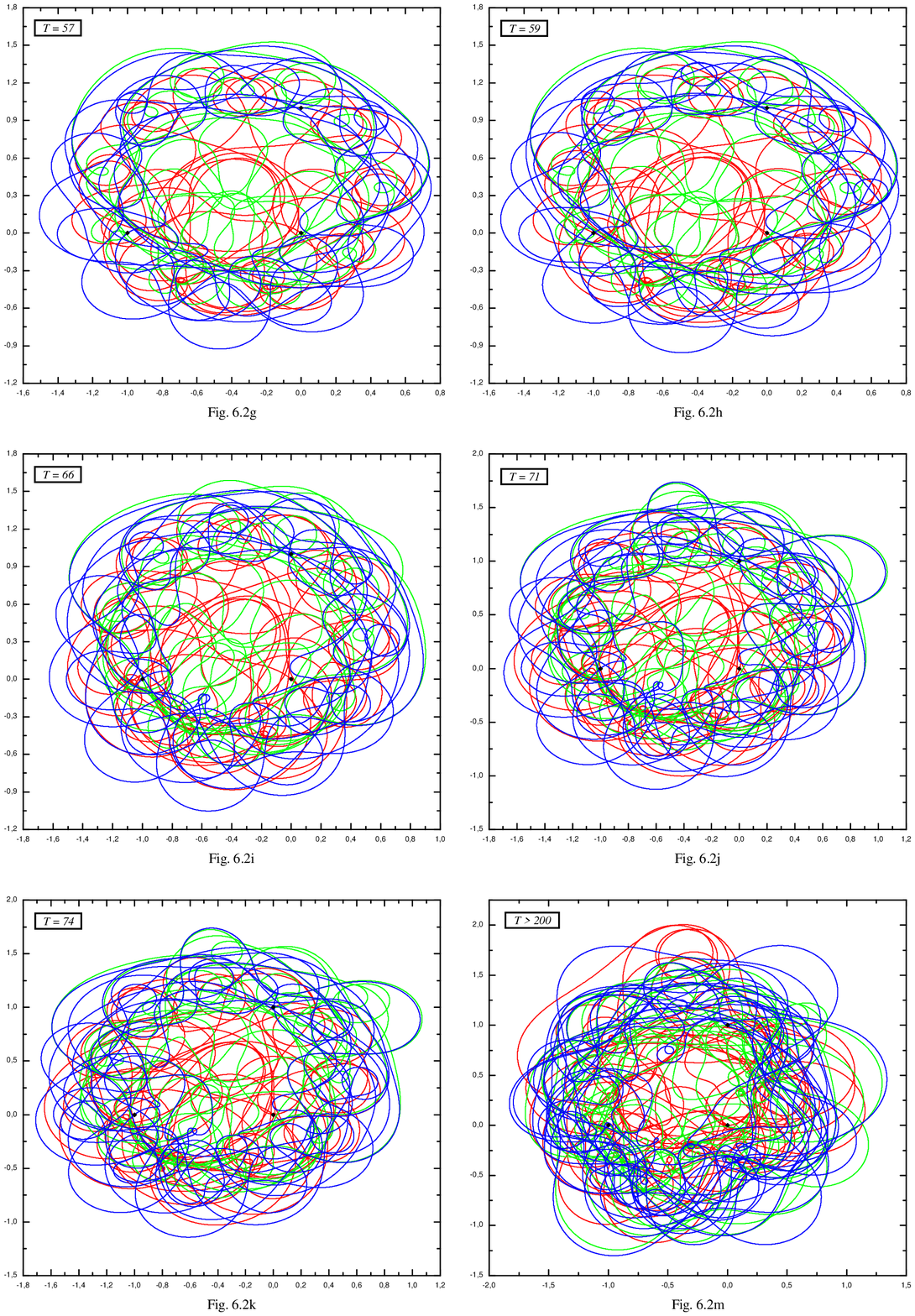}
\end {center}

\newpage

\begin {center}
\includegraphics[width=15cm] {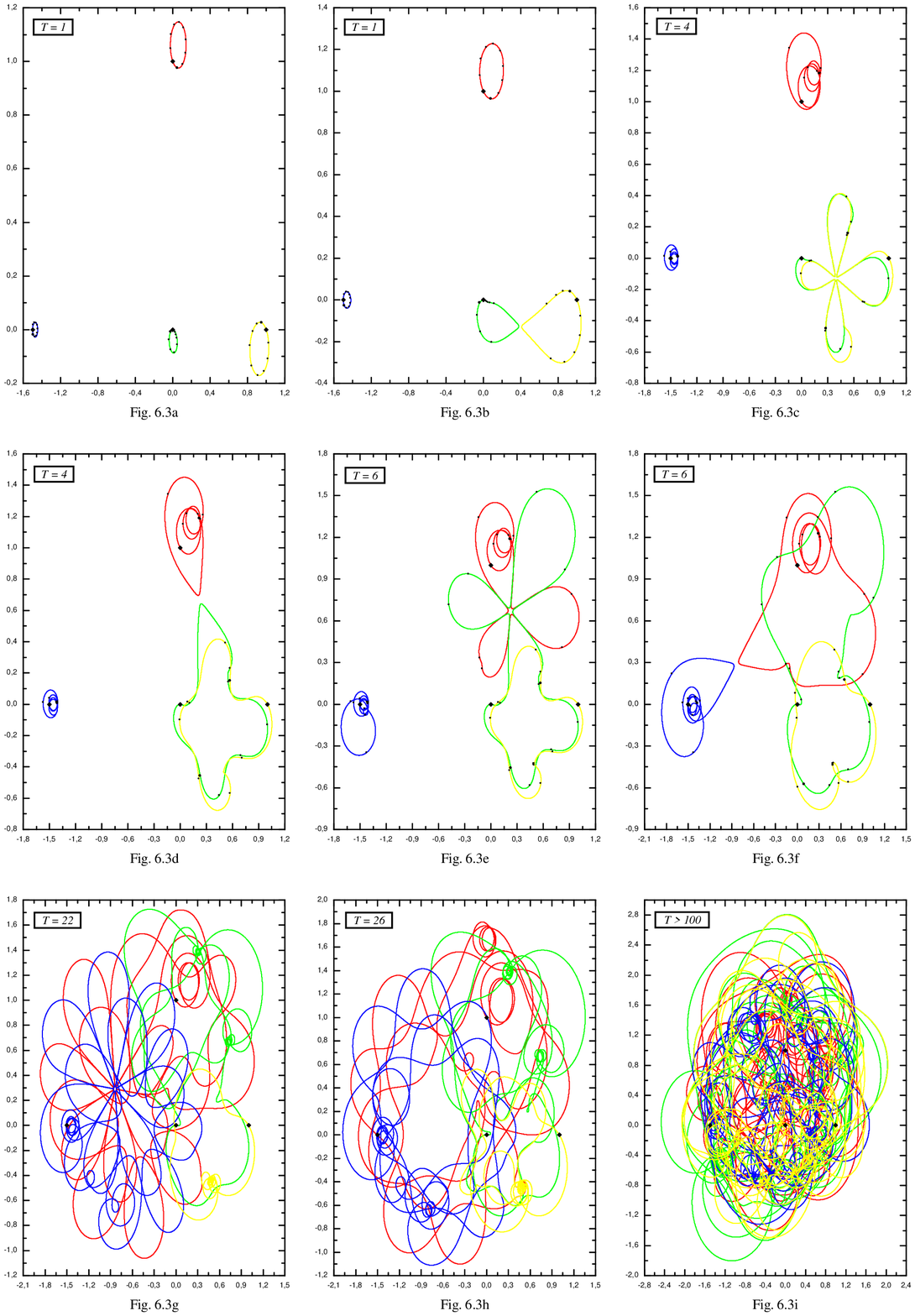}
\end {center}
\newpage

\begin {center}
\includegraphics[width=15cm] {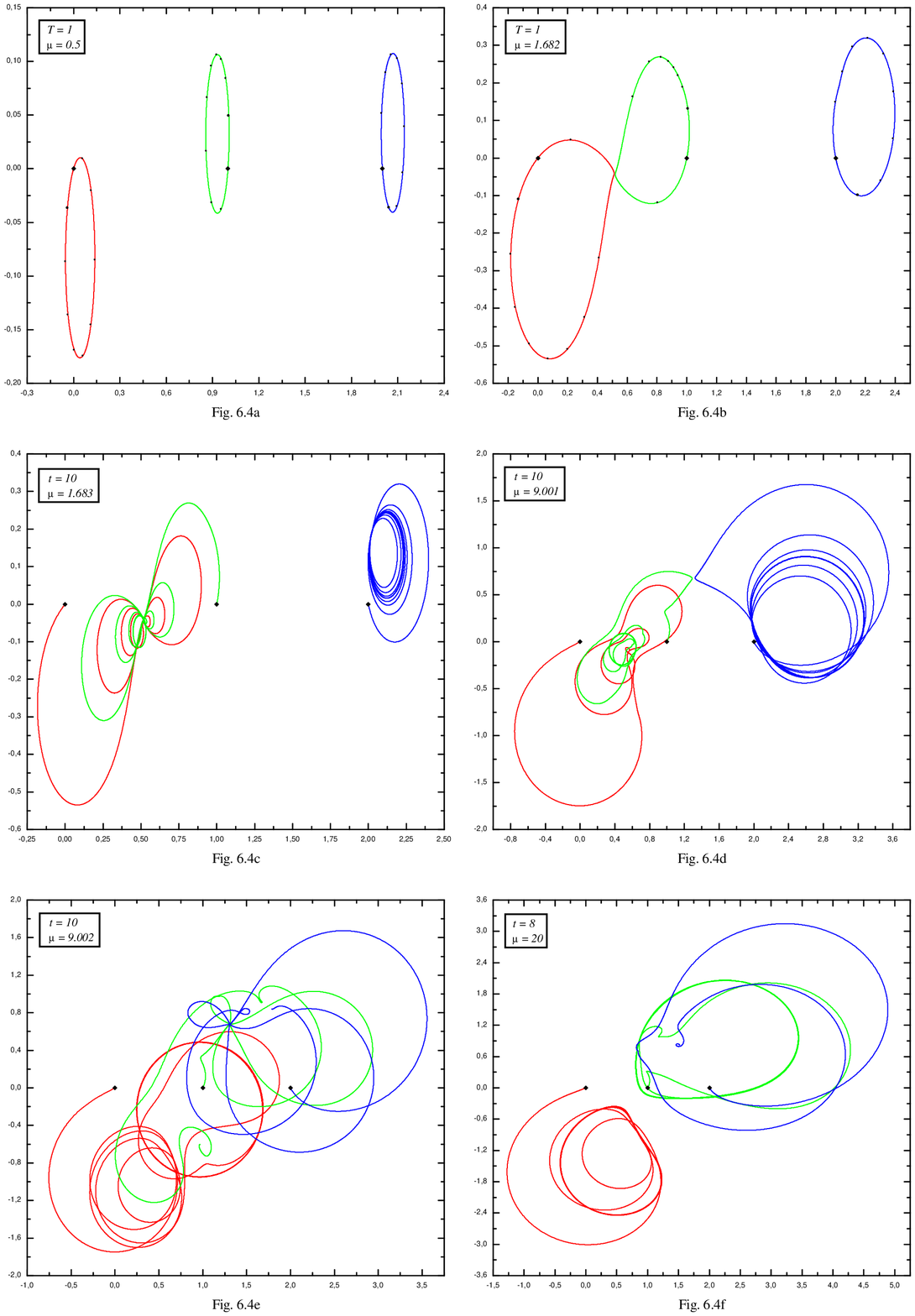}
\end {center}

\newpage

\begin {center}
\includegraphics[width=15cm] {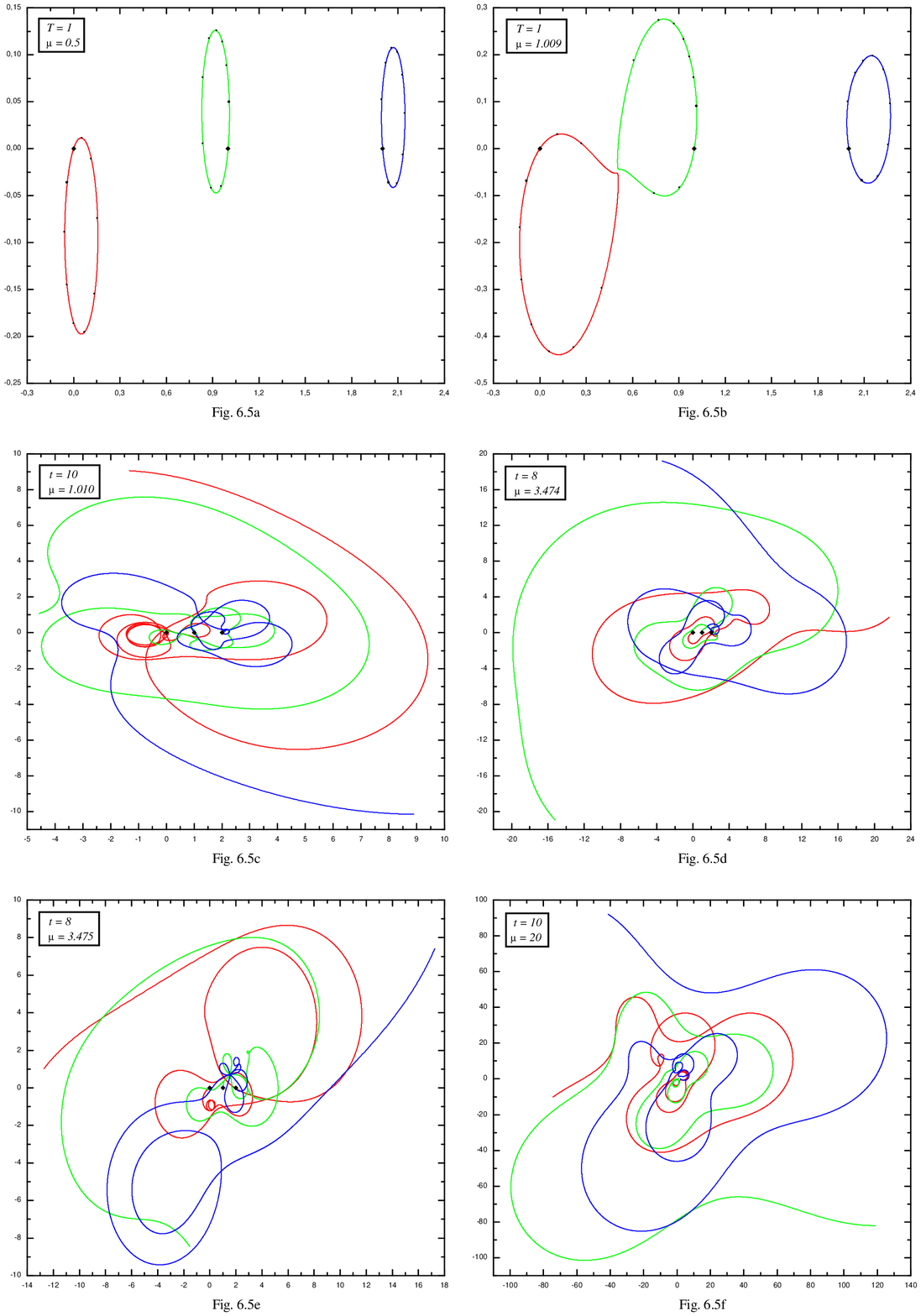}
\end {center}
\newpage

\section{Outlook}

The rich phenomenology featured by the solutions of the many-body
problem in the plane investigated in this paper led us to refer to
it, at least colloquially among us, as a ``goldfish'', thereby
extending to this model the terminology that was initially
suggested for its integrable variant~[12]. Much additional fun and
insight can certainly be gleaned from a more complete exploration
of its solutions than we were able to report here; let us
re-emphasize that the most illuminating way to pursue this study
shall be provided by the computer code that simulates -- by
solving numerically the relevant equations of motions, see (1.1)
-- the actual behavior of the many-body problem as it unfolds over
time; this computer code is indeed a remarkably efficient research
tool which we hope to make available soon to the scientific
community at large~[8]. As for the more theoretical aspect of the
investigation of this many-body problem, let us recall that in
this paper a rather satisfactory understanding of the behavior of
the system has been provided for the case in which the coupling
constants $a_{nm}$ are all \textit{nonnegative}, $a_{nm} \ge 0$,
see (1.1) (or, equivalently, $\mbox{Re}(a_{nm}) \ge 0$, see (1.5)), while
the case when this restriction does not hold still needs to be
fully investigated.

\subsection*{Acknowledgements}

While the results reported in this paper were under development we discussed
them with several colleagues, who often provided precious suggestions. In
particular we like to thank for these Robert Conte, Ovidiu Costin, Herman
Flaschka, Giovanni Gallavotti, Nalini Joshi, Martin Kruskal, Fran\c{c}ois
Leyvraz, Alexander Turbiner.

\renewcommand{\theequation}{A.\arabic{equation}}
\setcounter{equation}{0}
\section*{\large Appendix A}

In this Appendix we provide a derivation of (1.9). Our starting point is the
standard theorem that guarantees the analyticity of the solutions of
(systems of) analytic ODEs (see for instance Section 13.21 of [7]). Our
treatment is closely analogous to that of [10].

The standard formulation of this theorem refers to a system of first-order
ODEs, say
\begin{equation}
{w}'_{j} = f_{j} \left( {w_{1} ,w_{2} ,\ldots,w_{2N}}  \right),\qquad j = 1,2,\ldots,2N,
\end{equation}
which coincides with our system, see (1.7), via the assignments
\begin{subequations}
\begin{gather}
w_{n} \equiv w_{n} \left( {\tau}  \right) = \zeta _{n} \left( {\tau}
\right) - \zeta _{n} \left( {0} \right),\\
w_{N + n} \equiv w_{N + n} \left( {\tau}  \right) = c\left[ {{\zeta} '_{n}
\left( {\tau}  \right) - {\zeta} '_{n} \left( {0} \right)} \right] ,
\end{gather}
\end{subequations}
as well as
\begin{subequations}
\begin{gather}
f_{n} = c^{ - 1}w_{N + n} + {\zeta} '_{n} \left( {0} \right) ,\\
f_{N+n} = 2c\sum\limits_{m=1,m \ne n}^{N} a_{nm}
\frac{[c^{-1} w_{N+n} + \zeta'_{n}(0)]
[c^{-1} w_{N+m} + \zeta'_{m}(0)]}
{[w_{n} - w_{m} + \zeta_{n}(0) - \zeta_{m}(0)]}.
\end{gather}
\end{subequations}
Here of course (as in the rest of our paper) the index $n$ takes all integer
values in the range $1 \le n \le N$, and $N$ is the total number of
particles. Note that with these assignments all the $2N$ quantities $w_{j}
\left( {\tau}  \right)$ vanish at the origin, $w_{j} \left( {0} \right) =
0$, as appropriate to reproduce the notation of [7]. In these formulas $c$
is an arbitrary \textit{positive} constant (which we have introduced for
convenience, see below).

The standard result [7] provides then the following lower bound for the
radius $\rho $ of the circular disk $D$, centered at the origin $\tau = 0$
in the complex $\tau $-plane, within which the solutions $w_{n} \left( {\tau
} \right)$ of (A.1), hence (see (A.2a)) as well the solutions $\zeta _{n}
\left( {\tau}  \right)$ of (1.7), are holomorphic:
\begin{equation}
\rho > b/\left[ {\left( {2N + 1} \right)M} \right]
\end{equation}
(this formula coincides with the last equation of Section 13.21 of [7], with
the assignments $m = 2N$ and $a = \infty $, the first of which is
justified by the fact that (A.1) features $2N$ equations, the second of
which is justified by the \textit{autonomous} character of our equations of
motion, see (A.1) with (A.3)). The two \textit{positive} quantities $b$ and
$M$ in this equation, (A.4), are defined as follows. The quantity $b$ is
required to guarantee that the right-hand sides of (A.1) are
\textit{holomorphic} provided there hold the $2N$ inequalities
\begin{equation}
\left| {w_{j}}  \right| \le b;
\end{equation}
this entails, in our case, the single restriction
\begin{equation}
b < \tilde {\zeta} /2
\end{equation}
with $\tilde {\zeta} $ defined by (1.9b), because the only source of
singularities is in our case the vanishing of a denominator in the
right-hand side of (A.3b) (which is clearly excluded by (A.5) with (A.6) and
(1.9b)). The second quantity in (A.4), $M \equiv M\left( {b} \right)$, is
the upper bound of the right-hand sides of (A.1) when the quantities $w_{j}
$ satisfy the restriction (A.5) with (A.6); but of course the inequality
(A.4) holds \textit{a fortiori} if we overestimate $M$ (as we shall now do).

Indeed, the right-hand sides of (A.3) clearly entail, via (A.2) and
(1.9b,c),
\begin{equation}
M < \max\left[ c^{ - 1}b + {\tilde {\zeta
}}',2cN\alpha \left( {c^{ - 1}b + {\tilde {\zeta} }'}
\right)^{2}/\left( {\tilde {\zeta}  - 2b} \right) \right]
\end{equation}
where the maximum should be taken among the two terms, while it remains our
privilege to choose the values of the \textit{positive} quantity $c$, as
well as the value of the \textit{positive} quantity~$b$ (respecting of
course the inequality (A.6)). The \textit{positive} quantity $\alpha $ in
this formula is of course the maximal value of the moduli of the ``coupling
constants'' $a_{nm} $,
\begin{equation}
\alpha = \max\limits_{n,m = 1,\ldots,N;\,n \ne m} \left| {a_{nm}} \right|  .
\end{equation}

We now make the convenient choice
\begin{equation}
c = \left[ \tilde {\zeta}  - 2b\left( {1 + N\alpha}
\right) \right]/\left[ 2N\alpha {\tilde {\zeta
}}' \right],
\end{equation}
which equalizes the two terms inside the maximum function in (A.7). Note
that the requirement that $c$ be \textit{positive} entails that (A.6) must
now be replaced by the more stringent inequality
\begin{equation}
b < \tilde {\zeta} /\left[ {2\left( {1 + N\alpha}  \right)} \right].
\end{equation}

Hence from (A.7) we now get
\begin{equation}
M < {\tilde {\zeta} }'\left( {\tilde {\zeta}  - 2b} \right)/\left[
\tilde {\zeta}  - 2b\left( {1 + N\alpha}  \right)\right],
\end{equation}
and via (A.4) there thus obtains the formula
\begin{equation}
\rho > R\tilde {\zeta}/{\tilde {\zeta} }'
\end{equation}
with
\begin{equation}
R = x\left[ {1 - 2\left( {1 + N\alpha}  \right)x} \right]/\left[
{\left( {2N + 1} \right)\left( {1 - 2x} \right)} \right],
\end{equation}
where the choice of the \textit{positive} quantity $x = b/\tilde {\zeta} $
remains our privilege, provided it falls in the interval
\begin{equation}
0 < x < \left[ {2\left( {1 + N\alpha}  \right)} \right]^{ - 1}
\end{equation}
which corresponds to (A.10).

Since (A.12) coincides with (1.9a), this latter equation is now proven. It
would be easy to obtain the optimal choice of the constant $R$ by maximizing
the right-hand side of (A.13) in the interval (A.14). More simply we just
take $x$ in the middle of that interval (A.14), namely we set
\begin{equation}
x = \left[ {4\left( {1 + N\alpha}  \right)} \right]^{- 1},
\end{equation}
and we thereby obtain for $R$ the following simple expression:
\begin{equation}
R = \left[ {4\left( {1 + 2N} \right)\left( {1 + 2N\alpha}
\right)} \right]^{- 1}.
\end{equation}

\renewcommand{\theequation}{B.\arabic{equation}}
\setcounter{equation}{0}
\section*{\large Appendix B}

In this appendix we investigate the nature of the branch points of the
solutions of the first order ODE (4.4b),
\begin{subequations}
\begin{equation}
\left( {{\zeta} '} \right)^{2} = V^{2}\left[ 1 + \left( {\zeta /L}
\right)^{ - 2a}\right],
\end{equation}
with $V$ and $L$ arbitrary (nonvanishing) constants, obviously related to
the initial data:
\begin{equation}
V = {\zeta} '_{1} \left( {0} \right) + {\zeta} '_{2} \left( {0} \right)
\end{equation}
(see (4.2)), and
\begin{equation}
L = \zeta \left( {0} \right)\left\{\left[ {{\zeta} '\left( {0}
\right)/V} \right]^{2} - 1\right\}^{1/\left( {2a} \right)}
\end{equation}
\end{subequations}
(see (B.1a)).

Let us begin by setting
\begin{equation}
\zeta \left( {\tau}  \right) = Lf\left( {w} \right),\qquad
w = \left( {V/L} \right)\tau,
\end{equation}
so that the equation satisfied by $f\left( {w} \right)$ read
\begin{subequations}
\begin{equation}
\left[ {{f}'\left( {w} \right)} \right]^{2} = 1 + \left[ {f\left( {w}
\right)} \right]^{- 2a} = 1 + \left[ {f\left( {w} \right)}
\right]^{\beta}.
\end{equation}
Here and below
\begin{equation}
\beta = - 2a,
\end{equation}
\end{subequations}
and the prime appended to the function $f$ signifies of course
differentiation with respect to its argument $w$.

Two possible mechanisms are at play to yield a singularity of $f\left( {w}
\right)$, say at $w = w_{b} $: \textit{(i)}~the vanishing of $f\left( {w} \right)$,
\begin{subequations}
\begin{equation}
f\left( {w_{b}}  \right) = 0;
\end{equation}
\textit{(ii)} the divergence of $f\left( {w} \right)$,
\begin{equation}
f\left( {w_{b}}  \right) = \infty.
\end{equation}
\end{subequations}
Let us discuss them separately. But in both cases our approach will be based
on replacing (B.3a) by a recursive set of equations, namely
\begin{equation}
\left[ {{}^{\left( {j + 1} \right)}{f}'\left( {w} \right)} \right]^{2} = 1
+ \left[ {{}^{\left( {j} \right)}f\left( {w} \right)} \right]^{ - 2a} = 1
+ \left[ {{}^{\left( {j} \right)}f\left( {w} \right)} \right]^{\beta
},\qquad j = 0,1,2,\ldots,
\end{equation}
and by assigning appropriately the initial element, ${}^{\left( {0}
\right)}f\left( {w} \right)$, of this iteration, so that it account properly
for the leading behavior of the solution $f\left( {w} \right)$ in the
neighborhood of the singular point $w = w_{b} $ -- as demonstrated by the
preservation of this singular behavior through the iteration (see below).

Let us begin with mechanism \textit{(i)}, by therefore setting
\begin{equation}
{}^{\left( {j} \right)}f\left( {w_{b}}  \right) = 0,\qquad j = 0,1,2,\ldots .
\end{equation}
Firstly we consider the case characterized by the inequality
\begin{subequations}
\begin{equation}
\mbox{Re}\left( {a} \right) > 0,
\end{equation}
\end{subequations}
and we introduce correspondingly the exponent (see below)
\begin{subequations}
\begin{equation}
\gamma = 1/\left( {1 + a} \right),
\end{equation}
\end{subequations}
which clearly then satisfies the restriction
\setcounter{equation}{6}
\begin{subequations}
\setcounter{equation}{1}
\begin{equation}
0 < \mbox{Re}\left( {\gamma}  \right) < 1.
\end{equation}
\end{subequations}
Correspondingly we set (consistently with (B.6))
\setcounter{equation}{8}
\begin{equation}
{}^{\left( {0} \right)}f\left( {w} \right) = \gamma ^{- \gamma} \left(
{w - w_{b}}  \right)^{\gamma},
\end{equation}
and via (B.5) we then get
\begin{subequations}
\begin{equation}
\left[ {{}^{\left( {1} \right)}{f}'\left( {w} \right)} \right]^{2} =
\left[ {{}^{\left( {0} \right)}f\left( {w} \right)} \right]^{ -
2a}\left\{ 1 + \left[ {{}^{\left( {0} \right)}f\left( {w}
\right)} \right]^{ - 2a}\right\},
\end{equation}
namely
\begin{equation}
{}^{\left( {1} \right)}{f}'\left( {w} \right) = \left[ {{}^{\left( {0}
\right)}f\left( {w} \right)} \right]^{ - a}\left\{ 1 + \left[
{{}^{\left( {0} \right)}f\left( {w} \right)} \right]^{ - 2a}\right\}^{1/2},
\end{equation}
namely (via (B.9))
\begin{equation}
{}^{\left( {1} \right)}{f}'\left( {w} \right) = \gamma ^{1 - \gamma
}\left( {w - w_{b}}  \right)^{\gamma - 1}\left\{1 + \gamma
^{2\left( {\gamma - 1} \right)}\left( {w - w_{b}}  \right)^{2\left( {1
- \gamma}  \right)} \right\}^{1/2}.
\end{equation}
\end{subequations}
To get this expression we used the relation
\setcounter{equation}{7}
\begin{subequations}
\setcounter{equation}{1}
\begin{equation}
a\gamma = 1 - \gamma
\end{equation}
\end{subequations}
which is clearly entailed by (B.8a).

In the neighborhood of $w = w_{b} $ we can rewrite (B.10c) as follows:
\setcounter{equation}{9}
\begin{subequations}
\setcounter{equation}{3}
\begin{gather}
{}^{\left( {1} \right)}{f}'\left( {w} \right) = \gamma ^{1 - \gamma
}\left( {w - w_{b}}  \right)^{\gamma - 1}\nonumber\\
\qquad \times \left\{1 +
\sum\limits_{l = 1}^{\infty}  {} \left( {\begin{array}{c}
1/2\\
l
\end{array}} \right)\gamma ^{2l\left( {\gamma - 1} \right)}\left( {w -
w_{b}}  \right)^{2l\left( {1 - \gamma}  \right)}\right\}.
\end{gather}
\end{subequations}
Here and below we use of course the usual notation for the binomial
coefficient, so that
\begin{gather}
\left( {\begin{array}{c}
1/2\\
l
\end{array}} \right) = \Gamma \left( {3/2} \right)\left[ {\Gamma \left(
{3/2 - l} \right)\Gamma \left( {1 + l} \right)} \right]^{- 1} \nonumber\\
\phantom{\left( {\begin{array}{c}
1/2\\
l
\end{array}} \right)} {}= \left(
{ - 2} \right)^{ - \left( {1 + l} \right)}\left( {l - 1/2} \right)\left(
{2l - 1} \right)!!/l!.
\end{gather}
We then integrate (B.10d) (using (B.6) and (B.7b)) and we thereby get (using
(B.8a))
\begin{gather}
{}^{(1)}f(w) = \gamma^{-\gamma} (w - w_{b})^{\gamma}\nonumber\\
\phantom{{}^{(1)}f(w) =}\times\left\{ 1 + \sum\limits_{l = 1}^{\infty}
(1 + 2la)^{-1} \left( {\begin{array}{c}
1/2\\
l
\end{array}} \right) \gamma^{2l(\gamma-1)}
(w-w_{b})^{2l(1-\gamma)} \right\}.
\end{gather}
A comparison of this expression, (B.12), with (B.9) demonstrates the
preservation of the leading term characterizing the behavior in the
neighborhood of the singularity.

This result, as well as the following development, suggest setting
\begin{gather}
{}^{(j)}f(w) = \gamma^{-\gamma}(w - w_{b})^{\gamma}
\left\{ 1 + \sum\limits_{l = 1}^{\infty} \sum\limits_{k = 0}^{l}
{}^{(j)}g_{kl}(w - w_{b})^{k\gamma}(w - w_{b})^{2l(1-\gamma)} \right\},\nonumber\\
\phantom{{}^{(j)}f(w) =}
j = 0,1,2\ldots,
\end{gather}
of course with
\begin{equation}
{}^{\left( {0} \right)}g_{kl} = 0,\qquad {}^{\left( {1} \right)}g_{kl} =
\delta _{k0} \left( {1 + 2la} \right)^{ - 1}\left( {\begin{array}{c}
1/2\\
l
\end{array}} \right)\gamma ^{2l\left( {\gamma - 1} \right)}.
\end{equation}

We now show that this \textit{ansatz}, (B.13), is preserved by the iteration
(B.5), which we now rewrite in the following form (analogous to (B.10b)):
\begin{equation}
{}^{\left( {j + 1} \right)}{f}'\left( {w} \right) = \left[ {{}^{\left( {j}
\right)}f\left( {w} \right)} \right]^{ - a}\left\{1 + \left[
{{}^{\left( {j} \right)}f\left( {w} \right)} \right]^{ - 2a}\right\}^{1/2}.
\end{equation}
Indeed the insertion of (B.15) in (B.13) yields
\begin{gather}
{}^{(j+1)}f(w) = \gamma^{-\gamma}(w - w_{b})^{\gamma}
\left\{ 1 + \sum\limits_{l = 1}^{\infty} \sum\limits_{k = 0}^{l}
{}^{(j)}\tilde{g}_{kl}(w - w_{b})^{k\gamma}(w - w_{b})^{2l(1-\gamma)}\right\},\nonumber\\
\phantom{{}^{(j+1)}f(w) =}j = 0,1,2,\ldots,
\end{gather}
with the coefficients ${}^{\left( {j} \right)}\tilde {g}_{kl} $ related to
the coefficients ${}^{\left( {j} \right)}g_{kl} $ by the formula
\begin{gather}
\left\{ 1 + \sum\limits_{l = 1}^{\infty}  {} \sum\limits_{k =
0}^{l} {} {}^{\left( {j} \right)}g_{kl} x^{k\gamma} y^{l}\right\}^{ - a}
\left[1 + \gamma ^{2\left( {\gamma - 1}
\right)}y\left\{1 + \sum\limits_{l = 1}^{\infty}  {}
\sum\limits_{k = 0}^{l} {} {}^{\left( {j} \right)}g_{kl} x^{k\gamma
}y^{l}\right\}^{2a} \right]^{1/2}\nonumber\\
\qquad {}= 1 + \sum\limits_{l = 1}^{\infty}  {} \sum\limits_{k = 0}^{l} {}
{}^{\left( {j} \right)}\tilde {g}_{kl} x^{k\gamma} y^{l}.
\end{gather}
To get this formula, (B.16) with (B.17), we used again (B.8b) and we also
set $x = (w-w_{b})^{\gamma}$, $y = (w-w_{b})^{2(1-\gamma)}$.
A closed form expression of the coefficients ${}^{\left( {j}
\right)}\tilde {g}_{kl} $ in terms of the coefficients ${}^{\left( {j}
\right)}g_{kl} $ is not available, but there is no difficulty in principle
to compute them sequentially, by expanding the left-hand side of (B.17) in
powers of $x$ and $y$.

It is then clear that (B.16) can be integrated to yield (via (B.6) and (B.7b))
\begin{gather}
{}^{\left( {j + 1} \right)}f\left( {w} \right) = \gamma ^{ - \gamma
}\left( {w - w_{b}}  \right)^{\gamma} \nonumber\\
\phantom{{}^{\left( {j + 1} \right)}f\left( {w} \right) =}\times \left\{ 1 +
\sum\limits_{l = 1}^{\infty}  {} \sum\limits_{k = 0}^{l} {} {}^{\left( {j +
1} \right)}g_{kl} \left( {w - w_{b}}  \right)^{k\gamma} \left( {w -
w_{b}}  \right)^{2l\left( {1 - \gamma}  \right)} \right\},
\end{gather}
with
\begin{equation}
{}^{\left( {j + 1} \right)}g_{kl} = \left( {1 + k + 2la} \right)^{ -
1}{}^{\left( {j} \right)}\tilde {g}_{kl}.
\end{equation}

The consistency of (B.18) with (B.13) validates the \textit{ansatz} (B.13)
for all (nonnegative integer) values of the index $j$, and by assuming that
this continues to hold in the limit as this index diverges, $j \to \infty $,
and that in this limit ${}^{\left( {j} \right)}f\left( {w} \right)$ yields
the solution $f\left( {w} \right)$ of (B.3a), as indicated by a comparison
of (B.3a) with (B.5), we infer for $f\left( {w} \right)$ the expression
\begin{gather}
f(w) = \gamma^{-\gamma}(w - w_{b})^{\gamma}
\left\{ 1 + \sum\limits_{l = 1}^{\infty} \sum\limits_{k = 0}^{l}
g_{kl}(w - w_{b})^{k\gamma}(w - w_{b})^{2l(1-\gamma)} \right\},
\nonumber\\
\phantom{f(w)=} j = 0,1,2,\ldots ,
\end{gather}
Here the coefficients $g_{kl} $ are of course defined in terms of the above
iteration, $g_{kl} = {}^{\left( {\infty}  \right)}g_{kl} $, and they depend
only on the ``coupling constant'' $a$ (and of course, as indicated by the
notation, on the two indices $k$ and $l$), and we of course assume that, for
$w$ sufficiently close to $w_{b} $, namely for $\left| {w - w_{b}}  \right|$
sufficiently small, the infinite sum over the index $l$ in the right-hand
side of this expression, (B.20), does converge.

Of course we could have directly started from the \textit{ansatz} (B.20) and
then verified its consistency with the nonlinear ODE (B.3a) -- but the
iterative approach appeared to us preferable in order to explain here how
one arrives at such an \textit{ansatz}.

Let us now proceed and discuss, always in the context of mechanism
\textit{(i)}, the complementary case to (B.7a), namely let us now assume
that
\begin{subequations}
\begin{equation}
\mbox{Re}\left( {a} \right) < 0 ,
\end{equation}
entailing of course
\begin{equation}
\mbox{Re}\left( {\beta}  \right) > 0 .
\end{equation}
\end{subequations}
It is now convenient to rewrite the iteration formula (B.5) in the form
\begin{equation}
{}^{\left( {j + 1} \right)}{f}'\left( {w} \right) = \pm \left\{ 1
+ \left[ {{}^{\left( {j} \right)}f\left( {w} \right)} \right]^{\beta
} \right\}^{1/2},\qquad j = 0,1,2,\ldots,
\end{equation}
and to solve it -- consistently with (B.4a), again with the initial
conditions (B.6)-- by starting from the zeroth-order assignment
\begin{equation}
{}^{\left( {0} \right)}f\left( {w} \right) = \pm \left( {w - w_{b}}  \right).
\end{equation}
We then get
\begin{subequations}
\begin{gather}
{}^{\left( {1} \right)}{f}'\left( {w} \right) = \pm \left\{ 1 +
\left[ { \pm \left( {w - w_{b}}  \right)} \right]^{\beta} \right\}^{1/2},\\
{}^{\left( {1} \right)}{f}'\left( {w} \right) = \pm \left\{ 1 +
\sum\limits_{l = 1}^{\infty}  {} \left( {\begin{array}{c}
{1/2}\\
{l}
\end{array}} \right)\left[ { \pm \left( {w - w_{b}}  \right)}
\right]^{l\beta}  \right\},
\end{gather}
\end{subequations}
entailing, via (B.6),
\begin{equation}
{}^{\left( {1} \right)}f\left( {w} \right) = \pm \left( {w - w_{b}}
\right)\left\{ 1 + \sum\limits_{l = 1}^{\infty}  {} \left( {1 +
l\beta}  \right)^{ - 1}\left( {\begin{array}{c}
{1/2}\\
{l}
\end{array}} \right)\left[ { \pm \left( {w - w_{b}}  \right)}
\right]^{l\beta} \right\}.
\end{equation}

We now set, as suggested by this formula, (B.25),
\begin{equation}
{}^{\left( {j} \right)}f\left( {w} \right) = \pm \left( {w - w_{b}}
\right)\left\{ 1 + \sum\limits_{l = 1}^{\infty}  {} {}^{\left(
{j} \right)}g_{l} \left[ { \pm \left( {w - w_{b}}  \right)}
\right]^{l\beta}  \right\},\qquad j = 0,1,2,\ldots,
\end{equation}
and verify the consistency of this \textit{ansatz}, (B.26), with the
iteration (B.22). Indeed, by proceeding as above, we get
\begin{equation}
{}^{\left( {j + 1} \right)}f\left( {w} \right) = \pm \left( {w - w_{b}}
\right)\left\{ 1 + \sum\limits_{l = 1}^{\infty}  {} {}^{\left(
{j + 1} \right)}g_{l} \left[ { \pm \left( {w - w_{b}}  \right)}
\right]^{l\beta} \right\},
\end{equation}
with
\begin{equation}
{}^{\left( {j + 1} \right)}g_{l} = \left( {1 + l\beta}  \right)^{ -
1}{}^{\left( {j} \right)}\tilde {g}_{l}
\end{equation}
and
\begin{equation}
\left\{ 1 + x\left[ 1 + \sum\limits_{l = 1}^{\infty}
{} {}^{\left( {j} \right)}g_{l} x^{l} \right]
\right\}^{1/2} = 1 + \sum\limits_{l = 1}^{\infty}  {}{}^{\left( {j}
\right)}\tilde {g}_{l} x^{l}.
\end{equation}
So, in analogy to the preceding case, we infer now the following expression
for the solution of (B.3a) with (B.4a) and (B.21):
\begin{equation}
f\left( {w} \right) = \pm \left( {w - w_{b}}  \right)\left\{ 1 +
\sum\limits_{l = 1}^{\infty}  {} g_{l} \left[ { \pm \left( {w - w_{b}}
\right)} \right]^{l\beta}  \right\}.
\end{equation}
The significance of the choice among one or the other alternative for the $
\pm $ sign is explained in Section~4.

Let us now turn to the consideration of mechanism \textit{(ii)}, see (B.4b).
In this case the assumption (B.7a) is inconsistent with (B.4b), because
(B.4b) with (B.7a) entails, via (B.3a), that ${f}'\left( {w_{b}}  \right) =
\pm 1$ , while a finite value of ${f}'\left( {w_{b}}  \right)$ is clearly
inconsistent with (B.4b). It is moreover clear that the assumption (B.7a) is
consistent with (B.21) only if this inequality is strengthened to read
\begin{subequations}
\begin{equation}
\mbox{Re}\left( {a} \right) < - 1,
\end{equation}
entailing (see (B.8a))
\begin{equation}
\mbox{Re}\left( {\gamma}  \right) < 0 .
\end{equation}
\end{subequations}
Then the appropriate representation of the solution is given by (B.20),
which is then clearly consistent with (B.4b) due to (B.31b) (and note that
all the terms in the sums in the right-hand side of this representation,
(B.20), become vanishingly small for $w \approx w_{b} $, again thanks to
(B.31b), because the index $k$ is limited not to exceed the index~$l$).

This concludes our analysis, which validates the results reported in Section~4.

Let us end this Appendix with a final remark, based on (B.1), which we
conveniently rewrite in the following form (see (4.4a)):
\begin{subequations}
\begin{equation}
\left( {{\zeta} '} \right)^{2} - B\zeta ^{ - 2a} = V^{2},
\end{equation}
where of course $V$ is given in terms of the initial data by (B.1b) and $B$
by the formula (see (B.1b) and (4.1b))
\begin{equation}
B = - 4{\zeta} '_{1} \left( {0} \right){\zeta} '_{2} \left( {0}
\right)\left[ {\zeta _{1} \left( {0} \right) - \zeta _{2} \left( {0}
\right)} \right]^{2a}.
\end{equation}
\end{subequations}
Assume now that the coupling constant $a$ is a \textit{negative integer,} $a
= - p$, so that (B.32) reads
\begin{equation}
\left( {{\zeta} '} \right)^{2} + 4{\zeta} '_{1} \left( {0} \right){\zeta
}'_{2} \left( {0} \right)\left[ {\zeta _{1} \left( {0} \right) - \zeta
_{2} \left( {0} \right)} \right]^{- 2p}\zeta ^{2p} = V^{2},
\end{equation}
and assume moreover that the initial data, $\zeta _{1} \left( {0}
\right)$, $\zeta _{2} \left( {0} \right)$,
${\zeta} '_{1} \left( {0}
\right)$, ${\zeta} '_{2} \left( {0} \right)$, are \textit{all real}, so that
this equation of motion, (B.33), and its solution, are also \textit{real}
(for real $\tau $), and the solution of the equations of motion (1.7) is as
well \textit{real} for real $\tau $. It is then clear from (B.33) (which can
then be interpreted as the energy conservation formula for a one-dimensional
particle in a power-law potential which vanishes at the origin and diverges
at large distance) that, if the two initial ``velocities'' ${\zeta} '_{1}
\left( {0} \right)$, ${\zeta} '_{2} \left( {0} \right)$ have the \textit{same
sign}, so that the potential energy (i.e., the second term in the left-hand
side of (B.33)) diverges at large distances to \textit{positive} infinity,
then the solution $\zeta \left( {\tau}  \right)$ of this equation (B.33) is,
for real $\tau $, always \textit{periodic}, albeit with a period which
depends on the initial data. In this case, however, the motion of the two
``particles'' $\zeta _{1} \left( {\tau}  \right)$ and $\zeta _{2} \left( {\tau} \right)$
will \textit{not} be periodic, due to the drift associated with
the uniform motion of their center of mass (note that, if the two initial
``velocities'' ${\zeta} '_{1} \left( {0} \right)$, ${\zeta} '_{2} \left( {0}
\right)$ have the \textit{same sign}, the center of mass will indeed move).
Let us emphasize that this discussion refers to the solution of the
equations of motion (1.7) in the special case in which they can themselves
be interpreted to represent the (one-dimensional) motion of \textit{real}
particles. For explicit examples of this phenomenon see the treatments in
Section~4 of the two-body problems with pair coupling constants $a = -
1$, $a = - 2$ and $a = - 3$, respectively.

\renewcommand{\theequation}{C.\arabic{equation}}
\setcounter{equation}{0}
\section*{Appendix C}

In this Appendix we outline the treatment that justifies the main results of
Section~5. Hence we investigate the behavior of the (generic) solution
$\underline {\zeta}  \left( {\tau}  \right)$ of (1.7) in the neighborhood of
its singularities. In analogy to what we did in Appendix B our basic
approach is to replace (1.7) with the following recursive set of equations
\begin{equation}
{}^{\left( {j + 1} \right)}{\zeta} ''_{n} = 2\sum\limits_{m = 1;\,m \ne
n}^{N} {} a_{nm} {}^{\left( {j} \right)}{\zeta} '_{n} {}^{\left( {j}
\right)}{\zeta} '_{m} /\left( {{}^{\left( {j} \right)}\zeta _{n} -
{}^{\left( {j} \right)}\zeta _{m}}  \right),\qquad j = 0,1,2,\ldots,
\end{equation}
and to solve this recursive sequence of equations by starting from an
appropriate assignment for the zeroth-order term ${}^{\left( {0}
\right)}\underline {\zeta}  \left( {\tau}  \right)$ so that it capture the
leading behavior of the solution $\underline {\zeta}  \left( {\tau}
\right)$ in the neighborhood of the singularity whose nature we wish to
ascertain. Let us emphasize that this procedure is only supposed to work in
the neighborhood of the value, $\tau = \tau _{b} $, at which a singularity
occurs, and its effectiveness is predicated upon the successful
identification of the zeroth-order assignment ${}^{\left( {0}
\right)}\underline {\zeta}  \left( {\tau}  \right)$, as demonstrated by the
subsequent preservation of the leading part of its behavior for $\tau
\approx \tau _{b} $ throughout the iteration (see examples below). We
restrict our treatment to a demonstration of this fact; to qualify as a
complete proof our analysis should be complemented by two additional,
synergistically related, results (which will instead be taken for granted
here): \textit{(i)} a proof that the resulting representation of $\underline
{\zeta}  \left( {\tau}  \right)$ as an infinite series (see for instance
(5.4), (5.6) and (5.9)) does converge in a sufficiently small neighborhood
of the singular point $\tau = \tau _{b} $, and \textit{(ii)} a proof that
the sequence produced by the iteration (C.1) (supplemented by appropriate
``initial conditions'' at every iteration cycle: see examples below) does
converge to a solution $\underline {\zeta}  \left(
{\tau}  \right)$ of (1.7),
\begin{equation}
\lim\limits_{j \to \infty}  \left[{}^{\left( {j}\right)}\underline{\zeta} \right]
= \underline \zeta \left( {\tau}  \right).
\end{equation}

Throughout this Appendix C we use the notation of Section 5 without
reporting anew the corresponding definitions. Our first task is to justify
(5.4). To this end we set
\begin{subequations}
\begin{gather}
{}^{\left( {0} \right)}\zeta _{1} \left( {\tau}  \right) = b + c\left(
{\tau - \tau _{b}}  \right)^{\gamma}  + v\left( {\tau - \tau _{b}}
\right),\\
{}^{\left( {0} \right)}\zeta _{2} \left( {\tau}  \right) = b - c\left(
{\tau - \tau _{b}}  \right)^{\gamma}  + v\left( {\tau - \tau _{b}}
\right),\\
{}^{\left( {0} \right)}\zeta _{n} \left( {\tau}  \right) = b_{n} + v_{n}
\left( {\tau - \tau _{b}}  \right),\qquad n = 3,\ldots,N.
\end{gather}
\end{subequations}

From this assignment and (C.1) (with $j = 0$ and $n = 1$) we get
\begin{subequations}
\begin{gather}
{}^{(1)}{\zeta}''_{1} = 2a
[c \gamma  (\tau - \tau _{b})^{\gamma - 1} + v]
[- c\gamma (\tau - \tau _{b})^{\gamma - 1} + v]/
[2c(\tau - \tau_{b})^{\gamma}]
\nonumber\\
{}+ 2\sum\limits_{m = 3}^{N} a_{1m}
[c \gamma (\tau - \tau _{b})^{\gamma - 1} + v]v_{m}/
[b - b_{m} + c(\tau - \tau _{b})^{\gamma} + (v - v_{m})(\tau - \tau_{b})]
\\
{} = - a \gamma ^{2} c (\tau - \tau _{b})^{\gamma - 2}
[1-v^{2}(c\gamma)^{-2}(\tau - \tau _{b})^{2-2\gamma}]
\nonumber\\
{}+ 2 c \gamma (\tau - \tau_{b})^{\gamma - 1}
[1 + v (c \gamma)^{- 1}(\tau - \tau_{b})^{1 - \gamma} ]\nonumber\\
\times \sum\limits_{m = 3}^{N} a_{1m} v_{m} (b - b_{m})^{- 1}
\left \{ 1 + (b - b_{m})^{- 1}
[c (\tau - \tau_{b})^{\gamma} + (v - v_{m})(\tau - \tau_{b})] \right\}^{- 1}.
\end{gather}
\end{subequations}
The motivation to re-write (C.4a) in the form (C.4b) is of course based on
(5.2b), which is as well essential for the subsequent developments.

We now use the following formula:
\begin{subequations}
\begin{equation}
\left( {1 - \varepsilon - \delta}  \right)^{ - 1} = \sum\limits_{k = 0}^{\infty}
\sum\limits_{l = 0}^{k} \left( {\begin{array}{c}
k\\
l
\end{array}} \right)\varepsilon ^{k - l}\delta ^{l},
\end{equation}
where
\begin{equation}
\left( {\begin{array}{c}
k\\
l
\end{array}} \right) = k!/\left[ {l!\left( {k - l} \right)!} \right]
\end{equation}
\end{subequations}
is the binomial coefficient, which is hereafter understood to vanish if $l$
or $\left( {k - l} \right)$ is a \textit{negative} integer. This formula,
(C.5), is of course only applicable provided
\begin{subequations}
\begin{equation}
\left| {\varepsilon + \delta}  \right| < 1.
\end{equation}
\end{subequations}

Via (C.5) we rewrite (C.4b) as follows:
\setcounter{equation}{3}
\begin{subequations}
\setcounter{equation}{2}
\begin{gather}
{}^{(1)}{\zeta}''_{1} = \gamma  (\gamma - 1)  c (\tau - \tau_{b})^{\gamma - 2}
+ (a v^{2} / c)(\tau - \tau_{b})^{-\gamma}
\nonumber\\
\phantom{{}^{(1)}{\zeta}''_{1} =}+ \left[ 1 + v\left( {c\gamma}  \right)^{ - 1}\left(
{\tau - \tau _{b}}  \right)^{1 - \gamma} \right]
\sum\limits_{k = 0}^{\infty}  {} \sum\limits_{l = 0}^{k} {}
f_{kl}^{\left( {1} \right)} \left( {\tau - \tau _{b}}  \right)^{\left(
{k - l + 1} \right)\gamma + l - 1},
\end{gather}
where
\begin{gather}
f_{kl}^{(1)} = 2 c \gamma \left( {\begin{array}{c}
k\\
l
\end{array}} \right) \sum\limits_{m = 3}^{N} a_{1m} v_{m}
(b - b_{m})^{- 1}\nonumber\\
\phantom{f_{kl}^{(1)} =}{}\times [- c / (b - b_{m})]^{k - l}
[-(v - v_{m}) / (b - b_{m})]^{l},
\end{gather}
\end{subequations}
and we also took advantage of the relation
\setcounter{equation}{6}
\begin{equation}
\gamma - 1 = - a\gamma
\end{equation}
entailed by the definition (5.2a) of $\gamma $. Note that the step from
(C.4c) to (C.4d) is justified provided the following condition holds (which
corresponds to (C.6a)):
\begin{equation}
\left| c  (\tau - \tau_{b})^{\gamma} + (v-v_{m})(\tau-\tau_{b}) \right| <
\left| b-b_{m} \right|,
\end{equation}
which is certainly true (see (5.2b)) provided $\tau $ is sufficiently close
to $\tau _{b} $, namely $\left| {\tau - \tau _{b}}  \right|$ is sufficiently
small.

We now integrate twice (C.4c), adjusting the two integration constants so
that, for $\tau \approx \tau _{b} $, the three leading terms of ${}^{\left(
{1} \right)}\zeta _{1} \left( {\tau}  \right)$ coincide with ${}^{\left( {0}
\right)}\zeta _{1} \left( {\tau}  \right)$, see (C.3a) (note that the first
correction term next to the constant one, namely the second one of the
three, is \textit{automatically} OK) and we thereby get
\begin{subequations}
\begin{gather}
{}^{\left( {1} \right)}\zeta _{1} \left( {\tau}  \right) = b + c\left(
{\tau - \tau _{b}}  \right)^{\gamma}  + v\left( {\tau - \tau _{b}}
\right) + av^{2}\left[ {c\left( {2 - \gamma}  \right)\left( {1 -
\gamma}  \right)} \right]^{ - 1}\left( {\tau - \tau _{b}}  \right)^{2- \gamma}
\nonumber\\
\phantom{{}^{\left( {1} \right)}\zeta _{1} \left( {\tau}  \right) =}{}+
\sum\limits_{j = 1}^{\infty}  {} \sum\limits_{k = 1}^{\infty}  {}
{}^{\left( {1} \right)}g_{jk}^{\left( {1} \right)} \left( {\tau - \tau
_{b}}  \right)^{j}\left( {\tau - \tau _{b}}  \right)^{k\gamma },
\end{gather}
with
\begin{gather}
{}^{\left( {1} \right)}g_{jk}^{\left( {1} \right)} = \left( {j + k\gamma}
\right)^{ - 1}\left( {j - 1 + k\gamma}  \right)^{- 1}f_{j + k -
2,j - 1}^{\left( {1} \right)}
\nonumber\\
\phantom{{}^{\left( {1} \right)}g_{jk}^{\left( {1} \right)} =}{}
+ v\left( {c\gamma}  \right)^{ - 1}\left[ {j + 1 + \left( {k - 1}
\right)\gamma}  \right]^{ - 1}\left[ {j + \left( {k - 1}
\right)\gamma}  \right]^{ - 1}f_{j + k - 2,j}^{\left( {1} \right)} .
\end{gather}
\end{subequations}

In a completely analogous manner we get
\begin{subequations}
\begin{gather}
{}^{\left( {2} \right)}\zeta _{1} \left( {\tau}  \right) = b - c\left(
{\tau - \tau _{b}}  \right)^{\gamma}  + v\left( {\tau - \tau _{b}}
\right) - av^{2}\left[ {c\left( {2 - \gamma}  \right)\left( {1 -
\gamma}  \right)} \right]^{- 1}\left( {\tau - \tau _{b}}  \right)^{2
- \gamma}
\nonumber\\
\phantom{{}^{\left( {2} \right)}\zeta _{1} \left( {\tau}  \right) =}{}
+ \sum\limits_{j = 1}^{\infty}  {} \sum\limits_{k = 1}^{\infty}  {}
{}^{\left( {1} \right)}g_{jk}^{\left( {2} \right)} \left( {\tau - \tau
_{b}}  \right)^{j}\left( {\tau - \tau _{b}}  \right)^{k\gamma},
\end{gather}
with
\begin{gather}
{}^{\left( {1} \right)}g_{jk}^{\left( {2} \right)} = \left( {j + k\gamma}
\right)^{ - 1}\left( {j - 1 + k\gamma}  \right)^{ - 1}f_{j + k -
2,j - 1}^{\left( {2} \right)}
\nonumber\\
\phantom{{}^{\left( {1} \right)}g_{jk}^{\left( {2} \right)} =}{}
- v\left( {c\gamma}  \right)^{ - 1}\left[ {j + 1 + \left( {k - 1}
\right)\gamma}  \right]^{ - 1}\left[ {j + \left( {k - 1}
\right)\gamma}  \right]^{ - 1}f_{j + k - 2,j}^{\left( {2} \right)},
\end{gather}
where
\begin{gather}
f_{kl}^{(2)} = - 2c \gamma \left( {\begin{array}{c}
k\\
l
\end{array}} \right) \sum\limits_{m = 3}^{N} a_{2m} v_{m} (b - b_{m})^{- 1}\nonumber\\
\phantom{f_{kl}^{(2)} =}{}\times
[c / (b - b_{m})]^{k - l} [ -(v - v_{m}) / (b - b_{m}) ]^{l}.
\end{gather}
\end{subequations}

Also analogous is the derivation of ${}^{\left( {1} \right)}\zeta _{n}
\left( {\tau}  \right)$, $n = 3,\ldots,N$:
\begin{subequations}
\begin{gather}
{}^{\left( {1} \right)}\zeta _{n} \left( {\tau}  \right) = b_{n} + v_{n}
\left( {\tau - \tau _{b}}  \right)
\nonumber\\
\phantom{{}^{\left( {1} \right)}\zeta _{n} \left( {\tau}  \right) =}{}
+ \sum\limits_{j = 1}^{\infty}  {} \sum\limits_{k = \delta _{j1}} ^{\infty
} {} {}^{\left( {1} \right)}g_{jk}^{\left( {n} \right)} \left( {\tau -
\tau _{b}}  \right)^{j}\left( {\tau - \tau _{b}}  \right)^{k\gamma
},\qquad n = 3,\ldots,N,
\end{gather}
with
\begin{gather}
{}^{\left( {1} \right)}g_{jk}^{\left( {n} \right)} = \left[ {\left( {j +
k\gamma}  \right)\left( {j - 1 + k\gamma}  \right)} \right]^{ -
1}\Bigg\{ \left( {1 - \delta _{j1}}  \right)\left[ {v_{n}
/\left( {b_{n} - b} \right)} \right]\left[ {c/\left( {b_{n} - b}
\right)} \right]^{k} \nonumber\\
\phantom{{}^{\left( {1} \right)}g_{jk}^{\left( {n} \right)} =}{}
\times  [- (v_{n} - v) / (b_{n} - b)]^{j - 2}
[v + k (j - 1)^{- 1}\gamma (v - v_{n})]\nonumber\\
\phantom{{}^{\left( {1} \right)}g_{jk}^{\left( {n} \right)} =}{}
\times \left( {\begin{array}{c}
j + k - 2\\
k
\end{array}} \right) 2 \sum\limits_{s = 1,2} a_{ns} (- 1)^{(s - 1)k}\nonumber\\
\phantom{{}^{\left( {1} \right)}g_{jk}^{\left( {n} \right)} =}{}
+ \delta _{k,0} 2\sum\limits_{m = 3}^{N} {} a_{nm} v_{n} v_{m}
\left( {b_{n} - b_{m}}  \right)^{ - 1}\left[ { - \left( {v_{n} - v_{m}
} \right)/\left( {b_{n} - b_{m}}  \right)} \right]^{j - 1}\Bigg\} .
\end{gather}
\end{subequations}

At this point we introduce the \textit{ansatz}
\begin{subequations}
\begin{gather}
{}^{\left( {j} \right)}\zeta _{s} \left( {\tau}  \right) = b + \left( { - 1}
\right)^{s - 1}c\left( {\tau - \tau _{b}}  \right)^{\gamma}  +
v\left( {\tau - \tau _{b}}  \right)
\nonumber\\
\phantom{{}^{\left( {j} \right)}\zeta _{s} \left( {\tau}  \right) =}{}
+ \sum\limits_{k = 1}^{\infty} {} \sum\limits_{l,m = 0; \,l + m \ge
1}^{\infty}  {} {}^{\left( {j} \right)}g_{klm}^{\left( {s} \right)} \left(
{\tau - \tau _{b}}  \right)^{k + l\gamma + m\left( {1 - \gamma}
\right)},\qquad s = 1,2,\\
{}^{\left( {j} \right)}\zeta _{n} \left( {\tau}  \right) = b_{n} + v_{n}
\left( {\tau - \tau _{b}}  \right)
\nonumber\\
\phantom{{}^{\left( {j} \right)}\zeta _{n} \left( {\tau}  \right) =}{}
+ \sum\limits_{k = 1}^{\infty} {} \sum\limits_{l = \delta _{k1}} ^{\infty
} {} \sum\limits_{m = 0}^{\infty}  {} {}^{\left( {j} \right)}g_{klm}^{\left(
{n} \right)} \left( {\tau - \tau _{b}}  \right)^{k + l\gamma + m\left(
{1 - \gamma}  \right)},\qquad n = 3,\ldots,N,
\end{gather}
\end{subequations}
which is clearly consistent (for the appropriate choice of the coefficients
$g_{klm}^{\left( {n} \right)} $) with both (C.3) ($j = 0$) and (C.10,11) ($j
= 1$), and we show that it is moreover consistent with the iteration (C.1),
namely that insertion of this \textit{ansatz}, (C.12), in the right-hand
side of (C.1) yields, after appropriate expansions and integrations, its
validity at the iteration order $j + 1$. To this end one must of course use
the formal expansion formula
\begin{subequations}
\begin{equation}
\left\{ 1 + \sum\limits_{k = 1}^{\infty}  {} g_{k} \varepsilon
^{k} \right\}^{ - 1} = 1 + \sum\limits_{k = 1}^{\infty}  {}
\tilde {g}_{k} \varepsilon ^{k},
\end{equation}
where we assume that the coefficients $g_{k} $ are given and the
coefficients $\tilde {g}_{k} $ are obtained from them (clearly this can be
done sequentially starting from the lowest powers, although a closed form
expression is not generally available). Of course this formula can only hold
under the restriction
\begin{equation}
\left| {\varepsilon}  \right| < E,
\end{equation}
\end{subequations}
where the constant $E$ is required to be small enough so that all the sums
in (C.13a) converge (and in order that this be possible appropriate
restrictions must hold to begin with on the behavior of the coefficients
$g_{k} $ when their indices diverge, to guarantee that the sum in the
left-hand side of this formula, (C.13a), do converge when (C.13b) holds).

It is easy to convince oneself that all one in fact needs to prove is that
the following expansions,
\begin{subequations}
\begin{gather}
^{(j)} \zeta_{s} (\tau) = b + (- 1)^{(s - 1)}c(\tau - \tau _{b})^{\gamma}
+ v(\tau - \tau _{b}) \nonumber\\
\phantom{^{(j)} \zeta_{s} (\tau) =}{}+
O\left( |(\tau - \tau_{b})|^{1 + \gamma},
|(\tau - \tau_{b})|^{2-\gamma} \right), \qquad s = 1,2,\\
{}^{\left( {j} \right)}\zeta _{n} \left( {\tau}  \right) = b_{n} + v_{n}
\left( {\tau - \tau _{b}}  \right) + O\left( {\left| {\left( {\tau -
\tau _{b}}  \right)} \right|^{1 + \gamma} } \right),\qquad n =
3,\ldots,N,
\end{gather}
\end{subequations}
which are clearly entailed by (C.12) (and are of course valid for $\tau
\approx \tau _{b} $), are consistent with the iteration (C.1), namely that
their validity at the order $j$, see (C.14), entails their validity at the
order $j + 1$. Indeed this result (once proven) entails that the iteration
process keeps producing only the powers of $\left( {\tau - \tau _{b}}
\right)$ that appear in the right-hand side of (C.12) (and no logarithms!),
with the coefficients ${}^{\left( {j + 1} \right)}g_{klm}^{\left( {n}
\right)} $ uniquely determined via (C.1) by the coefficients ${}^{\left( {j}
\right)}g_{klm}^{\left( {n} \right)} $ and the other constants appearing in
the right-hand side of (C.12).

To prove the consistency of (C.14) with (C.1) we now note that these
formulas, (C.14), entail
\begin{subequations}
\begin{gather}
{}^{(j)}{\zeta}'_{s}(\tau) = \gamma(- 1)^{(s - 1)} c (\tau - \tau_{b})^{\gamma - 1}
+ v \nonumber\\
\phantom{{}^{(j)}{\zeta}'_{s}(\tau) =}{}
+ O\left( |(\tau - \tau_{b})|^{\gamma}|, |(\tau - \tau _{b})|^{1 - \gamma} \right)
,\qquad s = 1,2,\\
{}^{\left( {j} \right)}{\zeta} '_{n} \left( {\tau}  \right) = v_{n} +
O\left( {\left| {\left( {\tau - \tau _{b}}  \right)} \right|^{\gamma
}} \right),\qquad n = 3,\ldots,N ,
\end{gather}
\end{subequations}
hence
\begin{subequations}
\begin{gather}
{}^{\left( {j} \right)}{\zeta} '_{1} \left( {\tau}  \right){}^{\left( {j}
\right)}{\zeta} '_{2} \left( {\tau}  \right) = - \gamma
^{2}c^{2}\left( {\tau - \tau _{b}}  \right)^{2\left( {\gamma - 1}
\right)} + v^{2} \nonumber\\
\phantom{{}^{\left( {j} \right)}{\zeta} '_{1} \left( {\tau}  \right){}^{\left( {j}
\right)}{\zeta} '_{2} \left( {\tau}  \right) =}{}
+ O\left( {\left| {\left( {\tau - \tau _{b}}  \right)}
\right|^{2\gamma - 1},\left| {\left( {\tau - \tau _{b}}  \right)}
\right|} \right) ,\\
{}^{(j)}{\zeta}'_{s}(\tau){}^{(j)}{\zeta}'_{n}(\tau) =
(- 1)^{(s - 1)} \gamma  c  v_{n} (\tau - \tau_{b})^{\gamma - 1} + vv_{n}
+ O\left( |(\tau - \tau_{b})|^{\gamma} |(\tau - \tau_{b})|^{1 - \gamma}  \right),
\nonumber\\
\phantom{{}^{(j)}{\zeta}'_{s}(\tau){}^{(j)}{\zeta}'_{n}(\tau) =}{}
s = 1,2,\quad n = 3,\ldots,N,
\\
{}^{\left( {j} \right)}{\zeta} '_{n} \left( {\tau}  \right){}^{\left( {j}
\right)}{\zeta} '_{m} \left( {\tau}  \right) = v_{n} v_{m} + O\left(
{\left| {\left( {\tau - \tau _{b}}  \right)} \right|^{\gamma} }
\right),\qquad n,m = 3,\ldots,N,
\end{gather}
\end{subequations}
and
\begin{subequations}
\begin{gather}
{}^{\left( {j} \right)}\zeta _{1} \left( {\tau}  \right) - {}^{\left( {j}
\right)}\zeta _{2} \left( {\tau}  \right) = 2c\left( {\tau - \tau _{b}}
\right)^{\gamma}  + O\left( {\left| {\left( {\tau - \tau _{b}}  \right)}
\right|^{1 + \gamma} ,\left| {\left( {\tau - \tau _{b}}  \right)}
\right|^{2 - \gamma} } \right),\\
{}^{\left( {j} \right)}\zeta _{s} \left( {\tau}  \right) - {}^{\left( {j}
\right)}\zeta _{n} \left( {\tau}  \right) = b - b_{n} + O\left( {\left|
{\left( {\tau - \tau _{b}}  \right)} \right|^{\gamma} }
\right),\qquad s = 1,2,\quad n = 3,\ldots,N,\!\!\\
{}^{\left( {j} \right)}\zeta _{n} \left( {\tau}  \right) - {}^{\left( {j}
\right)}\zeta _{m} \left( {\tau}  \right) = b_{n} - b_{m} + O\left( {\left|
{\left( {\tau - \tau _{b}}  \right)} \right|^{1 + \gamma} }
\right),\qquad n,m = 3,\ldots,N,
\end{gather}
\end{subequations}
hence
\begin{subequations}
\begin{gather}
\left[ {}^{(j)}\zeta_{1}(\tau)-{}^{(j)}\zeta_{2}(\tau) \right]^{- 1}
= (2c)^{- 1}(\tau - \tau_{b})^{-\gamma}\nonumber\\
\phantom{\left[ {}^{(j)}\zeta_{1}(\tau)-{}^{(j)}\zeta_{2}(\tau) \right]^{- 1}=}\times
\left[1 + O\left( {\left| {\left( {\tau - \tau _{b}}  \right)} \right|,\left| {\left( {\tau - \tau _{b}}
\right)} \right|^{2\left( {1 - \gamma}  \right)}} \right) \right],\\
\left[ {}^{(j)}\zeta_{s}(\tau)-{}^{(j)}\zeta_{n}(\tau) \right]^{- 1}
= (b - b_{n})^{- 1}
\left[ 1 + O\left( {\left| {\left( {\tau - \tau _{b}}  \right)} \right|^{\gamma
}} \right) \right],\nonumber\\
\phantom{\left[ {}^{(j)}\zeta_{s}(\tau)-{}^{(j)}\zeta_{n}(\tau) \right]^{- 1}=}{}
s = 1,2;\quad n = 3,\ldots,N,\\
\left[ {}^{(j)}\zeta_{n}(\tau)-{}^{(j)}\zeta_{m}(\tau) \right]^{-1}
= (b_{n}-b_{m})^{-1}
\left[ 1 + O\left( {\left| {\left( {\tau - \tau _{b}}  \right)}
\right|^{1 + \gamma} } \right) \right],\nonumber\\
\phantom{\left[ {}^{(j)}\zeta_{n}(\tau)-{}^{(j)}\zeta_{m}(\tau) \right]^{-1}=}{}
n,m = 3,\ldots,N.
\end{gather}
\end{subequations}

It is therefore clear that (C.1) entails (via (C.16a,b) and (C.17a,b))
\begin{subequations}
\begin{gather}
{}^{(j + 1)}{\zeta}''_{s} \left( {\tau}  \right) = 2a
\left[ - \gamma ^{2}c^{2}\left( {\tau - \tau _{b}} \right)^{2(\gamma - 1)}
+ v^{2} + O\left( | ( {\tau - \tau _{b}}  ) |^{2\gamma - 1},
| ( {\tau - \tau _{b}}  ) | \right) \right]
\nonumber\\
\qquad {} \times \left( { - 1} \right)^{s - 1}\left( {2c} \right)^{ -
1}\left( {\tau - \tau _{b}}  \right)^{ -\gamma} \left[1 +
O\left( {\left| {\left( {\tau - \tau _{b}}  \right)} \right|,\left|
{\left( {\tau - \tau _{b}}  \right)} \right|^{2\left( {1 - \gamma}
\right)}} \right) \right]
\nonumber\\
\qquad {} + 2\sum\limits_{m = 3}^{N} {} a_{sm} \left[ \left( { - 1}
\right)^{\left( {s - 1} \right)}\gamma cv_{n} \left( {\tau - \tau _{b}
} \right)^{\gamma - 1} + vv_{n} + O\left( {\left| {\left( {\tau - \tau
_{b}}  \right)} \right|^{\gamma} ,\left| {\left( {\tau - \tau _{b}}
\right)} \right|^{1 - \gamma} } \right) \right] \nonumber\\
\qquad {}\times \left( {b - b_{n}}  \right)^{ - 1}\left[1  + O\left(
{\left| {\left( {\tau - \tau _{b}}  \right)} \right|^{\gamma} }
\right) \right],\qquad s = 1,2,
\end{gather}
namely
\begin{gather}
{}^{(j + 1)}{\zeta}''_{s}(\tau)
= (- 1)^{s - 1}\gamma(\gamma - 1)(\tau - \tau_{b})^{\gamma - 2} \nonumber\\
\phantom{{}^{(j + 1)}{\zeta}''_{s}(\tau)=}{}+
O \left( |(\tau - \tau_{b})|^{-\gamma} ,
|(\tau - \tau_{b})^{\gamma - 1}| \right), \qquad s = 1,2.
\end{gather}
\end{subequations}
Integration (with the appropriate choice of integration constants) of this
equation, (C.19b), yields
\begin{gather}
^{(j + 1)}\zeta_{s}(\tau)
= b + (-1)^{(s-1)} c (\tau-\tau_{b})^{\gamma} +
v (\tau-\tau_{b})
\nonumber\\
\phantom{^{(j + 1)}\zeta_{s}(\tau)=}{}
+ O \left( |(\tau-\tau_{b})|^{1+\gamma} ,
|(\tau-\tau_{b})|^{2-\gamma} \right),\qquad s=1,2,
\end{gather}
and the consistency of this expression, (C.20), with (C.14a) is plain.

It is likewise clear that (C.1) entails (via (C.16a,c) and (C.17a,c))
\begin{subequations}
\begin{gather}
{}^{\left( {j + 1} \right)}{\zeta} ''_{n} \left( {\tau}  \right) =
2\sum\limits_{s = 1,2} \Big\{ a_{ns} \Big[
\left( { - 1} \right)^{\left( {s - 1} \right)}\gamma cv_{n}
\left( {\tau - \tau _{b}}  \right)^{\gamma - 1} + vv_{n}
\nonumber\\
\qquad {}+ O\left( |(\tau - \tau_{b})|^{\gamma},
|(\tau - \tau_{b})|^{1 - \gamma}, |(\tau - \tau_{b})|^{2\gamma - 1} \right)
\Big] (b_{n}- b)^{- 1}\left[ 1 + O \left (|(\tau - \tau_{b})|^{\gamma} \right) \right]\Big\}
\nonumber\\
\qquad {}+2\sum\limits_{m = 3}^{N} \left\{ a_{nm} \left[ v_{n}v_{m} +
O \left(|(\tau - \tau_{b})|^{\gamma} \right) \right]
(b_{n} - b_{m})^{- 1} [ 1 + O\left( |(\tau-\tau_{b})|^{1 + \gamma} \right)]\right\}
, \nonumber\\
\qquad n = 3,\ldots,N,
\end{gather}
namely
\begin{equation}
{}^{\left( {j + 1} \right)}{\zeta} ''_{n} \left( {\tau}  \right) = O\left(
{\left| {\tau - \tau _{b}}  \right|^{\gamma - 1}} \right),\qquad n = 3,\ldots,N,
\end{equation}
\end{subequations}
and the integration (with the appropriate choice of integration constants)
of this equation, (C.21b), yields
\begin{equation}
{}^{\left( {j + 1} \right)}\zeta _{n} \left( {\tau}  \right) = b_{n} + v_{n}
\left( {\tau - \tau _{b}}  \right) + O\left( {\left| {\left( {\tau -
\tau _{b}}  \right)} \right|^{1 + \gamma} } \right),\qquad n = 3,\ldots,N,
\end{equation}
which is clearly consistent with (C.14b). (Note however that the estimate
(C.21b) can be replaced by more stringent estimates if $a_{n1} = a_{n2} $
and $v_{n} \ne 0$, or if $v_{n} = 0$; indeed it is easily seen from (C.21a)
that the last term in the right hand sides of (C.14b) and (C.22) can be
replaced by $O\left( {\left| {\tau - \tau _{b}}  \right|^{2}} \right)$ if
$a_{n1} = a_{n2} $ and $v_{n} \ne 0$, and by $O\left( {\left| {\tau - \tau
_{b}}  \right|^{2 + \gamma} ,\left| {\tau - \tau _{b}}  \right|^{3 -
\gamma} } \right)$ if $v_{n} = 0$. This entails in subsequent equations some
corresponding changes that the reader will easily ascertain, but which do
not modify the main conclusions, see Section 5, as regards the nature of the
branch point of the solution $\underline {\zeta}  \left( {\tau}  \right)$).

This concludes our treatment meant to justify the validity of the
representation (5.4).

Let us now proceed to justify the \textit{ansatz} (5.6), namely the
compatibility of the expansion
\begin{subequations}
\begin{gather}
\zeta _{1} \left( {\tau}  \right) = b + c\left( {\tau - \tau _{b}}
\right)^{1 + \beta}  + \sum\limits_{k,l = 1;\, k + l \ge 3}^{\infty}  {}
g_{kl}^{\left( {1} \right)} \left( {\tau - \tau _{b}}  \right)^{k +
l\beta},\\
\zeta _{2} \left( {\tau}  \right) = b + v_{2} \left( {\tau - \tau _{b}}
\right) - c\left( {\tau - \tau _{b}}  \right)^{1 + \beta}
 + \sum\limits_{k = 1}^{\infty}  {} \sum\limits_{l = 2\delta _{k1}} ^{\infty
} {} g_{kl}^{\left( {2} \right)} \left( {\tau - \tau _{b}}  \right)^{k +
l\beta},\\
\zeta _{n} \left( {\tau}  \right)
= b_{n} + v_{n} \left( {\tau - \tau _{b} } \right)
+ \sum\limits_{k = 2}^{\infty}  {} \sum\limits_{l = 0}^{\infty}  {}
g_{kl}^{\left( {n} \right)} \left( {\tau - \tau _{b}}  \right)^{k +
l\beta} ,\qquad n = 3,\ldots,N,
\end{gather}
where
\begin{equation}
\beta = - 2a,\qquad \mbox{Re}\left( {\beta}  \right) > 0,
\end{equation}
\end{subequations}
with the evolution equation (1.7) satisfied by the solution $\underline
{\zeta}  \left( {\tau}  \right)$. Of course this \textit{ansatz}, (C.23), is
again only applicable for $\tau \approx \tau _{b} $, namely in the
neighborhood of the value~$\tau _{b} $ of the independent variable where the
solution $\underline {\zeta}  \left( {\tau}  \right)$ features a ``two-body
collision'', characterized, see (C.23), by the equality and inequalities
\begin{equation}
\zeta _{1} \left( {\tau _{b}}  \right) = \zeta _{2} \left( {\tau _{b}}
\right) \ne \zeta _{n} \left( {\tau _{b}}  \right),\qquad n = 3,\ldots,N.
\end{equation}
And clearly in that neighborhood this \textit{ansatz}, (C.23), entails
\begin{subequations}
\begin{gather}
\zeta _{1} \left( {\tau}  \right) = b + c\left( {\tau - \tau _{b}}
\right)^{1 + \beta}  + O\left( {\left| {\tau - \tau _{b}}  \right|^{2 +
\beta} ,\left| {\tau - \tau _{b}}  \right|^{1 + 2\beta} } \right) ,\\
\zeta _{2} \left( {\tau}  \right) = b + v_{2} \left( {\tau - \tau _{b}}
\right) - c\left( {\tau - \tau _{b}}  \right)^{1 + \beta}  + O\left(
{\left| {\tau - \tau _{b}}  \right|^{2},\left| {\tau - \tau _{b}}
\right|^{1 + 2\beta} } \right),
\\
\zeta _{n} \left( {\tau}  \right) = b_{n} + v_{n} \left( {\tau - \tau _{b}
} \right) + O\left( {\left| {\tau - \tau _{b}}  \right|^{2}}
\right),\qquad n = 3,\ldots,N.
\end{gather}
\end{subequations}
Here we limit our treatment to show that this behavior is compatible with
the evolution equation (1.7), which is the key point of our argument (see
above -- it is instead left for the diligent reader to repeat, in close
analogy to what we did before, the iteration argument that leads to the
\textit{ansatz} (C.23)).

Clearly the expansions (C.25) entail
\begin{subequations}
\begin{gather}
{\zeta} '_{1} \left( {\tau}  \right) = \left( {1 + \beta}
\right)c\left( {\tau - \tau _{b}}  \right)^{\beta} \left[
1 + O\left( {\left| {\tau - \tau _{b}}  \right|,\left| {\tau - \tau
_{b}}  \right|^{\beta} } \right) \right] ,\\
{\zeta} '_{2} \left( {\tau}  \right) = v_{2}\left[ 1 + O\left(
{\left| {\tau - \tau _{b}}  \right|,\left| {\tau - \tau _{b}}
\right|^{\beta} } \right) \right] ,\\
{\zeta} '_{n} \left( {\tau}  \right) = v_{n} \left[1 + O\left(
{\left| {\tau - \tau _{b}}  \right|} \right) \right],\qquad n = 3,\ldots,N,
\end{gather}
\end{subequations}
hence
\begin{subequations}
\begin{gather}
{\zeta} '_{1} \left( {\tau}  \right){\zeta} '_{n} \left( {\tau}  \right) =
\left( {1 + \beta}  \right)cv_{n} \left( {\tau - \tau _{b}}
\right)^{\beta} \left[1 + O\left( {\left| {\tau - \tau _{b}}
\right|,\left| {\tau - \tau _{b}}  \right|^{\beta} } \right)\right],
\nonumber\\
\phantom{{\zeta} '_{1} \left( {\tau}  \right){\zeta} '_{n} \left( {\tau}  \right) =}{}n = 2,\ldots,N,
\\
{\zeta} '_{2} \left( {\tau}  \right){\zeta} '_{n} \left( {\tau}  \right) =
v_{2} v_{n} \left[1 + O\left( {\left| {\tau - \tau _{b}}
\right|,\left| {\tau - \tau _{b}}  \right|^{\beta} } \right)\right],\qquad n = 3,\ldots,N,\\
{\zeta} '_{n} \left( {\tau}  \right){\zeta} '_{m} \left( {\tau}  \right) =
v_{n} v_{m} \left[1 + O\left( {\left| {\tau - \tau _{b}}
\right|} \right) \right],\qquad n,m = 3,\ldots,N ,
\end{gather}
\end{subequations}
as well as
\begin{subequations}
\begin{gather}
\zeta _{1} \left( {\tau}  \right) - \zeta _{2} \left( {\tau}  \right) = -
v_{2} \left( {\tau - \tau _{b}}  \right)\left[1 + O\left(
{\left| {\tau - \tau _{b}}  \right|^{1 + \beta} } \right) \right],
\\
\zeta _{s} \left( {\tau}  \right) - \zeta _{n} \left( {\tau}  \right) =
\left( {b - b_{n}}  \right)\left[ 1 + O\left( {\left| {\tau -
\tau _{b}}  \right|} \right) \right],\qquad s = 1,2,\quad n = 3,\ldots,N,\\
\zeta _{n} \left( {\tau}  \right) - \zeta _{m} \left( {\tau}  \right) =
\left( {b_{n} - b_{m}}  \right)\left[1 + O\left( {\left| {\tau
- \tau _{b}}  \right|} \right) \right],\qquad n,m = 3,\ldots,N,
\end{gather}
\end{subequations}
hence
\begin{subequations}
\begin{gather}
\left[ {\zeta _{1} \left( {\tau}  \right) - \zeta _{2} \left( {\tau}
\right)} \right]^{ - 1} = - v_{2}^{ - 1} \left( {\tau - \tau _{b}}
\right)^{ - 1}\left[1 + O\left( {\left| {\tau - \tau _{b}}
\right|^{1 + \beta} } \right) \right],\\
\left[ {\zeta _{s} \left( {\tau}  \right) - \zeta _{n} \left( {\tau}
\right)} \right]^{ - 1} = \left( {b - b_{n}}  \right)^{ - 1}\left[
1 + O\left( {\left| {\tau - \tau _{b}}  \right|} \right)
\right],\nonumber\\
\phantom{\left[ {\zeta _{s} \left( {\tau}  \right) - \zeta _{n} \left( {\tau}
\right)} \right]^{ - 1} =}{}
 s = 1,2,\quad n = 3,\ldots,N,\\
\left[ {\zeta _{n} \left( {\tau}  \right) - \zeta _{m} \left( {\tau}
\right)} \right]^{ - 1} = \left( {b_{n} - b_{m}}  \right)^{ - 1}\left[
1 + O\left( \left| {\tau - \tau _{b}}  \right| \right)
\right],\qquad n,m = 3,\ldots,N.
\end{gather}
\end{subequations}

It is now easy, using these formulas, to verify the consistency of (C.25)
with (1.7). Indeed for $n = 1$ (1.7) yields (via (C.27a) and (C.29a,b))
\begin{subequations}
\begin{gather}
{\zeta} ''_{1} = - 2ac\left( {1 + \beta}  \right)\left( {\tau - \tau
_{b}}  \right)^{ - 1 + \beta} \nonumber\\
\phantom{{\zeta} ''_{1} =}{}
\times \left[ 1 + O\left( {\left| {\tau
- \tau _{b}}  \right|,\left| {\tau - \tau _{b}}  \right|^{\beta} }
\right) \right] + O\left( {\left| {\tau - \tau _{b}}
\right|^{\beta} } \right),
\end{gather}
namely, via (C.23d),
\begin{equation}
{\zeta} ''_{1} = c\beta \left( {1 + \beta}  \right)\left( {\tau - \tau_{b}}
\right)^{ - 1 + \beta}  + O\left( {\left| {\tau - \tau _{b}}
\right|^{\beta} ,\left| {\tau - \tau _{b}}  \right|^{ - 1 + 2\beta} } \right),
\end{equation}
\end{subequations}
which can be immediately integrated (with an appropriate choice of
integration constants) to yield (C.25a). Likewise for $n = 2$ (1.7) yields
(via (C.27a) and (C.29a,b))
\begin{subequations}
\begin{gather}
{\zeta} ''_{2} = 2ac\left( {1 + \beta}  \right)\left( {\tau - \tau_{b}}
\right)^{ - 1 + \beta} \left[1 + O\left( {\left| {\tau
- \tau _{b}}  \right|,\left| {\tau - \tau _{b}}  \right|^{\beta} }
\right) \right] + O\left( {1} \right),
\end{gather}
namely, via (C.23d),
\begin{equation}
{\zeta} ''_{2} = - c\beta \left( {1 + \beta}  \right)\left( {\tau -
\tau _{b}}  \right)^{ - 1 + \beta}  + O\left( {\left| {\tau - \tau _{b}}
\right|^{ - 1 + 2\beta} ,1} \right),
\end{equation}
\end{subequations}
which can be immediately integrated (with an appropriate choice of
integration constants) to yield (C.25b). And finally for $n = 3,\ldots,N$ (1.7)
yields (via (C.27b,c) and (C.29b,c))
\begin{equation}
{\zeta} ''_{n} = O\left( {1} \right),
\end{equation}
which can be immediately integrated (with an appropriate choice of
integration constants) to yield (C.25c).

\renewcommand{\theequation}{D.\arabic{equation}}
\setcounter{equation}{0}

\section*{Appendix D}

In this Appendix D we introduce the monic polynomials $\Pi _{N} \left(
{\zeta ,\tau}  \right)$ respectively $P_{N} \left( {z,t} \right)$, of
degree $N$ in the variable $\zeta $ respectively $z$, the zeros of which
coincide with the coordinates, $\zeta _{n} \equiv \zeta _{n} \left( {\tau}
\right)$ respectively $z_{n} \equiv z_{n} \left( {t} \right)$, of the
particles moving according to the $N$-body problems (1.7) respectively
(1.5),
\begin{gather}
\Pi _{N} \left( {\zeta ,\tau}  \right) = \zeta ^{N} + \sum\limits_{m =
1}^{N} {} \gamma _{m} \left( {\tau}  \right)\zeta ^{N - m} =
\prod\limits_{n = 1}^{N} {} \left[ {\zeta - \zeta _{n} \left( {\tau}
\right)} \right],\\
P_{N} \left( {z,t} \right) = z^{N} + \sum\limits_{m = 1}^{N} {} c_{m}
\left( {t} \right)z^{N - m} = \prod\limits_{n = 1}^{N} {} \left[ {z -
z_{n} \left( {t} \right)} \right].
\end{gather}
Via such a position a one-to-one correspondence is introduced among the
(ordered) sets of coefficients $\left\{ {\gamma _{m} \left( {\tau}
\right),\,\,m = 1,\ldots,N} \right\}$ respectively $\left\{ {c_{m} \left( {t}
\right),\,\,m = 1,\ldots,N} \right\}$ and the (unordered) sets of zeros
$\left\{ {\zeta _{n} \left( {\tau}  \right),\,\,n = 1,\ldots,N} \right\}$
respectively $\left\{ {z_{n} \left( {t} \right),\,\,n = 1,\ldots,N} \right\}$.
This correspondence turns out to be extremely convenient [5,~2] in the
\textit{integrable} case in which all the coupling constants $a_{nm} $ in
(1.5) and (1.7) equal unity,
\begin{equation}
a_{nm} = 1.
\end{equation}
Indeed in this \textit{integrable} case these polynomials satisfy extremely
simple linear PDEs, namely (note that throughout this Appendix~D subscripted
variables denote partial differentiations)
\begin{subequations}
\begin{gather}
\Pi _{N,\tau \tau}  = 0,\\
P_{N,tt} = i\omega P_{N,t},\tag{D.5a}
\end{gather}
\end{subequations}
entailing of course
\setcounter{equation}{3}
\begin{subequations}
\setcounter{equation}{1}
\begin{gather}
{\gamma} ''_{m} = 0,\qquad \gamma _{m} \left( {\tau}  \right) = \gamma _{m}
\left( {0} \right) + {\gamma} '_{m} \left( {0} \right)\tau,\\
\ddot {c}_{m} = i\omega \dot {c}_{m} ,\qquad c_{m} \left( {t} \right) =
c_{m} \left( {0} \right) + \dot {c}_{m} \left( {0} \right)\left[
{\exp\left( {i\omega t} \right) - 1} \right]/\left( {i\omega}
\right).\tag{D.5b}
\end{gather}
\end{subequations}
It is this connection that lies behind the resolvent formula (1.17).

In this Appendix~D we investigate the extent to which an analogous approach
is useful to treat the $N$-body problems (1.7) and (1.5) in the (generally
\textit{nonintegrable}) case when the coupling constants $a_{nm} $
are still all equal, but they need not be unity,
\setcounter{equation}{5}
\begin{equation}
a_{nm} = a,
\end{equation}
with $a$ an arbitrary (possibly complex) constant.

Hereafter we focus on the $N$-body problem (1.7), and only at the end, by
taking advantage of our by now usual trick (see (1.6)), we mention the
analogous results related to the $N$-body problem (1.5). We also drop,
whenever this is notationally convenient, the subscript $N$ attached to the
symbols $\Pi $ and $P$ (namely hereafter $\Pi _{N} \equiv \Pi$, $P_{N}
\equiv P$).

Our results hinge on the following basic result, the proof of which is
relegated to the very end of this Appendix.

\medskip

\noindent
{\bf Lemma D.1.} {\it If the $N$ zeros $\zeta _{n} \left(
{\tau}  \right)$ evolve according to (1.7) with (D.6), the (monic)
polynomial $\Pi \left( {\zeta ,\tau}  \right)$ of degree
$N$ in $\zeta $, see (D.1), satisfies the following
nonlinear PDE:
\begin{equation}
\Pi _{\tau \tau}  \left[ {\Pi _{\zeta} }  \right]^{2} + \left( {1 - a}
\right)\left\{ \Pi _{\zeta \zeta}  \left( {\Pi _{\tau} }
\right)^{2} - \Pi _{\zeta}  \left[ \left( {\Pi _{\tau} }
\right)^{2}\right]_{\zeta} \right\} = \Psi \Pi,
\end{equation}
where $\Psi \equiv \Psi \left( {\zeta ,\tau}  \right) \equiv \Psi
_{2N - 4} \left( {\zeta ,\tau}  \right)$ is a polynomial of degree
 $2N - 4$ in the variable $\zeta $,}
\begin{equation}
\Psi \left( {\zeta ,\tau}  \right) = \sum\limits_{k = 0}^{2N - 4} {}
\psi _{k} \left( {\tau}  \right)\zeta ^{N - k}.
\end{equation}

Note that this PDE, (D.7), does \textit{not} explicitly feature the
parameter $N$.

Let us note that (D.1) entails the identification
\begin{equation}
\gamma _{1} \left( {\tau}  \right) = - \sum\limits_{n = 1}^{N} {} \zeta _{n}
\left( {\tau}  \right) ,
\end{equation}
so that the coefficient $\gamma _{1} \left( {\tau}  \right)$ clearly, see
(1.7), satisfies the equation
\begin{subequations}
\begin{equation}
{\gamma} ''_{1} \left( {\tau}  \right) = 0.
\end{equation}
Hence $\Pi _{\tau \tau}  \left( {\zeta ,\tau}  \right)$ is a polynomial
in $\zeta $ of degree $N - 2$ (see (D.1)). It is therefore clear, see (D.1),
that the two sides of the PDE (D.7) are polynomials of degree $3N - 4$ in
$\zeta $, hence this PDE entails $3N - 3$ equations to be satisfied by the
$N$ coefficients $\gamma _{n} \left( {\tau}  \right)$, see (D.1), and the
$2N - 3$ coefficients $\psi _{k} \left( {\tau}  \right)$, see (D.8). Hence
the number of equations matches the number of unknowns ($3N - 3 = N +
\left( {2N - 3} \right)$). Moreover these equations are clearly all linear
and purely algebraic (not differential!) for the $2N - 3$ coefficients
$\psi _{k} \left( {\tau}  \right)$, see (D.7) and (D.8). It is therefore in
principle possible to eliminate the $2N - 3$ unknowns $\psi _{k} \left(
{\tau}  \right)$ and to thereby obtain a system of only $N$, generally
nonlinear, second-order ODEs for the $N$ unknowns $\gamma _{n} \left( {\tau
} \right)$, or rather a system of $N - 1$, generally nonlinear, second-order
ODEs for the $N - 1$ unknowns $\gamma _{n} \left( {\tau}  \right)$,
$n =2,\ldots,N$, since $\gamma _{1} \left( {\tau}  \right)$ shall turn out to
satisfy the linear second-order ODE (D.10a), which of course entails
\begin{equation}
\gamma _{1} \left( {\tau}  \right) = \gamma _{1} \left( {0} \right) +
{\gamma} '_{1} \left( {0} \right)\tau.
\end{equation}
\end{subequations}

A systematic algorithmic procedure (``Euclid's method'') to eliminate the
coefficients of the polynomial $\Psi $ from (D.7) and to thereby obtain the
(generally nonlinear) second-order ODEs satisfied by the coefficients
$\gamma _{n} \left( {\tau}  \right)$ can be based on the observation (which
is actually instrumental, see below, to obtain the PDE (D.7)) that the
right-hand side of (D.7) vanishes for $\zeta = \zeta _{n} $, see (D.1). One
can therefore lower the degree of the left-hand side of (D.7) by applying
sequentially the substitution
\begin{equation}
\zeta ^{N} = - \sum\limits_{m = 1}^{N} {} \gamma _{m} \zeta ^{N - m}
\end{equation}
to the term of highest degree, or to all terms of degree higher than $N$.
After applying this procedure $2N - 3$ times the left-hand side of (D.7)
becomes a polynomial in $\zeta $ of degree $N - 1$, since each application
lowers by one unit the degree of the polynomial. But since the polynomial of
degree $N - 1$ thus obtained must vanish at the $N$ points $\zeta = \zeta
_{n} $ (see (D.7)), it vanishes identically, namely all its $N$ coefficients
vanish: and this requirement yields precisely the $N$ equations satisfied by
the $N$ coefficients $\gamma _{n} \left( {\tau}  \right)$ (hence one of
these~$N$ equations will be precisely (D.10a)). For instance for $N = 2$ one
thereby obtains
\begin{equation}
{\gamma} ''_{2} = 2\left( {1 - a} \right)\left( {4\gamma
_{2} - \gamma _{1}^{2}}  \right)^{ - 1} \left({
{\gamma'}_{2}^{2} - {\gamma'}_{2} {\gamma'}_{1} \gamma _{1} +
{\gamma'}_{1}^{2} \gamma _{2} }\right),
\end{equation}
of course with $\gamma _{1} \left( {\tau}  \right)$ given by (D.10b), and
for $N = 3$ one obtains
\begin{subequations}
\begin{gather}
\gamma''_{2}=2(1-a)\delta ^{-1}\big[ \gamma _{1}^{\prime
2}\left( -6\gamma_2^2\gamma_1^2-4\gamma _{1}\gamma _{2}\gamma _{3}+\gamma _{2}^{3}+9\gamma
_{3}^{2}\right) +\gamma _{2}^{\prime 2}\left( -3\gamma _{1}\gamma _{3}+\gamma
_{2}^{2}\right)\nonumber\\
%
%
\phantom{\gamma''_{2}=}{}+\gamma _{3}^{\prime 2}\left( \gamma _{1}^{2}-3\gamma_{2}\right)
+\gamma _{1}^{\prime }\gamma _{2}^{\prime }
\left(-\gamma_{1}\gamma _{2}^{2}+4\gamma _{1}^{2}\gamma _{3}-3\gamma _{2}\gamma
_{3}\right) +2\gamma _{1}^{\prime }\gamma _{3}^{\prime }\left( -3\gamma
_{1}\gamma _{3}+\gamma _{2}^{2}\right)
\nonumber\\
\phantom{\gamma''_{2}=}{}
+\gamma _{2}^{\prime }\gamma
_{3}^{\prime }\left(
-3\gamma_3\gamma_1+\gamma_2^2
\right) \big] ,\\
\gamma'' _{3}=2(1-a)\delta ^{-1}\big[ \gamma _{1}^{\prime
2}\gamma _{3}\left( -3\gamma _{1}\gamma _{3}+\gamma _{2}^{2}\right) +\gamma
_{2}^{\prime 2}\gamma _{3}\left( \gamma _{1}^{2}-3\gamma _{2}\right) +\gamma
_{3}^{\prime 2}\left( \gamma _{1}^{3}-4\gamma _{1}\gamma _{2}+9\gamma
_{3}\right)
\nonumber\\
\phantom{\gamma'' _{3}=}{}
+\gamma _{1}^{\prime }\gamma _{2}^{\prime }\gamma _{3}\left(
-\gamma _{1}\gamma _{2}+9\gamma _{3}\right) +2\gamma _{1}^{\prime
}\gamma _{3}^{\prime }\gamma _{3}\left( \gamma _{1}^{2}-3\gamma _{2}\right)
\nonumber\\
\phantom{\gamma'' _{3}=}{}
+\gamma _{2}^{\prime }\gamma _{3}^{\prime }\left( -3\gamma _{1}\gamma
_{3}-\gamma _{1}^{2}\gamma _{2}+4\gamma _{2}^{2}\right) \big] ,\\
\delta = 27\gamma _{3}^{2} - 18\gamma _{3} \gamma _{2} \gamma _{1} +
4\gamma _{3} \gamma _{1}^{3} + 4\gamma _{2}^{3} - \gamma _{2}^{2}
\gamma _{1}^{2} ,
\end{gather}
\end{subequations}
of course with $\gamma _{1} \left( {\tau}  \right)$ given again by (D.10b).

Note the consistency of these equations with the general result ${\gamma}''_{n} = 0$,
valid for arbitrary $N$ in the \textit{integrable} ($a = 1$)
case [5,~2].

The diligent reader will also verify the consistency of these ODEs, (D.12),
(D.13) and (D.7), with the solutions in the \textit{free} ($a = 0$) case,
when of course, for all $n = 1,\ldots,N$ with arbitrary $N$, $\zeta _{n} \left(
{\tau}  \right) = \zeta _{n} \left( {0} \right) + {\zeta} '_{n} \left( {0}
\right)\tau $ (see (1.7) and (D.6), with $a = 0$), and the coefficients
$\gamma _{n} \left( {\tau}  \right)$ have the corresponding expressions
entailed by the well-known explicit formulas expressing the coefficients of
a polynomial in terms of its zeros, see (D.1). The diligent reader will also
obtain the solutions of (D.12) that correspond to the two-body problems
treated in Section~4, and the solutions of (D.13) respectively (D.7) that
correspond to the similarity solutions (with $N = 3$ respectively arbitrary~$N$) of Section~3.

The results of~[11] entail the \textit{conjecture} that, if $a = -
1/2$, the solutions of the two nonlinear coupled nonautonomous ODEs (D.13)
with (D.10b) possess the (``super Painlev\'e'') property to \textit{only}
feature solutions which are \textit{entire} functions of the independent
variable~$\tau $.

Let us now report tersely the analogous results relevant to the equations of
motion~(1.5) rather than (1.7). To this end we set
\begin{equation}
\Pi _{N} \left( {\zeta ,\tau}  \right) = P_{N} \left( {z,t}
\right),\qquad \zeta = z,\qquad \tau = \left[ {\exp\left( {i\omega t}
\right) - 1} \right]/\left( {i\omega}  \right),
\end{equation}
as suggested by (1.6), and we then note that the nonlinear PDE (D.7) can now
be re-written as follows:
\begin{equation}
\left( {P_{tt} - i\omega P_{t}}  \right)\left[ {P_{z}}  \right]^{2}
+ \left( {1 - a} \right)\left\{P_{zz} \left( {P_{t}}
\right)^{2} - P_{z} \left[\left( {P_{t}}  \right)^{2} \right]_{z}  \right\} = QP.
\end{equation}
Here we introduced via
\begin{equation}
Q \equiv Q\left( {z,t} \right) = \exp\left( { - 2i\omega t}
\right)\Psi \left( {\zeta ,\tau}  \right)
\end{equation}
the polynomial $Q\left( {z,t} \right)$ of degree $2N - 4$ in the
variable $z$,
\begin{equation}
Q\left( {z,t} \right) = \sum\limits_{k = 0}^{2N - 4} {} q_{k} \left( {t}
\right)z^{N - k}.
\end{equation}
We do not elaborate any further on the polynomial solutions of the PDE
(D.15), since with trivial modifications what is written above after (D.7)
remains applicable, and it is as well quite easy to apply the change of
(independent) variable $\gamma _{n} \left( {\tau}  \right) = c_{n} \left(
{t} \right)$, see (D.14) and (D.1,2) in order to obtain from (D.10a), from
(D.12) and from (D.13) the corresponding ODEs for the coefficients $c_{n}
\left( {t} \right)$. We only note that we expect (D.15) to possess generally
a lot of solutions that are \textit{completely periodic} with period $T$,
see (1.2), in the (\textit{real}) independent variable $t$. And in
particular it is natural (see above) to \textit{conjecture} that, if $a = -
1/2$, \textit{all} polynomial solutions $P\left( {z,t} \right)$ of degree
$N = 3$ of the nonlinear PDE (D.15) are \textit{completely periodic} with
period $T$ in the (\textit{real}) independent variable $t$.

Finally, let us obtain (D.7) from (D.1) and (1.7) with (D.6).

Differentiation of (D.1) with respect to $\tau $ yields
\begin{subequations}
\begin{equation}
\sum\limits_{m = 1}^{N} {} {\gamma} '_{m} \left( {\tau}  \right)\zeta ^{N
- m} = - \sum\limits_{n = 1}^{N} {} {\zeta} '_{n} \left( {\tau}
\right)\prod\limits_{l = 1;\, l \ne n}^{N} {} \left[ {\zeta - \zeta _{l} \left(
{\tau}  \right)} \right],
\end{equation}
and a further differentiation yields
\begin{gather}
\sum\limits_{m = 1}^{N} {\gamma}''_{m}(\tau)\zeta^{N-m} =
- \sum\limits_{n = 1}^{N} {\zeta}''_{n}(\tau) \prod\limits_{l = 1;\, l \ne n}^{N}
[\zeta-\zeta_{l}(\tau)]
\nonumber\\
\qquad {}+ \sum\limits_{n = 1}^{N}{\zeta}'_{n}(\tau)
\sum\limits_{l = 1;\, l \ne n}^{N}{\zeta}'_{l}(\tau)
\prod\limits_{k = 1;\, k \ne l,n}^{N} [\zeta-\zeta_{k}(\tau)].
\end{gather}
\end{subequations}
We now set $\zeta = \zeta _{n} $ in (D.18a) and we thereby get
\begin{equation}
{\zeta} '_{n} \left( {\tau}  \right) = - \sum\limits_{m = 1}^{N} {} {\gamma
}'_{m} \left( {\tau}  \right)\left[ {\zeta _{n} \left( {\tau}  \right)}
\right]^{N - m}/\Pi _{\zeta}  \left[ {\zeta _{n} \left( {\tau}
\right),\tau}  \right],
\end{equation}
where we used the relation
\begin{equation}
\Pi _{\zeta}  \left[ {\zeta _{n} \left( {\tau}  \right),\tau}  \right]
\equiv \partial \Pi \left( {\zeta ,\tau}  \right)/\partial \zeta
\left. {} \right|_{\zeta = \zeta _{n} \left( {\tau}  \right)} =
\prod\limits_{l = 1;\, l \ne n}^{N} {} \left[ {\zeta _{n} \left( {\tau}
\right) - \zeta _{l} \left( {\tau}  \right)} \right].
\end{equation}
Likewise, by setting $\zeta = \zeta _{n} $ in (D.18b) we get
\begin{gather}
-{\zeta}''_{n}(\tau) + 2{\zeta}'_{n}(\tau)
\sum\limits_{l = 1;\, l \ne n}^{N} {\zeta}'_{l}(\tau)/
[\zeta_{n}(\tau) - \zeta_{l}(\tau)] \nonumber\\
\qquad {}= \sum\limits_{m = 1}^{N} {\gamma}''_{m}(\tau)
[\zeta_{n}(\tau)]^{N - m}/
\Pi_{\zeta}[\zeta_{n}(\tau),\tau] ,
\end{gather}
where we used again (D.20).

We now use the equations of motion (1.7) and we thereby get
\begin{gather}
2{\zeta}'_{n}(\tau) \sum\limits_{l = 1;\, l \ne n}^{N}
(1 - a_{nl}){\zeta}'_{l}(\tau)/
[\zeta_{n}(\tau)-\zeta_{l}(\tau)] \nonumber\\
\qquad {}=\sum\limits_{m = 1}^{N}{\gamma}''_{m}(\tau)
[\zeta_{n}(\tau)]^{N - m}/\Pi_{\zeta}[\zeta_{n}(\tau),\tau],
\end{gather}
and via (D.19a) this yields
\begin{subequations}
\begin{gather}
2\sum\limits_{m = 1}^{N} {} {\gamma} '_{m} \left( {\tau}  \right)\left[
{\zeta _{n} \left( {\tau}  \right)} \right]^{N - m}\sum\limits_{l = 1;\,l
\ne n}^{N} {} \left( {1 - a_{nl}}  \right)\sum\limits_{k = 1}^{N} {}
{\gamma} '_{k} \left( {\tau}  \right)\left[ {\zeta _{l} \left( {\tau}
\right)} \right]^{N - k}\nonumber\\
\qquad {}\times{}
\left\{ {\left[ {\zeta _{n} \left( {\tau}
\right) - \zeta _{l} \left( {\tau}  \right)} \right]\Pi _{\zeta}  \left[
{\zeta _{l} \left( {\tau}  \right),\tau}  \right]} \right\}^{ - 1}
= \sum\limits_{m = 1}^{N} {} {\gamma} ''_{m} \left( {\tau}  \right)\left[
{\zeta _{n} \left( {\tau}  \right)} \right]^{N - m}.
\end{gather}

From now on we restrict consideration to the equal-coupling-constants case,
see (D.6), and we re-write (D.23a) as follows:
\begin{gather}
2\sum\limits_{m,k = 1}^{N} {} {\gamma} '_{m} \left( {\tau}
\right){\gamma} '_{k} \left( {\tau}  \right)\left[ {\zeta _{n} \left(
{\tau}  \right)} \right]^{N - m}\sum\limits_{l = 1;\,l \ne n}^{N} {} \left( {1
- a} \right)\left[ {\zeta _{l} \left( {\tau}  \right)} \right]^{N -
k}\nonumber\\
\qquad{}\times
\left\{ {\left[ {\zeta _{n} \left( {\tau}  \right) - \zeta _{l}
\left( {\tau}  \right)} \right]\Pi _{\zeta}  \left[ {\zeta _{l} \left(
{\tau}  \right),\tau}  \right]} \right\}^{ - 1}
= \sum\limits_{m = 1}^{N} {} {\gamma} ''_{m} \left( {\tau}  \right)\left[
{\zeta _{n} \left( {\tau}  \right)} \right]^{N - m}.
\end{gather}
\end{subequations}

We now note that the rational function of $\zeta $,
\begin{equation}
F_{nl} \left( {\zeta}  \right) = \zeta ^{N - k}\left[ {\left( {\zeta
_{n} - \zeta}  \right)\Pi \left( {\zeta}  \right)} \right]^{ - 1},
\end{equation}
where $\Pi \left( {\zeta}  \right)$ is the polynomial of degree $N$ that has
the $N$ zeros $\zeta _{n} $ (see (D.1)) and $k$ is an arbitrary
\textit{positive integer} not larger than $N$, $k = 1,\ldots,N$, vanishes
faster than $\zeta ^{ - 1}$ as $\zeta \to \infty $, so that the sum of all
its residues vanishes. Since this rational function has $N - 1$ simple poles
(at $\zeta = \zeta _{m}$, $m = 1,\ldots,N$, $m \ne n$) and a double pole (at
$\zeta = \zeta _{n} $), this implies the \textit{identity}
\begin{gather}
\sum\limits_{l = 1;\,l \ne n}^{N} {} \zeta _{l} ^{N - k}\left[ {\left(
{\zeta _{n} - \zeta _{l}}  \right)\Pi _{\zeta}  \left( {\zeta _{l}}
\right)} \right]^{ - 1}
\nonumber\\
\qquad {}= (N-k)  \zeta_{n}^{N-k-1}  [\Pi_{\zeta}(\zeta_{n})]^{-1}
-(1/2)  \zeta_{n}^{N-k}  \Pi_{\zeta \zeta}(\zeta_{n})
[\Pi_{\zeta} (\zeta_{n})]^{-2} .
\end{gather}

Via this identity (D.23b) yields
\begin{gather}
\left( {1 - a} \right)\sum\limits_{m,k = 1}^{N} {} {\gamma} '_{m} \left(
{\tau}  \right){\gamma} '_{k} \left( {\tau}  \right)
 \Big\{ 2\left( {N - k} \right)\left[ {\zeta _{n}
\left( {\tau}  \right)} \right]^{2N - m - k - 1}\Pi _{\zeta}  \left[
{\zeta _{n} \left( {\tau}  \right),\tau}  \right]\\
\qquad {}- \left[ {\zeta _{n}
\left( {\tau}  \right)} \right]^{2N - m - k}\Pi _{\zeta \zeta}  \left[
{\zeta _{n} \left( {\tau}  \right),\tau}  \right] \Big\}
= \left( {\Pi _{\zeta}  \left[ {\zeta _{n} \left( {\tau}  \right),\tau}
\right]} \right)^{2}\sum\limits_{m = 1}^{N} {} {\gamma} ''_{m} \left(
{\tau}  \right)\left[ {\zeta _{n} \left( {\tau}  \right)} \right]^{N - m}. \nonumber
\end{gather}

We now note that the definition (D.1) of the polynomial $\Pi \left( {\zeta
,\tau}  \right)$ implies the relations
\begin{subequations}
\begin{gather}
\sum\limits_{m,k = 1}^{N} {} {\gamma} '_{m} \left( {\tau}  \right){\gamma
}'_{k} \left( {\tau}  \right)\left[ {\zeta _{n} \left( {\tau}  \right)}
\right]^{2N - m - k} = \left[ {\Pi _{\tau}  \left( {\zeta ,\tau}
\right)} \right]^{2}\left. {} \right|_{\zeta = \zeta _{n} \left( {\tau}
\right)},\\
\sum\limits_{m,k = 1}^{N} {} {\gamma} '_{m} \left( {\tau}  \right){\gamma
}'_{k} \left( {\tau}  \right)2\left( {N - k} \right)\left[ {\zeta _{n}
\left( {\tau}  \right)} \right]^{2N - m - k - 1}
\\
\qquad = \sum\limits_{m,k = 1}^{N} {\gamma}'_{m} \left( {\tau}  \right)
{\gamma}'_{k} \left( {\tau } \right)\left( {2N - m - k} \right)
\left[ {\zeta _{n} \left( {\tau} \right)} \right]^{2N - m - k - 1}
=  \left\{ \left[ {\Pi _{\tau}  \left( {\zeta,\tau}  \right)}
\right]^{2} \right\}_{\zeta}  \Big|_{\zeta = \zeta_{n}
\left( {\tau}  \right)}\!,\nonumber
\end{gather}
\end{subequations}
where the first step in (D.27b) is of course justified by the possibility to
exchange the dummy indices $m$ and $k$. Hence from (D.26) and (D.1) we infer
that the polynomial
\begin{equation}
\left( {1 - a} \right)\left\{ \Pi _{\zeta}  \left( {\left[
{\Pi _{\tau} }  \right]^{2}} \right)_{\zeta}  - \Pi _{\zeta \zeta}
\left[ {\Pi _{\tau} }  \right]^{2} \right\} - \Pi _{\tau \tau}
\left[ {\Pi _{\zeta} }  \right]^{2},
\end{equation}
of degree $3N - 4$ in $\zeta $, vanishes at the $N$ points $\zeta = \zeta
_{n} $, namely that it has the same zeros as the polynomial of $N$ degree
$\Pi $, see (D.1), and this clearly implies the validity of (D.7).

\label{calogero-lastpage}

\end{document}